\documentclass[structabstract]{aa}  
\usepackage{graphicx}
\usepackage{txfonts}
\usepackage{natbib}
\usepackage{longtable}
\usepackage{lscape}
\usepackage{url}
\bibpunct{(}{)}{;}{a}{}{,} 
\begin{document}
\title{The optically bright post-AGB population of the LMC}


\author{E. van Aarle\inst{1}
 \and H. Van Winckel\inst{1}
 \and T. Lloyd Evans\inst{2}
 \and T. Ueta\inst{3}
 \and P. R. Wood\inst{4}
 \and A. G. Ginsburg\inst{5}
}

\institute{Instituut voor Sterrenkunde, K.U. Leuven, Celestijnenlaan 200D bus 2401, B-3001 Leuven, Belgium \\
 \email{els.vanaarle@ster.kuleuven.be}
 \and SUPA, School of Physics \& Astronomy, University of St Andrews, North Haugh, St Andrews KY16 9SS, Scotland, UK
 \and Department of Physics and Astronomy, University of Denver, 2112 E. Wesley, Denver, CO 80208, USA
 \and Mount Stromlo Observatory, Cotter Road, Weston Creek, ACT 2611, Australia
 \and Department of Astrophysical and Planetary Sciences, University of Colorado, Boulder, CO 80309, USA
}

\date{Received September 29, 2010; accepted March 16, 2011}

 
\abstract
{The detected variety in chemistry and circumstellar shell morphology of the limited sample of Galactic post-Asymptotic Giant Branch (AGB) stars is so large that there is no consensus yet on how the different objects are linked by evolutionary channels. The evaluation is complicated by the fact that their distances and hence luminosities remain largely unknown.
}
{We construct a catalogue of the optically bright post-AGB stars in the Large Magellanic Cloud (LMC). The sample forms an ideal testbed for stellar evolution theory predictions of the final phase of low- and intermediate-mass stars, because the distance and hence luminosity and also the current and initial mass of these objects is well constrained.
}
{
Via cross-correlation of the Spitzer SAGE catalogue with optical catalogues we selected a sample of LMC post-AGB candidates based on their $[8]-[24]$ colour index and estimated luminosity. We determined the fundamental properties of the central stars of 105 of these objects using low-resolution, optical spectra that we obtained at Siding Spring Observatory and SAAO.
}
{We constructed a catalogue of 70 high probability and 1337 candidate post-AGB stars that is available at the CDS\thanks{via anonymous ftp to cdsarc.u-strasbg.fr (130.79.128.5) or via http://cdsweb.u-strasbg.fr/cgi-bin/qcat?J/A+A/}. About half of the objects in our sample of post-AGB candidates show a spectral energy distribution (SED) that is indicative of a disc rather than an expanding and cooling AGB remnant. Like in the Galaxy, the disc sources are likely associated with binary evolution.  Important side products of this research are catalogues of candidate young stellar objects, candidate supergiants with circumstellar dust, and discarded objects for which a spectrum was obtained. These too are available at the CDS.
}
{}


\maketitle

\section{Introduction}

During their final evolution, stars of low- and intermediate-mass transit rapidly from the asymptotic giant branch (AGB) along the post-AGB passage towards the planetary nebula (PN) phase, before the stellar remnant cools down as a white dwarf (WD). In these late phases, it is the external mass-loss that governs stellar evolution and it is well known that the accumulated mass-loss of lower-mass stars is one of the major contributions to the interstellar medium (ISM) enrichment in gas and dust.  This recycling of gas and dust between the ISM and stars is one of the strong evolutionary drivers of a galaxy's visible matter and its chemical and spectral characteristics \citep[e.g.,][]{meixner06}. 

Although this scheme of the late phases of stellar evolution may be generally acknowledged, there is no understanding from first principles of several important physical processes that govern these evolutionary phases. The main shortcomings are related to the lack of understanding of the mass-loss mechanisms and mass-loss rate along the AGB ascent \citep[e.g.,][]{hoefner09}, the subsequent shaping processes of the circumstellar shells \citep[e.g.,][]{demarco09}, and the lack of fundamental understanding of the internal chemical evolution of these stars \citep[e.g.,][]{herwig05}. \newline

During the last few years it has been repeatedly shown that research on post-AGB stars can lead to surprising results regarding chemical evolution. Post-AGB stars are chemically much more diverse than previously thought. Some objects are the most s-process-enriched objects known to date \citep[e.g.,][and references therein]{reyniers04} while others are not enriched at all \citep[e.g.,][]{luck90,takeda07}. In addition, the photospheric composition of some post-AGB stars shows some degree of depletion of refractory elements: elements with a high dust condensation temperature are systematically underabundant \citep[e.g.,][]{giridhar05, maas05, hinkle07}. In almost all depleted post-AGB objects, there is observational evidence of a stable Keplerian circumbinary disc, which resembles protoplanetary discs around young stellar objects (YSOs) \citep{deruyter06}. The discs are spatially resolved, but this is only detectable using interferometric techniques \citep{deroo06,deroo07}, and they typically have a radius of only 50~AU at 10~$\mu$m.  The high dust-grain processing is an observational indication that the compact discs are stable \citep{deruyter05,gielen08,gielen09b}. Since the discovery of the binary nature of a small sample of the most depleted objects \citep{vanwinckel95}, this binary nature of the disc sources has been further confirmed \citep{vanwinckel09}. The orbital elements imply a phase of strong interaction when the evolved star was at giant dimensions. It is postulated that the disc was formed during this phase \citep{waters92}. 

Another key question in stellar evolution is the understanding of the physical processes driving and shaping the whole circumstellar environment (CSE). This is in part because of the high spatial and dynamical complexity of many of these structures, as illustrated abundantly by the optical images of PNe in scattered light and in nebular optical emission lines \citep[e.g.,][and references therein]{balick02}. A major discussion topic in the PNe community is the growing evidence that PN physics is in fact largely determined by processes only found in binary stars \citep[e.g.,][]{demarco09}.

The terms 'proto-planetary nebula' (PPN) and 'post-AGB star' are often used as synonyms, but in this contribution we restrict the use of 'PPN' to objects with a resolved circumstellar shell. The PPNe display a surprisingly wide variety in shape and structure in thermal emission and in scattered light \citep[e.g.,][and references therein]{meixner99, ueta00, sahai07, siodmiak08, verhoelst09}, which allows direct probing of the shells. They are better suited to study the AGB mass-loss process than are PNe because their structures are not yet affected by the ionisation front. In PNe, this front has swept through the bulk of the optically bright regions, and therefore the density structure in these regions does not necessarily reflect the history of AGB mass-loss and the shell structure shaping processes that might have occurred before the sweeping by the ionisation front. Most PPNe have axisymmetric structures with a wide range of pole-to-equator density gradients \citep{meixner02, ueta03}. Moreover, in many objects, molecular collimated outflows are present very early after the AGB evolution. These winds carry a linear momentum that is too big to be driven by the radiation pressure on the dust particles \citep{bujarrabal01}. Symmetry breaking of the spherical winds must hence occur at the end of the AGB phase, but the primary driving forces remain unclear.

In addition to gaining insight into their own evolutionary phase \citep[e.g.,][]{vanwinckel03}, post-AGB star studies can also help us to better understand the AGB-evolution of which they are the final product. Thanks to the broad  spectral range of their electromagnetic spectrum, a simultaneous study of the stellar photosphere and the circumstellar environment is possible: the central star is responsible for the UV and optical emission while the cool circumstellar envelope radiates in the infrared (IR). \newline

Post-AGB stars evolve on a very fast track \citep{vassiliadis93, blocker95} and they are rare as a consequence. In the web catalogue of Tor\'un  \citep{szczerba07} only 326 Galactic objects are listed as good candidate post-AGB stars. One of the biggest problems in confronting the wide variety in observational characteristics of well-studied individual objects with stellar evolution theory is that distances and hence intrinsic luminosities are largely unknown. Therefore, there is no clear consensus on how the many detailed studies of individual objects might be linked by evolutionary channels. Thanks to the recent IR surveys we can now overcome this Galactic limitation by studying the post-AGB population in other galaxies.  In this paper we focus on the Large Magellanic Cloud (LMC).  Because of its proximity \citep[50~kpc,][]{feast99} and favourable angle \citep[$35^{\circ}$,][]{vandermarel01}, all LMC stars can be considered to have a similar distance from the Sun, hence, their luminosities and associated masses are directly measurable.  

The paper is organised as follows: In Sect.~\ref{sec:sampleselection} we explain our selection criteria to obtain a catalogue of post-AGB candidates inspired by the spectral energy distributions (SEDs) of Galactic sources. The diversity of Galactic post-AGB stars is so large that IR colour selection alone, even when combined with a luminosity criterion, does not allow us to unambiguously identify post-AGB stars without spectroscopic confirmation.  We therefore obtained a large number of low-resolution, optical spectra to investigate the central objects of the brightest post-AGB candidates. The results of this survey are discussed in Sect.~\ref{sec:spectra}. In Sect.~\ref{sec:analysis} we analyse different characteristics of the entire sample. We subsequently evaluate our endeavour in Sect.~\ref{sec:evaluation}. The catalogue of post-AGB candidates is available at the CDS.

\section{Sample selection} \label{sec:sampleselection}

\subsection{SAGE dataproducts}

Post-AGB stars are expected to be bright IR emitters because the relic of the mass-loss on the AGB is still able to convert the photospheric radiation efficiently into warm, thermal emission.  Hence, systematic searches to identify post-AGB stars became possible only when mid- to far-IR experiments were started in the mid 80's \citep[e.g.,][]{kwok93}. The IRAS mission launched in 1983 was most successful in enabling systematic identification of post-AGB candidates, and the ($[12]-[25]$, $[25]-[60]$) colour-colour diagram has been used extensively as a basis for ground-based follow-up to identify objects between the locus of the PNe and the late-type AGB stars \citep[e.g.,][]{kwok87, volk89, vanderveen89, manchado89, hu93, garcialario97, vandesteene00}. Some of these objects are optically fairly faint because the selection criteria focus on the expected properties of the expanding dust shell. Other systematic searches focused on cross-correlating optical catalogues with the IRAS point-source  catalogue in search of luminous objects with an IR excess due to thermal radiation of the circumstellar dust \citep[e.g.,][]{hrivnak89, pottasch88, trams91, Oudmaijer92}. These objects display a wider variety of IR-colours and are scattered over the IRAS colour-colour diagram as a consequence.

These large-scale mid-IR surveys were, however, only useful for Galactic post-AGB stars because the resolution and sensitivity remained too limited to study objects in the LMC.  The Spitzer \citep{werner04} \textit{Surveying the Agents of a Galaxy's Evolution} (SAGE) project for the first time allows a global study of the LMC at high enough spatial resolution and sensitivity to resolve individual objects at a range of mid-IR wavelengths. SAGE mapped a $7^{\circ} \times 7^{\circ}$ region of the LMC using the IRAC camera at wavelengths of 3.6, 4.5, 5.8 and 8.0~$\mu$m \citep{fazio04} and the MIPS camera at 24, 70 and 160~$\mu$m \citep{rieke04}. The survey was performed over two epochs in 2005 separated by three months, to diminish the instrumental artefacts and trace the source variability. The details of the survey are described in \citet{meixner06}. In total, over 6.9 million sources were detected with IRAC and about 40\,000 with MIPS. In this paper, we use the SAGE second release data products (September 2009) that combines both visits of the survey.  The Spitzer IRAC sources are band-merged with the Two Micron All Sky Survey (2MASS) in filters $J$, $H$ and $K_s$ (1.24, 1.66, and 2.16~$\mu$m) \citep{skrutskie06} and can be found at the websites of the Spitzer Science Centre (SSC) and NASA/IPAC \footnote{\texttt{http://irsa.ipac.caltech.edu/data/SPITZER/SAGE/}}.

\subsection{Colour criteria: optically bright post-AGB stars.}\label{ssec:colcrit}

\begin{figure}
\vspace{0cm}
\hspace{0cm}
\resizebox{\hsize}{!}{\includegraphics{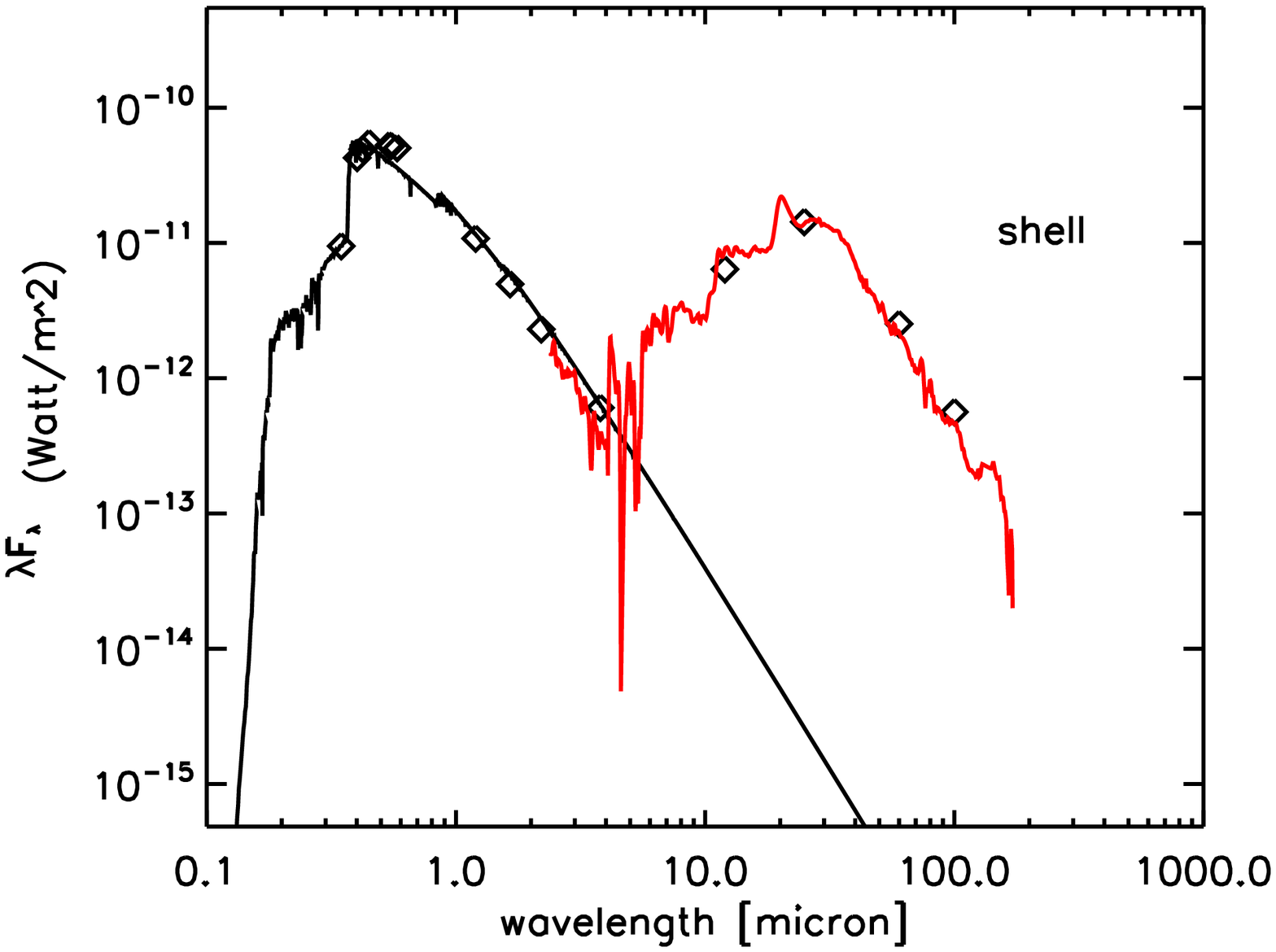}}
\resizebox{\hsize}{!}{\includegraphics{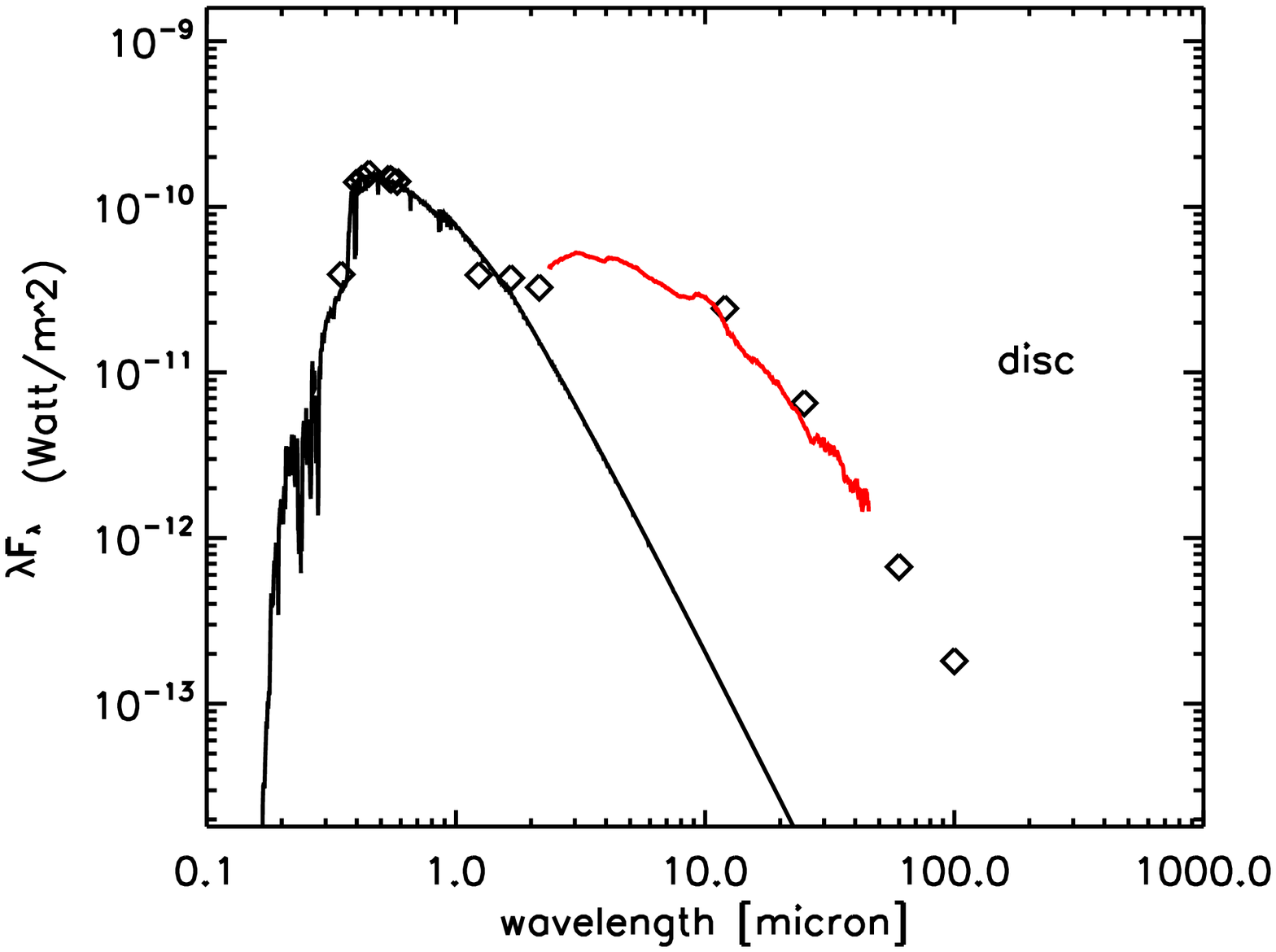}}
\caption{\label{fig:SEDGalactic} SEDs of the Galactic post-AGB stars HD~56126 (upper panel) and 89~Her (lower panel). The former is a carbon-rich post-AGB star and the latter an optically bright post-AGB star surrounded by a compact disc. The diamonds represent the dereddened broadband photometry, the full line at lower wavelengths the appropriate photospheric model, and the full line in the IR is the ISO spectrum.}
\end{figure}

Based on the shape of their SED and more specifically on their behaviour in the near-IR, we can roughly divide the Galactic post-AGB stars into two distinct classes. The first one consists of objects with a double-peaked SED of which the peak at longer wavelengths corresponds to the emission of an expanding, detached shell of cooling dust and gas. This shell is the remnant of the AGB mass-loss. Depending on the strength of the circumstellar reddening, the central star is directly detectable at shorter wavelengths or obscured by the circumstellar material. Some Galactic post-AGB stars are strongly extincted by a strong equatorial density enhancement, or torus, in the CSE \citep[e.g.,][]{meixner99, ueta00}. These objects lack a direct photospheric contribution, and the SED is completely dominated by thermal emission of dust.

The SEDs of the second type of object show a broad excess with an onset in the near-IR, which indicates that there is still hot dust in the system. These SEDs are indicative of a disc rather than an expanding envelope \citep{deruyter06}. Because these objects turned out not to be rare at all in our Galaxy \citep{vanwinckel07}, we incorporate their typical IR colours in our initial selection criteria. 

In Fig.~\ref{fig:SEDGalactic} we show some illustrative examples of Galactic post-AGB stars. In the upper panel, the SED of HD~56126, a carbon-rich post-AGB star \citep{vanwinckel00} with a dust excess typical of a carbon-rich, expanding shell is displayed \citep{hony03}. The lower panel shows the SED of 89~Her, an optically  bright post-AGB star surrounded by a compact disc. Around this object, an hourglass-shaped nebula has been detected in CO emission \citep{bujarrabal07}. \newline

\begin{figure}
\resizebox{\hsize}{!}{ \includegraphics{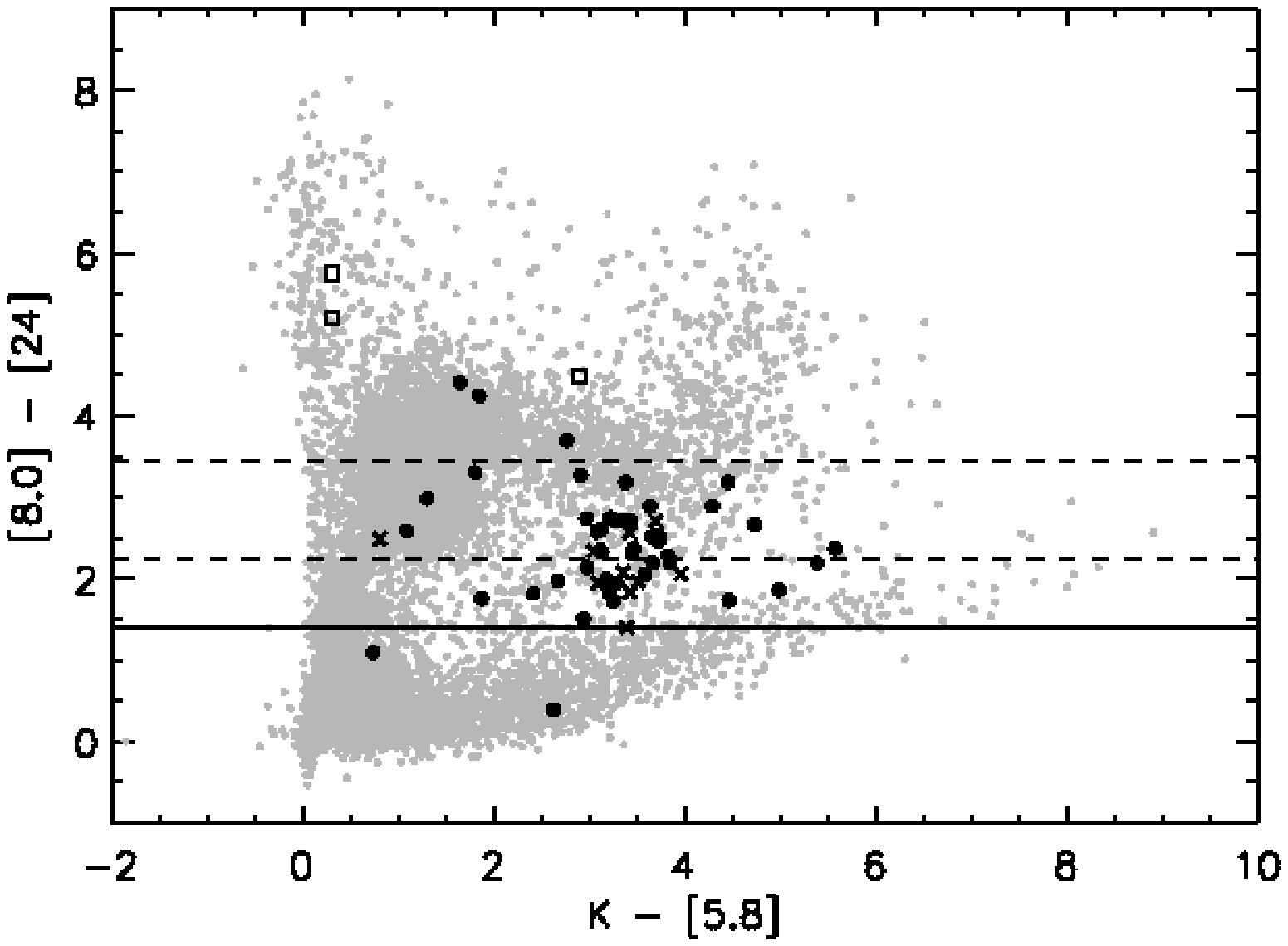}}
\caption{Colour-colour plot of $K-[5.8]$ versus $[8]-[24]$. The small grey dots in the background are the sources of the SAGE catalogue, the black crosses represent the 11 RV~Tauri stars in the LMC from the OGLE-III catalogue \citep{soszynski08} that are detected at 8 and 24~$\mu$m and the black dots and squares are the Galactic disc sources from \citet{deruyter06} and some Galactic shell sources observed with the IR spectrometer SWS onboard ISO transformed onto the Spitzer filters. The black solid line indicates the colour-cut we are using to select likely post-AGB candidates, the lower dashed line indicates the lower limit for post-AGB candidates with a shell and the upper dashed line the upper limit for post-AGB candidates with a disc (see Sect.~\ref{ssec:subsamples}). 
}
\label{fig:rvtau}
\end{figure}

To select the candidate post-AGB stars in the SAGE catalogue, we adopted the following strategy. First we cross-correlated the epoch 1 and epoch 2 MIPS 24~$\mu$m catalogues with a maximal position shift of 3\arcsec, and we only retained those objects that were detected in both epochs. For these sources, the average of both flux measurements was taken with the error corresponding to the standard deviation of both values with respect to this mean. We then selected all those objects with a valid 8~$\mu$m detection from the IRAC archive catalogue and subsequently cross-matched the IRAC sources with the MIPS selection, again allowing a maximal shift of 3\arcsec \  between the IRAC position and either MIPS position. When multiple matches were found, we only retained the closest one. At this stage we had 25\,194 sources.

Next, we narrowed our sample by applying a colour selection inspired by what is known from Galactic sources. We transformed the SEDs of the known Galactic disc sources \citep{deruyter06} onto the filters of the IRAC and MIPS cameras and displayed them in a $K - [5.8]$ versus $[8] - [24]$ colour-colour plot (see Fig.~\ref{fig:rvtau}). Evidently, the background SAGE data points bifurcate into two populations at $[8.0] - [24]$ $\sim$ 3 to 5 and $\sim$ 0 to 2. According to \citet{blum06}, the former corresponds to mainly background galaxies while the latter consists of extreme AGB stars and red supergiants. The bulk of the known Galactic disc sources lies above this second branch, but overlaps with the first. Based on this figure, we impose that $F(24) > 0.4 \times F(8)$, with $F(24)$ the flux at 24~$\mu$m in mJy, or $[8]-[24] > 1.384$ as a further selection criterion (the solid black line in Fig.~\ref{fig:rvtau}). This way we select suspected shell sources, which have typically colder dust with $F(24) > F(8)$, as well as disc sources and discard the branch at $[8.0] - [24]$ $\sim$ 0 to 2 which consists of extreme AGB stars and red supergiants. This colour criterion is deliberately meant not to be very restrictive, and 16\,570 objects in our constructed SAGE catalogue fulfil these criteria. \newline

The sample was subsequently cross-correlated with three optical catalogues using a search radius of 1.5\arcsec \ from the IRAC position, and only retaining the objects of which an optical magnitude in the $U$, $B$, $V$, $R$ or $I$-filter was listed in at least one of the catalogues. We used Massey's $UBVR$ CCD survey of the Magellanic Clouds \citep{massey02a}, which covers a 14.5 deg$^2$ region of the LMC, is complete to $V \sim 15.7$, and has a magnitude limit of 18 in $V$. This cross-correlation brought forth 566 matches. The LMC stellar catalogue of \citet{zaritsky04} is more extensive than Massey's: it covers the central 64 deg$^2$ area of the LMC and is complete to $V \sim 20$ with a magnitude limit of 23 in this filter. This database contained 6767 mutual sources with our selection. The third and final catalogue we used is the Guide Star Catalogue Version 2.3.2 (GSC2.3.2) (STScI, 2006) \citep{GSC2.3.2}, which is an all-sky database that is complete to magnitudes of $R \sim 20$ with a magnitude limit of 20.5 in $R$. The correlation with this catalogue resulted in 4759 common objects. 

There are 7568 objects of our initial IR sample that are not cross-identified in the optical catalogues. The general trend in their SEDs suggests that this is mainly because they are too faint at optical wavelengths. Another 376 objects do occur in the optical catalogues, but have no listing of a $U$, $B$, $V$, $R$, or $I$ magnitude; most of these are found in the Guide Star Catalogue. After the removal of all objects without a listing of at least one of the required magnitudes, our total sample contains 8626 stars.

\subsection{Luminosity-cut}\label{ssec:lumcut}

\begin{figure}
\resizebox{\hsize}{!}{ \includegraphics{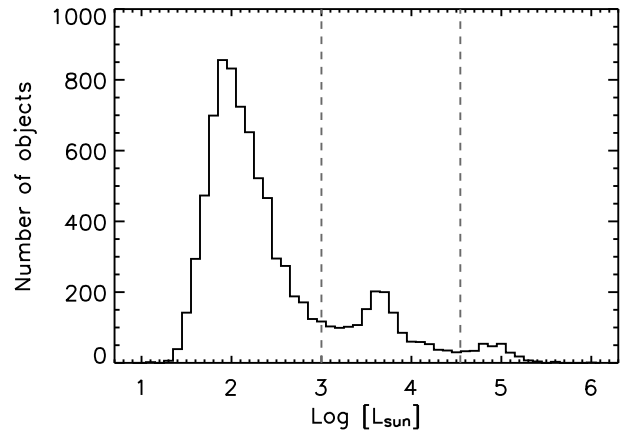}}
\caption{Histogram of the integrated luminosities based on black-body fittings of all objects in our sample before the luminosity-cut. The dashed grey lines indicate a lower and upper limit for the luminosity of post-AGB stars and are located at 1000 and 35\,000~$L_{\odot}$ respectively.}
 \label{fig:sghist}
\end{figure}

So far, all objects with an IR excess fulfilling the colour criteria and with optical photometry in one of the above three catalogues are selected. Obvious intruders in the sample are supergiants and YSOs, because these objects can also have a large IR excess. To refine our selection further we therefore imposed a luminosity criterion. Because we have no information on the spectral types of the different central objects, we simply integrated all flux in the system by fitting up to three black bodies (BB) to all available photometry without applying dereddening. To compute the luminosity from this total flux, we made use of a distance modulus of 18.54 \citep[e.g.,][]{keller06}. The black-body curves were fitted using a $\chi^2$ minimisation and by imposing that the flux up to the J-band belongs to the photospheric emission represented by the first BB. When a single BB fails to reproduce the IR part of the SED as well, a second and eventually third component is added with ever decreasing temperature. To account for the lack of photometry at wavelengths lower than 3600~\AA{} and to avoid fitting the Rayleigh-Jeans tail of the SED, we imposed an upper limit of 15000~K for the temperature of the hottest black body. This integral is a rough approximation of the true luminosity because it assumes spherical symmetry, the absence of ISM reddening, and an optically thin shell and imposes an upper limit to the photospheric temperature. A histogram of the results is given in Fig.~\ref{fig:sghist}.

The histogram of the calculated luminosities displays three clearly distinct peaks of which the luminosity values of the middle one agree with what we expect for the luminosity range of post-AGB stars based on the evolutionary tracks of \citet{blocker95}: they range from $\sim$ 1000 to 35\,000~$L_{\odot}$ (see Fig.~\ref{fig:HRdiagrdisc}). The peak at lower values corresponds to luminosities typical of bright YSOs, and we removed the bulk of them from the sample by introducing a luminosity-cut at 1000~$L_{\odot}$. This is not a strictly exclusive condition: as can be seen in Fig.~\ref{fig:sghist} we kept the tail of the first peak, and may consequently expect that some luminous YSOs are still present in our sample. Yet we attempt to construct a catalogue of optically bright post-AGB stars that is as complete as possible, so instead of removing practically all bright YSOs by introducing a higher lower-limit for the luminosity, we will eventually discard the remaining ones based on their spectra. The third peak at the highest luminosities corresponds to typical luminosities of supergiants. We disposed of the bulk of them by using a luminosity-cut at 35\,000~$L_{\odot}$. This corresponds to a $\mathrm{Log}(L/L_{\odot})$ of about 4.5 and is hence an upper limit for the luminosity of a 7~$M_{\odot}$ post-AGB star as has been computed by \citet{vassiliadis93} and \citet{blocker95}. After rejection of the objects of which the estimated luminosity does not fall within the predicted range for post-AGB stars, our sample still contains 1517 objects. 

We list the candidate YSOs and supergiants in two catalogues that are available at the CDS and contain the following information. For each catalogue, Column 1 gives the name of the source, Columns 2 and 3 contain the right ascension and declination, Columns 4-7 list the $U$ magnitude and its error from respectively \citet{zaritsky04} and \citet{massey02a}, the $B$ magnitudes and errors from these two catalogues and the Guide Star Catalogue \citep{GSC2.3.2} can be found in Columns 8-13 and the $V$ magnitudes and errors in Columns 14-19. Columns 20-23 give the $R$ magnitude from \citet{massey02a} and the $R_F$ magnitude from \citep{GSC2.3.2} with their errors, Columns 24 and 25 contain the $I$ magnitude and error from \citet{zaritsky04}, and Columns 26-31 list the 2MASS $J$, $H$ and $K_s$ magnitudes and errors. The 3.6, 4.5, 5.8 and 8.0~$\mu$m IRAC fluxes and errors are given in columns 32-39 and the MIPS 24, 70 and 160~$\mu$m fluxes and errors can be found in Columns 40-45. Column 46 lists the black-body-based luminosity.

\subsection{MIPS 70 and 160~$\mu$m data}

A cross-correlation with the MIPS 70 and 160~$\mu$m catalogues, SAGELMCfullMIPS70 and SAGELMCfullMIPS160, with a search radius of 3\arcsec \ reveals that 220 objects in our sample have a detection at 70~$\mu$m of which four are also listed in the 160~$\mu$m catalogue. With these extra data points, it can be seen that the SED of six of these sources is monotonically increasing towards wavelengths redder than 70~$\mu$m, which suggests that the bulk of the energy is radiated beyond this point. This is not typically seen in Galactic post-AGB stars, regardless of whether they have a disc or a shell SED. These objects are therefore probably either background galaxies or heavily obscured stars of another type, like YSOs with very cool optically thick dust. After their removal, our sample contains 1511 optically bright post-AGB candidates.

\subsection{Subsamples}\label{ssec:subsamples}

To assess the number of objects with a disc and those with a shell in our final sample, we introduced an upper limit of $F(24) < 3 \times F(8)$ for the flux at 24~$\mu$m for the former (see Fig.~\ref{fig:rvtau}). This implies that the star emits more flux at 8~$\mu$m than it does at 24~$\mu$m, and hence the dust will be closer to the photosphere of the central star. A lower limit of $F(24) > F(8)$ for objects with a shell was already introduced in Sect.~\ref{ssec:colcrit}.  We find that according to this extra cut, 665 objects are suspected to have a disc, 608 to have a shell and 238 objects fall in the grey zone where $F(8) < F(24) < 3 \times F(8)$ and no conclusion on their circumstellar geometry is possible based on these colour criteria. This shows that disc sources indeed form a significant fraction of the optically bright post-AGB stars with an IR excess. Note that we focus on objects detected in the optical, which means that heavily obscured post-AGB stars will be missing in our sample.

\section{Spectral information}

The colour and luminosity criteria we used are not very restrictive, so obviously not only post-AGB stars will occupy the sample. Similar IR colours can be detected from, amongst others, background galaxies, compact \ion{H}{II} regions, PNe, luminous YSOs and some remaining supergiants, as well as foreground stars. 

We therefore cross-correlated our sample of post-AGB candidates with catalogues of other known types of objects and obtained low-resolution, optical spectra of the brighter stars. This allowed us to eliminate these objects and to make an initial study of those which appear to be post-AGB stars.

\subsection{Other known types of objects} \label{ssec:remart}

Some objects in our sample have been studied before and are known not to be post-AGB stars. Many catalogues of other types of objects in the LMC exist, based on either colour criteria, the presence of certain features on photometric plates, or spectra in the infrared or the optical. We scanned the literature for publications that give spectral types of LMC objects that agree with the spectra taken by us (see Sect.~\ref{sec:spectra}) if these exist for some of the objects both catalogues have in common. Some other catalogues we could find are referenced in our final catalogue of post-AGB candidates (see Sect.~\ref{ssec:croscorr} and Appendix~\ref{app:catsample}).

Only three catalogues of LMC objects fulfilled these strict criteria: the list of massive stars of \citet{bonanos09}, the catalogue of cool carbon stars from \citet{kontizas01}, and the list of AGB stars classified by \citet{trams99}. \citet{kontizas01} detected 7760 carbon stars based on their strong Swan bands, the presence of which indicates that the stars are too cool to be post-AGB stars and hence must be still on the AGB. We unknowingly obtained spectra for two of these and a comparison of this type of carbon star with two genuine post-AGB carbon stars can be found in Sect.~\ref{ssec:Cstar}. Our sample contains 43 cool carbon stars, which are discarded.

Of the 57 objects classified as AGB stars by \citet{trams99}, seven are part of our sample. For only three of these, the AGB status is confirmed by optical spectra: we found two additional cool carbon stars and an AGB star with spectral type M5e. These three objects were removed from our sample.

Twenty-four post-AGB candidates are listed in the catalogue of 1750 massive stars of \citet{bonanos09}, of which 23 are reconcilable with a post-AGB star. Nineteen of these belong to the subclass of O and early B stars, three are B[e] or Be stars and the last source has a spectrum similar to an Of/Of?p object. The remaining object is a possible X-ray binary and is removed from our sample of post-AGB candidates. No luminous blue variables, Wolf-Rayet stars, or red supergiants were found. Our sample now contains 1464 optically bright post-AGB candidates.

\subsection{Low-resolution optical spectra}\label{sec:spectra}

In addition to this literature study, we obtained low-resolution, optical spectra of the brighter stars in our sample. So far, we have spectra for 105 of the brightest objects. The bulk of these spectra were obtained with the ANU 2.3~m telescope at the Siding Spring Observatory (SSO) in Australia during the fall of 2007 and the spring and fall of 2008. The spectra were obtained using the Dual Beam Spectrograph at a resolution of $\sim$~4~\AA{} from $\sim$ 3300 - 9500~\AA{}. In some cases only the red arm of the spectrograph was used at resolution of $\sim$~8~\AA{} from $\sim$~5300~-~10\,000~\AA{}. These 87 spectra of 86 objects were reduced following the standard recipes for long-slit spectral reduction. 

An additional 28 spectra of 24 objects were observed at SAAO with the 1.9~m Radcliffe telescope in late 2007, using the grating spectrograph with grating 8 in the red (6400 - 9200~\AA{}) with resolution 4~\AA{} or with grating 7 in the blue (3200 - 7200~\AA{}) with resolution 5~\AA{}. Most stars were only observed in one region. The spectra were reduced with IRAF, again according to the standard recipes.

The spectra were classified on the MK system, using the published spectra of \citet{jacoby84} as templates. For the objects observed at SAAO, we had a library of standard stars that were observed with the same instrumentation. The red SAAO spectra have fairly even sensitivity with wavelength, and the strengths of the classification features were measured. The dependence of these strengths on spectral type and luminosity class was established from the spectral type standards observed with the same instrument, and hence this subset of our LMC spectra were classified in a semi-quantitative way as well as by direct comparison with standards. The S/N of the blue SAAO spectra declines severely towards shorter wavelengths in the case of stars of type G and later, so that some features were of very low S/N or even missing. Classification of the blue SAAO spectra, as well as of the SSO spectra for which we did not have a set of standard stars taken with the same instrumental setup, was by visual comparison of the spectra with those of the standard stars. For galaxies, non-membership of the LMC was confirmed by a too high radial velocity.\newline

The low-resolution, optical spectra allow us to eliminate certain objects from the sample and enhance the probability that the remaining objects are indeed post-AGB stars. Of the 105 observed objects in our final sample, only 13 are rejected because they clearly cannot be post-AGB stars. We identified nine galaxies, two PN-like objects and two WC-like objects. These objects are discribed in Appendix ~\ref{app:spectra}. The remaining 92 objects are good post-AGB candidates and their spectra are discussed here.

\begin{figure}
\resizebox{\hsize}{!}{\includegraphics{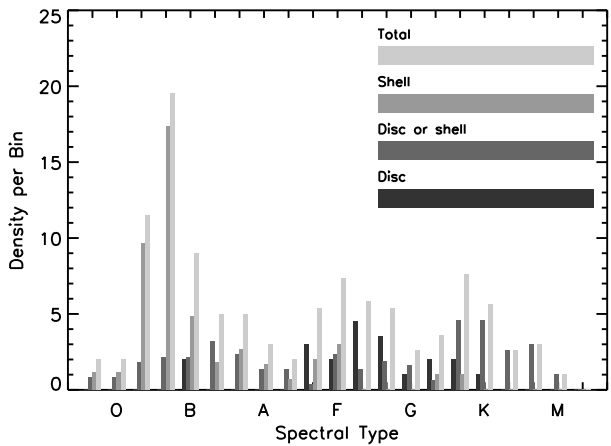}}
\caption{Histogram of the spectral types of the different groups of objects. Each spectral class is subdivided into three groups corresponding to objects of early, mid, and late type. If a post-AGB candidate's spectral type coincided with more than one bin, its weight was equally spread over all bins involved.}
\label{fig:sptdistr}
\end{figure}

The 92 spectral types determined by us together with the 23 additional spectral types from \citet{bonanos09} (see Sect.~\ref{ssec:remart}) cover a range conforming to what we expect from post-AGB stars. We detect all spectral types from late O to mid M, with number density peaks at B, F and K that correspond to shell sources, disc sources, and the group of objects in between (see Fig.~\ref{fig:sptdistr}). The spectral types of the disc sources range from F to mid K with two additional mid B[e] stars from the catalogue of \citet{bonanos09}, while those of the objects with a shell span the entire range from O to early~K with number density peaks among the earlier spectral types. The group of objects of which the SED shape does not allow us to conclude on the circumstellar environnement, cover all spectral types from O to M with a slight peak around the later types. For five objects some puzzling features in the spectra prevent us from determining the correct spectral type. Furthermore, we detect three carbon stars that remain good post-AGB candidates, and 23 objects with emission lines in their spectra. Both of these latter two groups of objects will now be discussed in more detail.

\subsubsection{Carbon stars} \label{ssec:Cstar}

\begin{figure*}
\centering
\includegraphics[width=17cm]{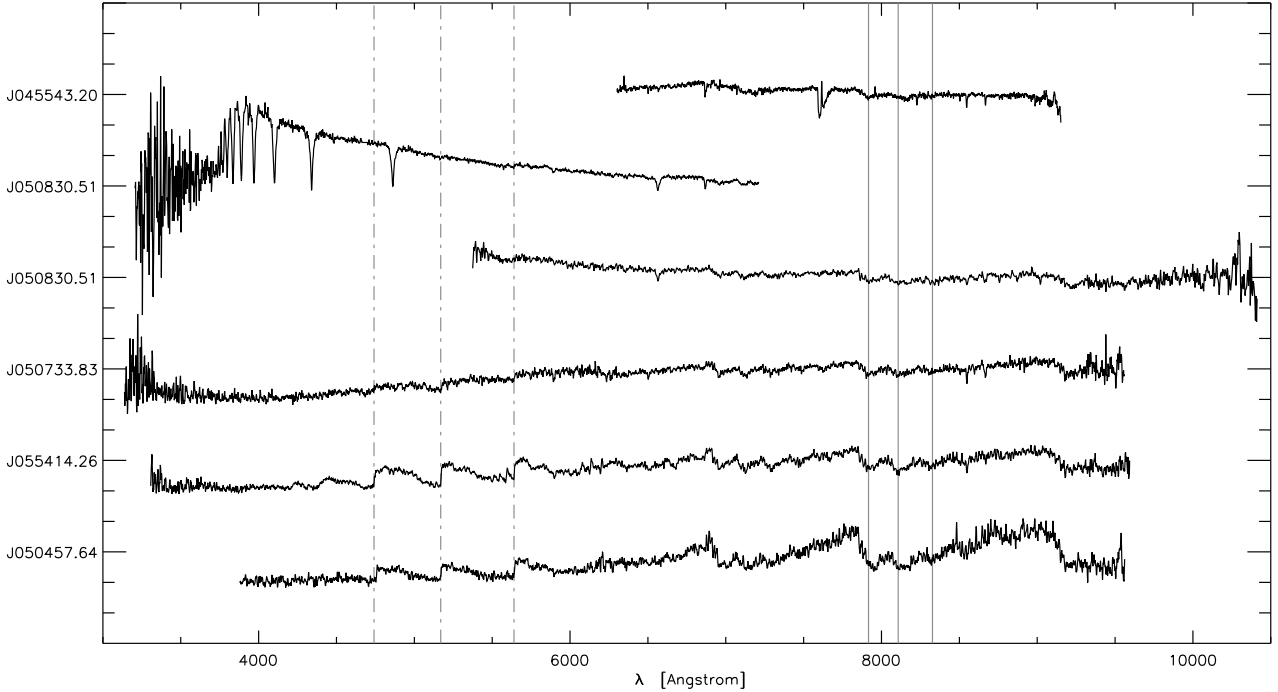}
\caption{Spectra of the five carbon stars for which a spectrum was obtained. J045543.20-675110.1, J050830.51-692237.4 and J050733.83-692119.9 remain part of our sample of post-AGB stars, but J055414.26-684020.4 and J050457.64-670511.9 are still on the AGB. All spectra were translated vertically in steps of 3 units to increase the visibility of the plot and normalised with respect to the region from 7500-8500~\AA{}. The upper spectrum of J050830.51-692237.4, which does not include this region, is scaled to emit the same amount of flux in the region from 6000-7000~\AA{} as the lower spectrum of this star. The vertical grey lines indicate the wavelengths at which spectral features of \element[][][][]{CN} (solid lines) and \element[][][][2]{C} (dot-dashed lines), indicators of \element[][][][]{C}-chemistry, are expected, corrected for the average radial velocity of the LMC \citep[262~km/s,][]{vandermarel02}.}
\label{fig:Cspectra_all}
\end{figure*}

We confirmed the spectral type of the known C-rich RV~Tauri star J045543.20-675110.1 \citep{pollard00} and found two new carbon stars -- J050733.83-692119.9 and J050830.51-692237.4 -- with strong \element[][][][]{CN} bands at 7910, 8100 and 8320~\AA{} in their spectra (see Fig.~\ref{fig:Cspectra_all}). In J050733.83-692119.9, these bands and the bands of \element[][][][2]{C} and \element[][][][]{CN} at shorter wavelengths are weaker than in the spectra of the two AGB stars -- J050457.64-670511.9 and J055414.26-684020.4 -- from the catalogue of \citet{kontizas01} (see Sect.~\ref{ssec:remart}), for which we have spectra. The bandstrengths are comparable with those in J045543.20-675110.1, the C-rich RV Tauri star, which is consistent with its status as a candidate post-AGB star of higher effective temperature than these AGB stars.

The red \element[][][][]{CN} bands in the spectrum of J050830.51-692237.4 are of a similar strength  as those in J050733.83-692119.9, but the carbon-star features become progressively weaker at shorter wavelengths as the spectrum rises steeply to the blue, as seen in Fig.~\ref{fig:Cspectra_all}. The strong bandhead at 5635~\AA{} is visible, though relatively weak, on the Siding Spring spectrum. The blue SAAO spectrum shows the band structure of the carbon star only at the extreme red end, near 7100~\AA{}. \element[][]{H}$\alpha$ is prominent in absorption and the blue spectral region shows only the spectrum of an A0-1IV star. These spectra thus appear to be of two different stars. The assigned luminosity class of the A star suggests that this is a Galactic foreground star, but the radial velocity of more than 200 km/s indicates LMC membership. The uncertain contribution of this star to the spectra prevents a decision as to whether the carbon star is an AGB star or a likely post-AGB star.

\begin{table}
\caption{Bolometric magnitudes and types of the carbon stars in our sample.}             
\label{table:cstars}      
\centering                          
\begin{tabular}{lcl}        
\hline\hline                 
Object Name (IRAC) & $M_{\mathrm{bol}}$ & Type\\    
\hline                        
J045543.20-675110.1 & -5.1 & G2-8(R)Ibe: \\
J050457.64-670511.9 & -4.7 & C-J: \\
J050733.83-692119.9 & -3.6 & C \\
J050830.51-692237.4 & -4.8 & C \\
J055414.26-684020.4 & -4.0 & C \\
\hline                                   
\end{tabular}
\end{table}

The bolometric magnitudes, derived from the luminosities calculated in Sect.~\ref{ssec:lumcut}, and types of the five carbon stars are given in Table~\ref{table:cstars}. The J classification of J050457.64-670511.9 is uncertain because the S/N at the diagnostic features is low. A comparison of these bolometric magnitudes with the observational carbon star luminosity function as published by \citet{marigo99} shows that two of these objects occur at the low-luminosity end and three are near the peak of this distribution, which extends from approximately $M_{\mathrm{bol}} = -3$ up to $M_{\mathrm{bol}} = -6.5$, with the peak located at about $M_{\mathrm{bol}} = -4.875$ (centre of the bin).

The three carbon stars that remain in our sample -- J045543.20-675110.1, J050733.83-692119.9 and J050830.51-692237.4 -- belong to the subgroup of post-AGB candidates of which the shape of the SED does not allow us to conclude on whether they have a disc or a shell. Carbon-rich Galactic post-AGB stars are mainly associated with shells \citep[e.g.,][]{kwok89}, while Galactic RV~Tauri stars with circumstellar dust generally have circumstellar discs \citep[e.g.,][]{lloydevans95, lloydevans97, vanwinckel99, gielen07}.

\subsubsection{Post-AGB candidates with an emission-line spectrum} \label{subsubsection:emlinespec}

\begin{figure*}
\centering
\includegraphics[width=17cm]{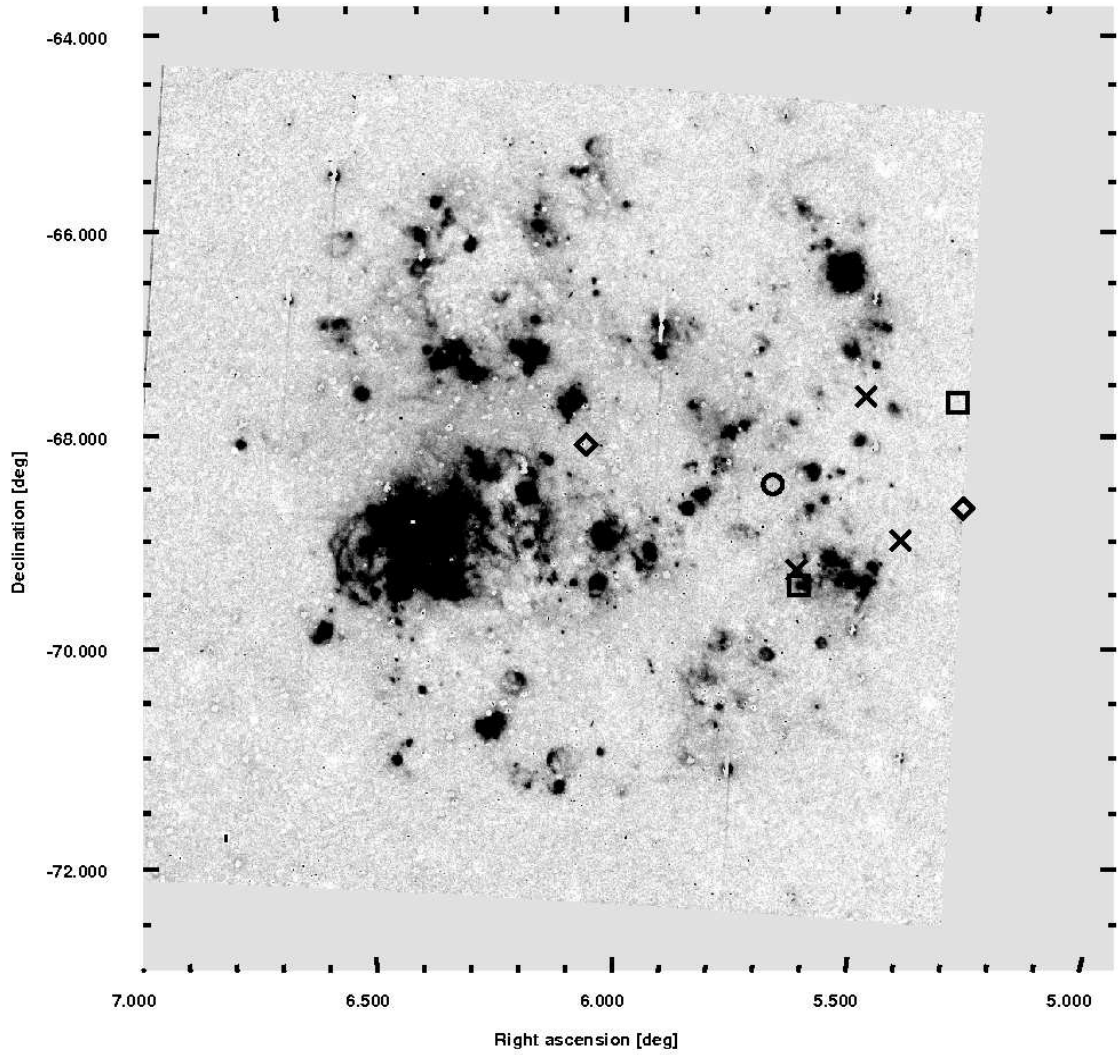}
\caption{Post-AGB candidates with emission lines in an \element[][]{H}$\alpha$ image of the LMC \citep{kennicutt95}. The circle denotes the object with the [\ion{O}{III}] features at 4959 and 5007~\AA{} in emission, they are in absorption for the two sources indicated with a square and are not present in the spectra of the two stars denoted with a diamond. The three crosses indicate the position of the other objects with an emission-line spectrum that did not cover the [\ion{O}{III}] feature.}
\label{fig:halpha_lmc}
\end{figure*}

Twenty-three objects display emission lines in their spectrum. The spectral type of these objects ranges from O to F-G, although for some objects the peculiarity of their spectrum did not permit us to determine the appropriate type. In Table~\ref{table:em} we summarise some of the most prominent features and Table~\ref{table:prinfeat_em} lists the principal features in the spectra of J045433.85-692036.2, J045729.38-661837.3, J054034.00-692509.8, J051525.45-661905.0, and J060323.20-680437.5, the five objects with the most emission features, and their identification. 

Two objects -- J054034.00-692509.8 and J045433.85-692036.2 -- display the B[e] phenomenon in their spectra: they have strong Balmer emission lines in combination with many emission lines of \ion{Fe}{II} and [\ion{Fe}{II}], as well as of [\ion{O}{I}] \citep{lamers98}. Our initial colour selection criterion already implies that these stars have a strong near- or mid-IR excess due to hot circumstellar dust, the fourth diagnostic feature of a B[e] star. J045729.38-661837.3 has many \ion{Fe}{II} emission lines, but only six possible [\ion{Fe}{II}] lines, of which four are unblended with \ion{Fe}{II} lines, which renders its status as a B[e] star doubtful. The strength of the [\ion{O}{I}] emission in these three stars increases markedly with increasing [\ion{Fe}{II}]/\ion{Fe}{II}.

Some of the emission-line spectra have the [\ion{O}{III}] feature at 4959 and 5007~\AA{} present, either in emission or absorption. The latter must arise from excess sky-subtraction of emission from a surrounding emission nebula. Given especially that one of our problems is to distinguish between young stars, which would probably tend to lie within emission nebulae, and much older, randomly distributed stars, we overplotted all post-AGB candidates with an emission-line spectrum on an \element[][]{H}$\alpha$ image of the LMC from \citet{kennicutt95}. The result of this can be seen in Fig.~\ref{fig:halpha_lmc}. Fifteen objects fall out of the spatial range of the image and are therefore not shown. Of the remaining eight, two objects display the [\ion{O}{III}] feature at 4959 and 5007~\AA{} in absorption and one of these -- J054412.04-682727.9 -- is located in an area where diffuse \element[][]{H}$\alpha$ emission is present. This object therefore lies in a starforming region and is likely a YSO. We also marked the positions of all objects for which the spectrum did not cover the spectral range around the [\ion{O}{III}] feature, but none of these seem to lie in a region where \element[][]{H}$\alpha$ is present.

Fifteen of the 23 emission-line stars are excluded from the final post-AGB sample because their SED-based final luminosity (see Sect.~\ref{ssec:finlum}) is higher than 35\,000~$L_{\odot}$. These objects are noted in Table~\ref{list_delobj}. The remaining eight emission-line stars fall in the luminosity range of post-AGB stars and are listed in Tables~\ref{catalogue_specdisc}, \ref{catalogue_specoutflow} and \ref{catalogue_specundet}. For each emission-line object, the respective table in which it can be found is cited in the final column of Table~\ref{table:em}.

\begin{landscape}
\begin{table}
\caption{Most prominent features in the emission-line stars we observed. We list the properties of \element[][]{H}${\alpha}$ (6563~\AA{}), \element[][]{H}${\beta}$ (4861~\AA{}), \element[][]{H}${\gamma}$ (4341~\AA{}), \element[][]{H}${\delta}$ (4102~\AA{}), \element[][][][]{K} (7699~\AA{}), \ion{O}{I} (7774 and 8446~\AA{}), [\ion{O}{I}] (6300 and 6363~\AA{}), [\ion{O}{II}] (3727~\AA{} and 7319 and 7330~\AA{}), [\ion{O}{III}] (4959 and 5007~\AA{}), \ion{He}{I} (5876~\AA{} and 6678~\AA{}),  [\ion{S}{II}] (6717 and 6731~\AA{}), [\ion{S}{III}] (9069~\AA{}), [\ion{N}{II}] (6548 and 6584~\AA{}), \ion{Ca}{II} (8498, 8542 and 8662~\AA{}), the Paschen lines (Pa, up to 8200~\AA{}) and \element[][][][]{Na} (5890 and 5896~\AA{}). In this table, 'e' stands for emission, 'a' for absorption, '0' indicates that the line is not present and '-' denotes a line outside of the spectral coverage or one where the spectrum is defective. The intensity of absorption and emission is noted on a qualitative scale of 1 to 6 with the notation '++' referring to excessively strong absorption. The additional symbols give some information on the shape of the spectral line with 'b' meaning that the feature is unusually broad, 'P' that a P Cygni absorption component is present, 's' that the feature is sharp and ':' indicating some kind of uncertainty, most likely doubt as to whether the feature is real when the S/N is low. Some notes are included in the last column, where 'S/N' denotes that the S/N of the spectrum is low and 'Pa' that an unusual Paschen decrement is present. For all objects, the reference to the table in which they appear is denoted. Many objects are excluded from the final post-AGB sample because the SED-based final luminosity (see Sect.~\ref{ssec:finlum}) was higher than 35\,000~$L_{\odot}$ and can be found in Table~\ref{list_delobj}. }
 \label{table:em}      
 \centering          
 \begin{tabular}{lcccccccccccccccccccl}     
 \hline\hline       
 Object & \element[][]{H}${\alpha}$ & \element[][]{H}${\beta}$ & \element[][]{H}${\gamma}$ & \element[][]{H}${\delta}$ & \element[][][][]{K} & \multicolumn{2}{c}{\ion{O}{I}} & [\ion{O}{I}] & \multicolumn{2}{c}{[\ion{O}{II}]} & [\ion{O}{III}] & \multicolumn{2}{c}{\ion{He}{I}} & [\ion{S}{II}] & [\ion{S}{III}] & [\ion{N}{II}] & \ion{Ca}{II} & Pa &  \element[][][][]{Na} & Notes\\
   & & & & & & 7774 & 8446 & & 3727 & 7319 & & 5876 & 6678 & & 9069 & & & & & \\
 \hline
J044555.34-705041.7 &    e5 &  e3 &   e2 &  e2 &   0 &   a4 &   e2 &  e1 &   - &  0 &   e1 &   0 &     0 &  0 &    0 &    0 &   e3 &   e2 &  a2 & \ref{catalogue_specundet}\\
J044904.66-690455.0 &    e4 &   - &    - &   - &   - &    0 &   e1 &   0 &   - &  0 &    - &   0 &     0 &  0 &    0 &    0 & e1-2 &    0 & a1: & \ref{list_delobj}\\
J044943.46-691955.3 &    e4 &   - &    - &   - &   - &    0 &   e2 &   0 &   - &  0 &    - &   0 &     0 &  0 &    0 &    0 &    0 &    0 &   0 & \ref{list_delobj}\\
J045433.85-692036.2 &    e5 &  e3 &   e2 &  e2 &  e2 &   a2 &   e1 &  e2 & 0+: &  0 &    0 &   0 &     0 &  0 &    0 & e1-2 &  e3+ &  e1+ &  e2 & \ref{list_delobj}, B[e] \\  
J045729.38-661837.3 &    e5 &  e3 &  e2P & e2P &  a1 &   e2 &   e4 &  e1 &   0 &  0 &    0 &   0 &     0 &  0 &    0 &  e1: & e1-2 &  e2+ &   0 & \ref{list_delobj}\\  
J050338.93-690158.6 &    e3 &  a2 &  a++ & a++ & a1: &    0 &    0 &   0 &   - &  0 &    0 &   0 &     0 &  0 &    0 &    0 &   e2 &    0 &  a2 & \ref{catalogue_specundet}\\
J050431.84-691741.4 &    e4 & e1P &   a3 &  a3 & a1: &   a2 & e1-2 & e1: &   0 &  0 &    0 &   0 &     0 &  0 &    0 &  e1: &   e2 &    0 &  a2 & \ref{catalogue_specundet}\\
J050451.70-663807.5 &   e4+ &  e3 &   e2 &  e1 &   0 &   e1 &   e4 &  e1 &   - &  0 &    0 &   0 &     0 &  0 &    0 &    0 &  e1: & e2-3 &   0 & \ref{list_delobj}  \\
J050535.21-685928.9 &    e3 &   - &    - &   - &   - &   a1 &  e0+ &   0 &   - &  0 &    - &  e1 &    e1 &  0 &    0 &   e1 &    0 &    0 &   0 & \ref{list_delobj} \\
J051123.63-700157.4 &   e6s &  e4 &   e3 &  e2 &   0 &    0 &   e2 &  e3 &  e3 & e3 &   e5 &  e4 &  e2-3 & e1 &   e4 &   e3 &    0 &   e2 &   0 & \ref{catalogue_specundet}\\
J051333.74-663418.9 &   e3  &  a2 &   a2 &  a2 &  a4 &   a2 &    0 &   0 &   0 &  0 &    0 &  e3 &  e1-2 &  0 &    0 & e1-2 &  a5b &   a2 &  a3 & \ref{catalogue_specdisc}\\
J051525.45-661905.0 & e4-5P & e3P &   a3 &  a3 & e0+ &    0 &   e2 &   0 &   0 &  0 &    0 &   0 &     0 &  0 &    0 &    0 &   e3 & e1-2 &   0 & \ref{catalogue_specoutflow}, S/N \\ 
J051648.08-692222.0 &    e3 &   - &    - &   - &   - &    0 &    0 & e1: &   - &  0 &    - &   0 &     0 &  0 &    0 &  e0+ &    0 &  e1: & a1: & \ref{list_delobj}, S/N \\
J052031.12-680057.2 &    e2 &   - &    - &   - &   - &    0 &   e1 &   - &   - &  0 &    - &   - &     0 &  0 &    0 &    0 &    0 &   e1 &   - & \ref{catalogue_specoutflow}, Pa \\
J052131.77-682058.8 &    e4 &   - &    - &   - &   - &    0 &   e1 &   0 &   - &  0 &    - &  e1 &     0 & e1 &    0 &  e1: &  e1: &    0 &   0 & \ref{list_delobj}\\
J052414.62-684110.0 &   e5s &  e3 &   e2 &  e1 &   0 &    0 &   e3 &   0 &   0 &  0 &    0 &   0 &     0 & e1 &    0 & e2-3 &    0 &    0 &   0 &  \ref{catalogue_specoutflow}\\
J052630.65-674036.6 &   e5s & e4P &   e3 &  e2 & a2? &    0 &   e3 &   0 &  e4 & e2 &   a3 &   0 &     0 & e2 & e3-4 &   e3 &   e2 &   e1 &  a2 & \ref{list_delobj} \\
J053017.49-685926.6 &    e2 &   - &    - &   - &   - &    0 &    0 &   0 &   - &  0 &    - & e1: &     0 &  0 &    0 &   e1 &    0 &    0 &  a1 & \ref{list_delobj} \\
J053539.76-673705.0 & e4-5s &   - &    - &   - &   - &    0 &    0 &   0 &   - &  0 &    - & e1+ & e0-1: & e1 &   e2 &   e1 &    0 &    0 &   0 & \ref{list_delobj}, S/N \\
J054034.00-692509.8 &   e5s &  e3 &   e2 &  e2 &   0 &    0 &   e4 &  e3 &  e4 & e1 &   a3 &   0 &     0 & e2 &    0 &   e3 &   e3 &   e1 &   0 & \ref{list_delobj}, B[e] \\ 
J054055.84-691614.5 &   e5s &   - &    - &   - &   - &    0 &   e1 &   0 &   - &  0 &    - &   0 &     0 & e1 &   e3 &   e4 &    0 &    0 &   0 & \ref{list_delobj}\\
J054412.04-682727.9 &  e4-5 &  e3 & e1-2 &   0 &   0 &    0 &   e2 &   0 &  e2 &  0 & e1-2 & e1: &    e1 & e1 &    0 &   e1 &   e2 &  e1: & e1: & \ref{list_delobj} \\
J060323.20-680437.5 &  e4-5 &  e3 &   e1 &  a1 &   0 &    0 &   e3 & e1: &   0 &  0 &    0 &   0 &    a1 &  0 &    0 &    0 &   e2 &   e2 &   0 & \ref{list_delobj} \\ 
\hline                  
\end{tabular}
\end{table}
\end{landscape}

\addtocounter{table}{1}

\section{Analysis of the sample}\label{sec:analysis}

\subsection{Spectral energy distributions} \label{subsection:SEDs}

Using the spectral types we obtained from the low-resolution optical spectra and the literature, we are able to constrain the appropriate atmosphere models for the different objects and hence determine their total reddening. We assumed that the ISM reddening law can be used to approximate the total (circumstellar + ISM) reddening. The luminosities were then computed by integrating the appropriate stellar model that fits the dereddened optical photometry. With a small adaptation of the procedure, we were able to estimate these parameters for the sources without known spectral type, too. The luminosities allow us to determine the positions of the different objects in the HR~diagram and lend us a view on their evolutionary stage by comparison with the evolutionary tracks of \citet{blocker95}.

\subsubsection{Objects with known spectral type}

We developed a method based on a Monte Carlo simulation to minimise simultaneously for the effective temperature as well as for the total reddening of the star for objects of which the spectral type is known. This results in better values for the luminosity than was possible with black-body fitting (Sect.~\ref{ssec:lumcut}) and estimates of the effective temperature and total reddening of these objects.

For each object, we only allowed models within a certain temperature range constrained by the spectral type \citep{appG}. The effective temperatures of all models that were used in the routine range from 3000 up to 10\,000~K in steps of 250~K, from 10\,000 up to 13\,000~K in steps of 500~K, from 13\,000 to 35\,000~K in steps of 1000~K and from 35\,000 up to 50\,000~K in steps of 2500~K. For atmosphere models with effective temperatures higher than or equal to 3500~K, we used Kurucz models with a metallicity of $-0.3$ \citep[][and updates]{kurucz79}. The models we utilised for the range of $T_{\mathrm{eff}}$ from 3000 to 3250~K were {\sc MARCS}-models with metallicity $-1.0$ \citep{marcs08}. The metallicity of the stars in the LMC span the large range from [\element[][][][]{Fe}/\element[][][][]{H}] $= -2.0$ up to [\element[][][][]{Fe}/\element[][][][]{H}] $= -0.3$ \citep{geisler09}. Lower metallicity objects will be brighter than the models, especially in the UV, but our data are not good enough to constrain this parameter. The same holds true for $\log g$, which was also not a free parameter. We only used one gravity for each temperature and chose log $g$ to be as low as possible within the model grids. It ranges up from 1.0 in steps of 0.5 for the coolest models and does not go above 2.0 for stars with a spectral type cooler than B.

The best model is determined by minimising the $\chi^2$ between the scaled atmosphere model and the dereddened photometric data. We assume that the photometry is dominated by the photosphere only up to 2.5~$\mu$m. This is a compromise between retaining enough data points to do the minimisation procedure and the fact that at wavelengths longer than 1.3~$\mu$m, the thermal radiation of the circumstellar material may dominate already in case of disc objects. The spectra themselves were left out of this process since the uncertainty on their absolute flux level is evaluated to be too large. If other photometric data were available, the magnitudes from the Guide Star Catalogue were not used in the procedure, because of their larger errors and sometimes significant difference from magnitudes found in the other two optical catalogues. This procedure allows us to determine the appropriate value of luminosity, effective temperature and total reddening.

The corresponding errors for those parameters -- luminosity, effective temperature, and total reddening -- were computed by use of the Monte Carlo Method. We constructed 250 alternative photometric datasets by using a Gaussian distribution of the errors on the photometric points and repeated the procedure described in the previous paragraph for each of these datasets. This yielded 250 possible models from which we removed the ones without convergence. For about 26~\% of the objects, the solutions cluster in two distinct groups. In case of a bimodal distribution around two significantly different temperatures, we only retained the group of models with temperatures around the one found for the original photometry.

Finally, the errors on the different parameters were computed as the standard deviation of the remaining solutions with respect to the values based on the original photometry. For the temperature, a minimal error equal to the maximal gridstep at that temperature was taken into account because the models were not interpolated onto a finer grid. 

There was one exception to this minimisation process: the photometric data of J050830.51-692237.4 are dominated by the stellar photospheres of two stars, which is why we only used the data up to 8000~\AA{} in the minimisation. \newline

\begin{figure}
\resizebox{\hsize}{!}{\includegraphics{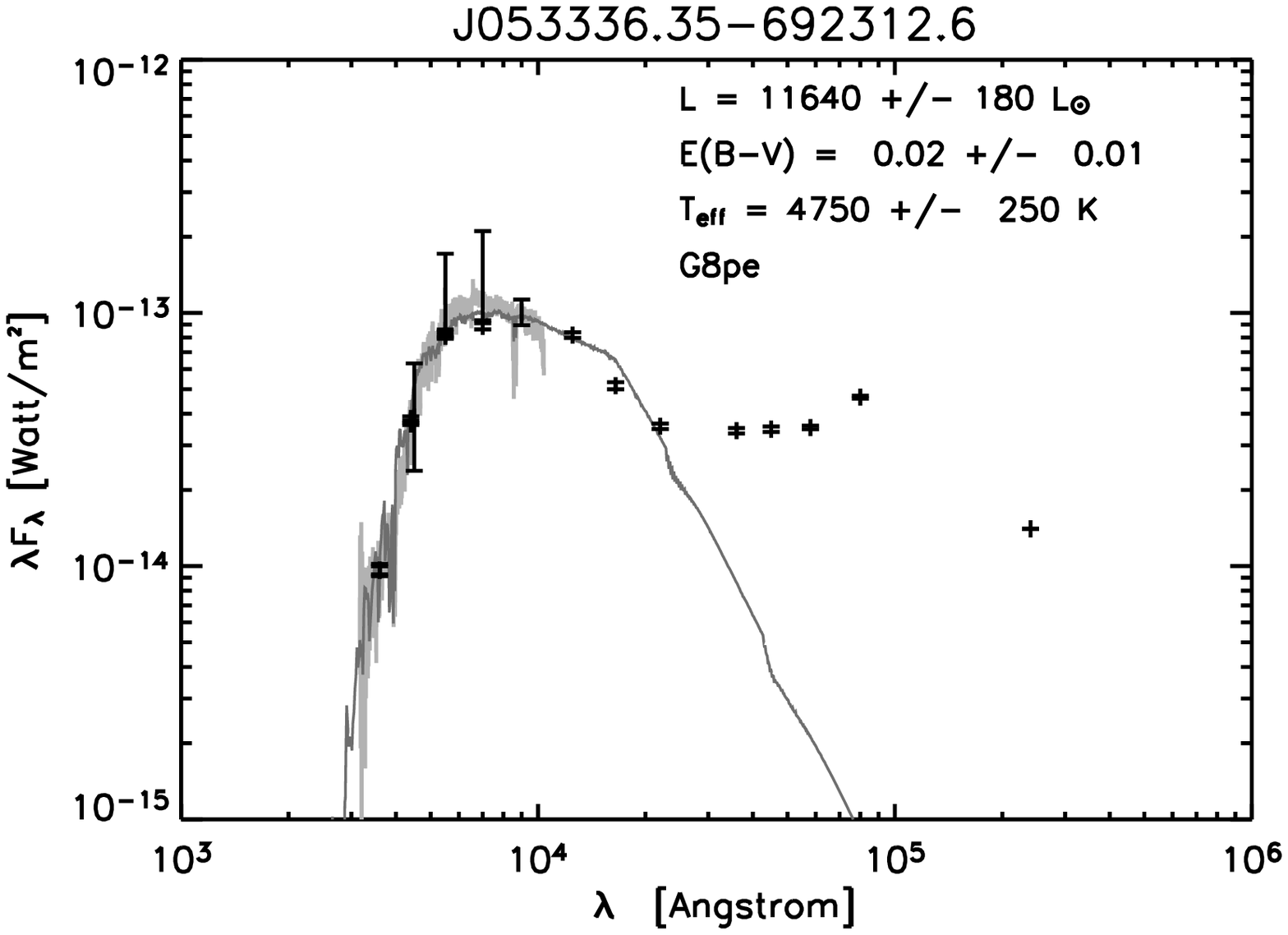}}
\resizebox{\hsize}{!}{\includegraphics{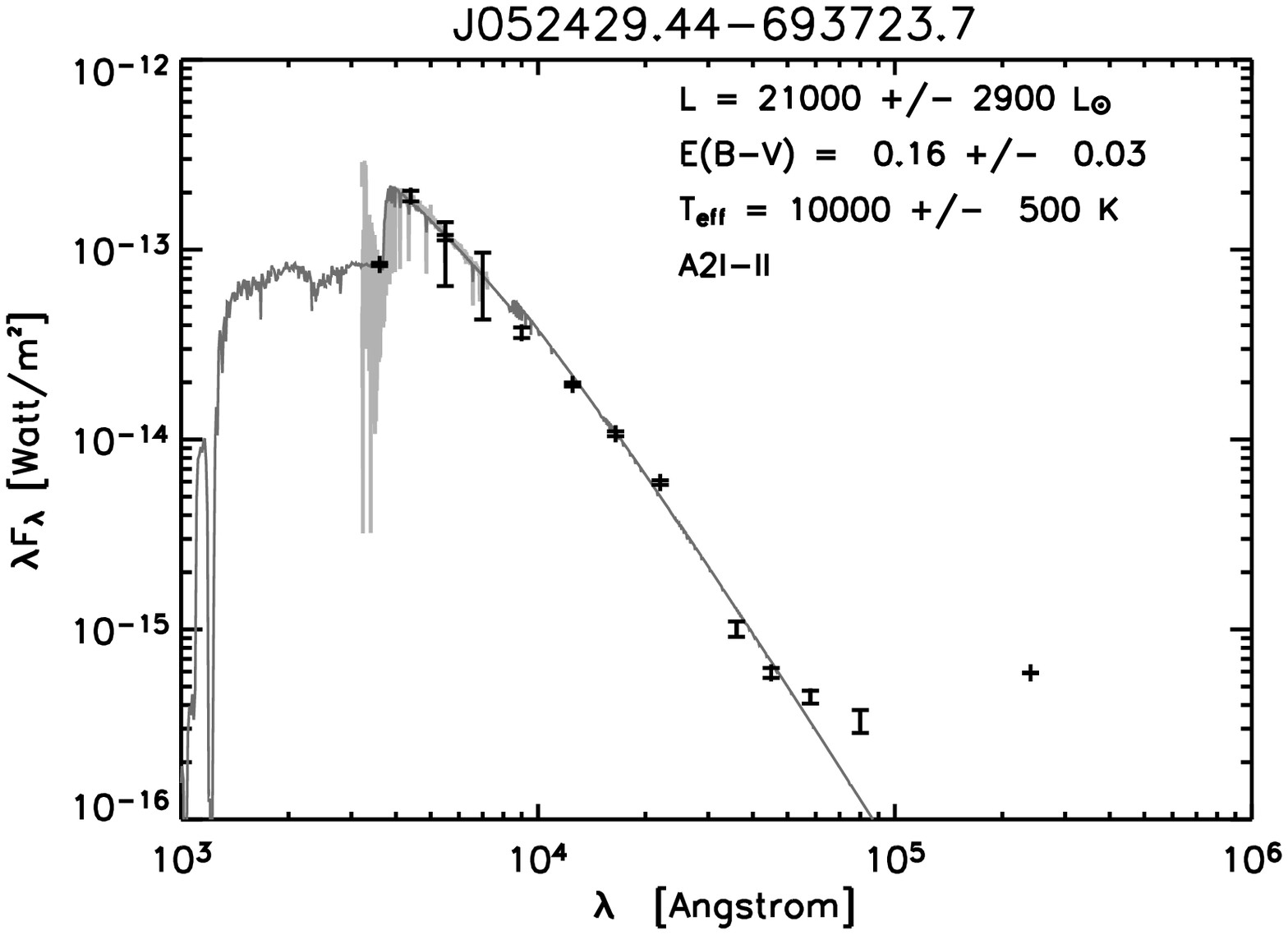}}
\caption{Example SEDs of post-AGB candidates with a disc (top) and a shell (bottom). The dark and light grey line are the Kurucz atmosphere model used and the low-resolution, optical spectrum. All photometric datapoints are dereddened.
}
 \label{fig:SEDs}
\end{figure}

\begin{figure}
\resizebox{\hsize}{!}{ \includegraphics{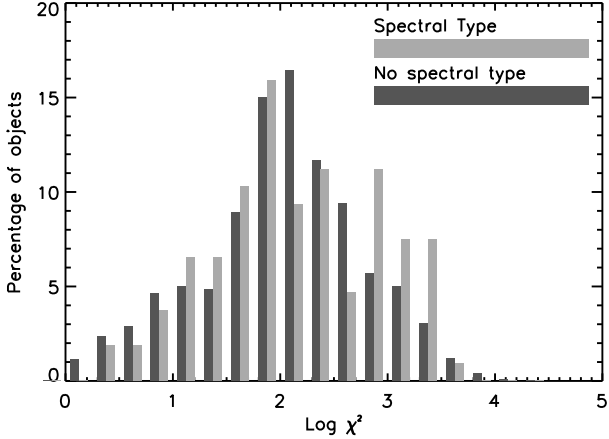}}
\caption{Distribution of the reduced $\chi^2$ values of the final SED solutions. Shown are the histogram of the final solutions for all post-AGB candidates with a known spectral type of the central object and the histogram of those without.}
 \label{fig:chi2distr}
\end{figure}

This minimisation procedure renders very convincing SED fits (e.g., Fig.~\ref{fig:SEDs}). In Fig.~\ref{fig:chi2distr} the histogram in light grey shows the distribution of the reduced $\chi^2$ values of the final models for all post-AGB candidates with known spectral type. Note that the reduced $\chi^2$ are still high because the quoted formal errors on the photometric data are fairly small. Because these are not the main contributors to the differences between the model and the dereddened photometric data, this does not imply the quality of our fits is low: e.g., the objects in Fig.~\ref{fig:SEDs}, which yield convincing SEDs, have reduced $\chi^2$ values of 104 and 70 respectively.

For objects with spectral types from O to B, the errors we find on the different parameters are large, owing to the lack of photometric data at wavelengths below 3600~\AA{}, where still a significant part of the luminosity is radiated. For all other objects most parameters are fairly stable, but the calculated values of the luminosities remain very sensitive to changes of the atmosphere model, the value of $E(B-V)$, and the variability of the photometry. A full catalogue of all SEDs can be found in Appendix~\ref{app:SEDs}.

\subsubsection{Objects without known spectral type}

We broadened this method for objects without known spectral type by allowing the entire grid of atmosphere models to be used in the minimisation procedure. The best fit will then yield both an estimate of the effective temperature of the central star and the total reddening. Without spectral information, the SED minimisation procedure is, however, very degenerate. To assess the reliability of such a method, we let this broader procedure run on all objects of which we know the spectral type. We then compared these results with the ones we obtained when taking into account the spectral types.

We find that 41~\% of the objects fall into the temperature range predicted by the spectral type. The effective temperatures of 73~\% of the post-AGB candidates differ by less than 25~\% and of 83~\% by less than 50~\% from our original solution. The large differences in the temperatures with and without use of the spectral type for the remaining 18 objects (17~\% of the total number) are mainly because we lack data at wavelengths below 3600~\AA{}, and the existing photometry of these objects is unable to prevent models with a higher temperature than follows from the spectral type from being chosen.

Changes in the adopted value of $E(B-V)$ impact on the model parameters, because an atmosphere model with a higher temperature typically needs a higher reddening value to fit a given dataset. Small differences in the temperature of the atmosphere model require relatively large changes of the colour excess, which indicates that this parameter will be less stable than the model temperature. We find that 46~\% of the reddening values are still in the error interval found with the use of the spectral types, as could be expected from the similar number of good matches in temperature. Only 55~\% and 66~\% differ, however, respectively less than 25~\% and  50~\% from this original solution for $E(B-V)$. For the remaining 34~\% of the objects the colour excess changes even more owing to the reasons indicated above and the fact that for low reddening values, a large relative error can correspond to a fairly small absolute error.

As the reddening is more poorly constrained, the results are even less stable for the computed luminosity. Again the objects with an early spectral type are most affected. While the bulk of their luminosity is radiated at wavelengths below 3600~\AA{}, our lack of photometry at these wavelengths allows a broad range in possible atmosphere models and hence luminosities. However, we still find that 47~\% of the values fall inside the 1~$\sigma$ errors of the original luminosity estimate, 50~\% differ less than 25~\% from this original solution and 67~\% less than 50~\%. For 33~\% of the objects the difference is bigger.

We can conclude that for about 40~\% of the objects, values for the effective temperature, colour excess, and luminosity can be found that fall inside the 1~$\sigma$ error interval on these parameters computed when taking into account the spectral type, despite the fact that spectral information on the central source is missing. The temperature is restricted the best in general, followed by the value of $E(B-V)$. The computed value of the luminosity remains the most uncertain because of its great dependence on the other two parameters. The results are the least reliable for objects with an early spectral type because of the lack of photometric data at short enough wavelengths. \newline

We applied this broader version of our minimisation procedure to the total sample with unknown spectral type. In Fig.~\ref{fig:chi2distr} we depict a histogram of the reduced $\chi^2$ values in dark grey. The histogram is very similar to the histogram obtained for objects with spectral types for objects with $\log \chi^2 \leq 2$, which shows that there is no systematic difference. The smaller relative number of objects without a known spectral type at $\log \chi^2 > 2$ stems from the fact that the effective temperature is an additional free parameter in these cases, unlike for the objects with known spectral type. The dip in the distribution of the latter at $2 < \log \chi^2 < 2.25$ is simply caused by a binning effect and by the much smaller number of objects in this group when compared to the former.

\subsection{Final luminosities} \label{ssec:finlum}

\begin{figure}
\resizebox{\hsize}{!}{ \includegraphics{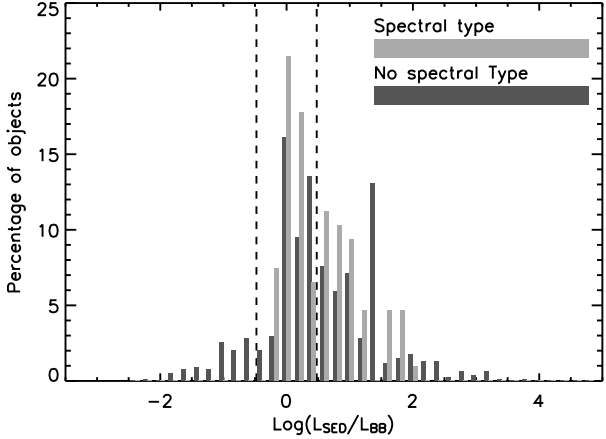}}
\caption{Histogram of the logarithm of the ratio of the luminosities computed in Sects.~\ref{subsection:SEDs} and \ref{ssec:lumcut}. The distributions of the objects with and without a known spectral type are displayed separately and the dashed lines indicate where ratios of $\frac{1}{3}$ and 3 are reached.}
 \label{fig:lumdiff}
\end{figure}

We now have two luminosities for all objects: the one calculated in Sect.~\ref{ssec:lumcut} where we fitted up to three black bodies to the raw photometry (the 'black-body-based luminosity' below) and the one from Sect.~\ref{subsection:SEDs} based on the integral of the stellar model, which is scaled to the dereddened photometric values (the 'SED-based luminosity' below). The latter luminosity calculation is the most reliable in most, but not all, cases because it is a better approximation of reality. In Fig.~\ref{fig:lumdiff} we show the ratio of the  luminosities determined by our preferred method to the black-body-based ones. There is a peak in the histogram corresponding to a ratio of 1.0, and 42~\% of the data lie in between $\frac{1}{3}$ and 3. \newline

For some objects with a known spectral type, the SED-based luminosities tend to be significantly higher than the black-body-based ones. These objects are without exception of an early spectral type. The maximum temperature of 15\,000~K for the hottest black body in the luminosity fit we imposed in Sect.~\ref{ssec:lumcut} because of the lack of data at wavelengths below 3600~\AA{} can account for this difference. We therefore favour the higher SED-based luminosities for all objects with a known spectral type. \newline

For the sources without spectral information, the situation is a bit more complicated. If there are enough photometric data to describe the photosphere well, the SED-based luminosity is likely to be better than the one based on the black-body fit. Our minimisation procedure has, however, two important shortcomings. 

\begin{figure}
\resizebox{\hsize}{!}{ \includegraphics{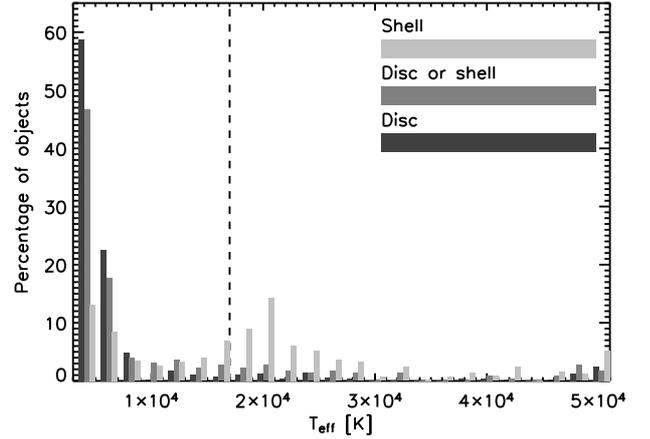}}
\caption{Histogram of the effective temperatures of the different types of objects based on SED fitting (see Sect.~\ref{subsection:SEDs}). The dashed line at 17\,000~K indicates the upper temperature limit imposed by us (see text).}
 \label{fig:tefftypedistr}
\end{figure}

First of all, the lack of photometric data at wavelengths below 3600~\AA{} implies that atmosphere models with a too high effective temperature can be chosen as the best fit. For these models, the available photometry only samples the Rayleigh-Jeans tail of the SED, which is obviously not very constraining and leads to a generous overestimation of the luminosity and total reddening. The luminosity thus remains poorly constrained for hot central stars, and we prefer the black-body-based luminosity, as calculated in Sect.~\ref{ssec:lumcut}. Especially for shell sources, this effect is noticeable (see Fig.~\ref{fig:tefftypedistr}). This type of object will only appear in our optically bright sample when the circumstellar envelope is sufficiently diluted by expansion. Given the very fast evolutionary timescales of the central star, the shell sources will therefore have hotter central stars on average. For the disc sources, the timescale is more difficult to compute because it depends on the lifetime of the compact discs, which is yet to be constrained. It is possible that disc evaporation is stimulated by the heating of the central star, which causes the object to disappear from our selection. The disc sample will therefore contain cooler central stars, on average.

We chose to use the SED-based luminosity only if the temperature of the final atmosphere model is below 17\,000~K. For all cooler objects, either the bulk of the luminosity is radiated at wavelengths redder than 3600~\AA{}, or the Balmer jump is prominent enough to confine the atmosphere model. It turns out that for 406 or 31~\% of the objects without a spectral type, an effective temperature of at least 17\,000~K is deduced from our fitting procedure. Hence for these objects, the black-body-based luminosity was used.

The second problem we encounter is that our $\chi^2$ method based only on photometry does not deal well with high reddening. The assumption that the interstellar extinction law is a good representation of the total reddening in the line of sight towards the photosphere is unproven and can accordingly cause problems when the extinction is dominated by the circumstellar component. We therefore discard all SED-based luminosities if the final atmosphere model has an $E(B-V)$ that exceeds the arbitrary value of 1.5. This is the case for 215 or 16~\% of the objects without a spectral type.

Next to the shortcomings of our minimisation procedure, the photometry itself is in some cases incapable of indicating a preferred model. We consequently also discard the SED-based luminosity if it is lower than its error. This occurs for 133 objects or 10~\% of the objects without a spectral type.

In total, we used the black-body-based luminosity for 659 objects or 50~\% of the objects without known spectral type.\newline

For some objects, the SED-based luminosity estimate is higher than the upper limit of 35\,000~$L_{\odot}$ we introduced in Sect.~\ref{ssec:lumcut}. This is the case for 69 objects of which 44 have a known spectral type. An additional 113 objects without a spectral type have an SED-based luminosity below 1000~$L_{\odot}$. Because of the big uncertainty in the different parameters calculated by fitting an atmosphere model to the dereddened photometry for objects without a known spectral type, we decided to leave these 138 objects in the final sample and changed their luminosities back to the original values based on the black-body fit. The 44 objects with known spectral type that fall out of the imposed luminosity range were removed from our catalogue and can be found in Table~\ref{list_delobj}. 

Our final catalogue now contains 1407 objects. It is available at the CDS and contains the following information. Column 1 lists the name of the source, Columns 2 and 3 give the right ascension and declination, the source type based on the shape of the SED can be found in Column 4, Columns 5-8 contain the $U$ magnitude and error from respectively \citet{zaritsky04} and \citet{massey02a}, the $B$ magnitude and error from these two catalogues and the Guide Star Catalogue \citep{GSC2.3.2} is listed in Columns 9-14 and the $V$ magnitude and error in Columns 15-20. Columns 21 and 22 gives the $R$ magnitude and error from \citet{massey02a}, Columns 23 and 24 the $R_F$ magnitude from \citep{GSC2.3.2} and its error, Columns 25 and 26 contain the $I$ magnitude and error from \citet{zaritsky04} and Columns 27-32 give the 2MASS $J$, $H$ and $K_s$ magnitudes and errors. The 3.6, 4.5, 5.8 and 8.0~$\mu$m IRAC fluxes and errors are listed in columns 33-40 and the MIPS 24, 70 and 160~$\mu$m fluxes with their errors can be found in Columns 41-46. Column 47 lists the black-body-based luminosity, Columns 48-53 contain the SED-based luminosity, effective temperature, and reddening with their respective errors and Column 54 gives some remarks on cross-correlation with catalogues of other types of objects (see Sect.~\ref{ssec:croscorr}). We also list the spectral type in Column 55 and variability information (Column 56), when these data are available.

\subsection{Luminosity distribution}

\begin{figure}
\resizebox{\hsize}{!}{ \includegraphics{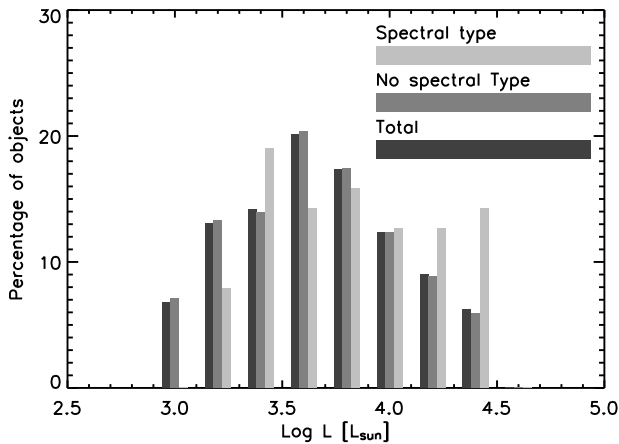}}
\caption{Histogram of the luminosities of all objects in our final
  sample. We distinguish between objects with and without a determined spectral type.}
 \label{fig:lumhistspec}
\end{figure}

In Fig.~\ref{fig:lumhistspec} we show the histogram of all final luminosities while distinguishing between objects with and without a determined spectral type. The distribution of the total sample is also shown. As expected, we see a peak in the luminosity distribution of the total sample around the middle of the imposed luminosity range. The anticipated bias towards lower-mass objects and hence lower luminosities, due to their slower evolution and higher number density as predicted by the initial mass function, is also observed. The histogram drops slowly towards higher values of the luminosity, suggesting that the luminosity-cut we introduced to discard supergiants is effective. Some AGB stars are, however, probably still polluting the sample towards the high-luminosity range. Towards lower luminosities, a steeper drop can be observed. This indicates that also the luminosity-cut intended to remove YSOs is probably effective, although some bright YSOs and objects on the upper part of the Red Giant Branch (RGB) with circumstellar dust might still be contained in the sample. 

When we compare the histograms of the objects with and without a known spectral type, it is clear that so far, the emphasis of our observations has been on visually brighter stars. A full spectral survey of all objects is needed, however.\newline

\begin{figure}
\resizebox{\hsize}{!}{ \includegraphics{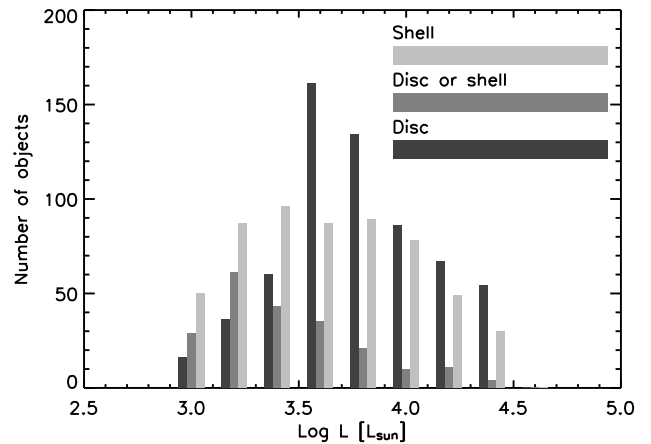}}
\caption{Histogram of the luminosity distributions of the different subsamples of our total list of post-AGB candidates.}
 \label{fig:lumhisttype}
\end{figure}

In Fig. \ref{fig:lumhisttype} we compare the luminosity distributions of the three different subsamples based on the shape of the SED. Obviously the histograms of both the objects with a disc and those with a shell span the luminosity range we observed for the entire sample and therefore represent a wide range of different initial masses. The distinction between both types of objects is more easily made at higher luminosities, as can be seen from the distribution of the objects for which the type of SED is undecided: at higher luminosities, the number of objects in this group becomes relatively small compared to the number of objects in the other two. The distribution of objects with a shell seems to be more constant over the entire luminosity range with possibly a small peak towards the lower values, while the sources with a disc display a clear peak in the middle of the luminosity range. This implies that objects with a shell are slightly more abundant at low luminosities while post-AGB candidates with a disc lie at average luminosities. Both are equally represented at high luminosities. However, this effect at low luminosities may also be partly introduced by the pollution of our sample with YSOs, which will mainly fall in our shell class because YSOs have a typical peak in their SED at wavelengths longer than 24~$\mu$m.

\subsection{HR~diagrams}

\begin{figure}
\resizebox{\hsize}{!}{ 
\includegraphics{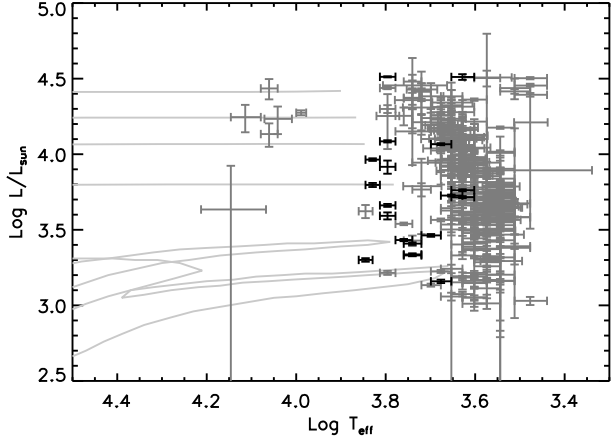}}
\caption{HR~diagram containing all post-AGB candidates with an SED indicative of a disc of which the final luminosities were computed using the SED (Sect.~\ref{subsection:SEDs}). The evolutionary tracks are from \citet{blocker95} and correspond to stars with, from bottom to top, an initial mass of 1, 3, 4, 5 and 7~$M_{\odot}$.
The 1~$M_{\odot}$ model suffers two late thermal pulses. Note that the objects shown here are a subset of the entire sample of disc sources as the objects for which the final luminosity was computed with the BB fit (Sect.~\ref{ssec:lumcut}) lack an effective temperature estimate and are hence missing in this plot.
}
 \label{fig:HRdiagrdisc}
\end{figure}

\begin{figure}
\resizebox{\hsize}{!}{ 
\includegraphics{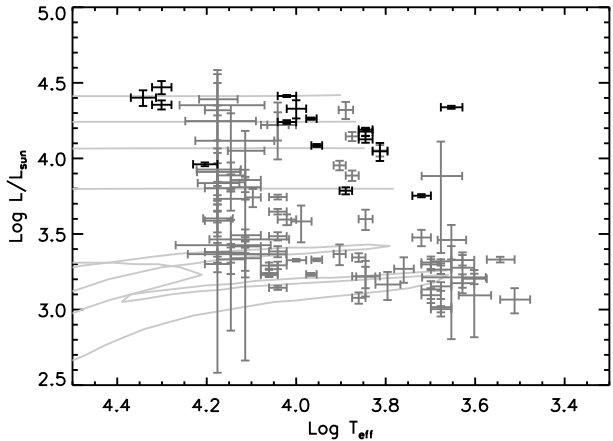}}
\caption{Same as Fig.~\ref{fig:HRdiagrdisc}, but for all post-AGB candidates with a shell.}
 \label{fig:HRdiagroutflow}
\end{figure}

\begin{figure}
\resizebox{\hsize}{!}{ 
\includegraphics{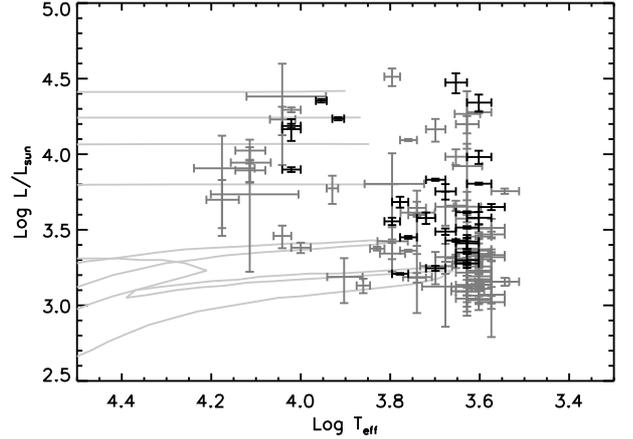}}
\caption{Same as Fig.~\ref{fig:HRdiagrdisc}, but for all post-AGB candidates with an SED that does not allow us to conclude on the circumstellar shape.}
 \label{fig:HRdiagrundet}
\end{figure}

In Figs.~\ref{fig:HRdiagrdisc}, \ref{fig:HRdiagroutflow}, and \ref{fig:HRdiagrundet} we show the position in the HR~diagram of all post-AGB candidates for which the final luminosity was computed based on the integral of the stellar atmosphere model that was scaled to the dereddened photometric values (Sect.~\ref{subsection:SEDs}). To increase the clarity of the plot, we distinguish between the different shapes of SED. The reliability of the displayed temperatures is only limited because we used atmosphere models with effective temperatures up to 50\,000~K for objects with a known spectral type and only retained solutions based on models with a maximum temperature of 17\,000~K for sources without. 

We can, however, conclude that the post-AGB candidates with a shell as well as those with a disc show a large spread in temperatures and that on average, disc candidates show a lower effective temperature than the other objects. 
For post-AGB candidates with a shell, a clear line from the upper left to the lower right is observed, to the right of which practically no objects are detected. This is probably because this type of objects only appears in the optical when the opacity of the circumstellar envelope declined sufficiently for the central star to become visible. Owing to the faster evolutionary timescales for more massive objects, these objects are already closer to the PNe stage when this happens. The spread in effective temperatures is lower for objects with a disc, which seem to accumulate at effective temperatures of 3000-6000~K. This might be a consequence of the evaporation of the circumstellar disc due to the heating of the central star, at which point the object would no longer show an IR excess and disappear from our sample. More research is needed to confirm this, however.

For objects with a shell we observe a bias towards lower luminosities (lower masses). This is predicted by the initial mass function and the fact that massive, single stars lose more mass than less massive stars, which causes them to be obscured by the optically thick wind they generate. There is little direct evidence for the binary nature of shell sources in the Galaxy.
Many disc sources appear to be 3-7~$M_{\odot}$ rather than 1~$M_{\odot}$ stars, which is unexpected. Maybe some of the disc sources are not really post-AGB stars with a circumstellar disc, but AGB stars with clumpy ejections. In this case, one can see the hotter dust near the central star as well as more remote clumps, which reside in general in a non-spherical wind. These clumpy winds seem to occur earlier in the AGB evolution when the mass-loss rate is lower \citep{feast03, olofsson04, monnier04} and will be more detectable in the SAGE survey for the more luminous stars.

\section{Evaluation} \label{sec:evaluation}

\subsection{Cross-correlation with catalogues of other types of objects}\label{ssec:croscorr}

We cross-correlated the list of post-AGB candidates with available catalogues of other types of objects in the LMC that are likely to display SEDs with characteristics similar to those of post-AGB stars. For this, we invariably used a search radius of 2\arcsec \ unless noted otherwise. Here, we discuss the results of these comparisons.

\subsubsection{RV~Tauri stars}\label{sssec:rvt}

\begin{table}
\caption{Names, spectral types, MACHO numbers and OGLE-III Cepheid IDs of the previously-known RV~Tauri stars in our sample.}             
\label{table:rvt}      
\centering                          
\begin{tabular}{llll}        
\hline\hline                 
Object Name (IRAC) & Spectral Type & MACHO  & OGLE-III \\     
\hline                        
J045543.20-675110.1 &     G2-8(R)Ibe: &  47.2496.8 &  T2CEP-015 \\
J050304.95-684024.7 &        F7-9Ibpe & 19.3694.19 &  T2CEP-029 \\
J050356.28-672724.8 &                 &            &  T2CEP-032 \\
J051418.09-691234.9 &        F7-9Ibpe & 79.5501.13 &  T2CEP-067 \\
J051845.47-690321.8 &        F2-4Ibpe &            &  T2CEP-091 \\
J052149.11-700434.2 &                 & 78.6698.38 &  T2CEP-104 \\
J052519.48-705410.0 &            G0Ib &            &  T2CEP-119 \\
J053150.98-691146.4 &       F4-G1Ibpe & 82.8405.15 &  T2CEP-147 \\
J053932.79-712154.4 &                 &  14.9582.9 &  T2CEP-169 \\
J054312.86-683357.1 &           F7Ibp &            &  T2CEP-180 \\
J055122.52-695351.4 &        F6-8Ibpe &            &  T2CEP-191 \\
\hline                                   
\end{tabular}
\end{table}

The post-AGB tracks cross the high-luminosity end of the Population~II Cepheid instability strip and consequently harbour the subclass of RV~Tauri stars, objects that display complex light curves with alternating deep and shallow minima with often intermittent lower amplitude irregularities. Hence, we expect the known RV~Tauri stars to be part of our sample.

The recent OGLE-III Catalogue of Variable Stars \citep{soszynski08} lists 37 of these objects of which 11 also belong to our sample. This is less than we anticipated, but only these 11 objects are detected by SAGE at both 8 and 24~$\mu$m. None of the selection criteria we used discarded any of them from the final sample, confirming the quality of these criteria. Six of the RV~Tauri stars in our sample are also in the MACHO catalogue of RV Tauri stars \citep{alcock98}. A list of the names of all RV~Tauri stars in our sample, their spectral types and MACHO numbers can be found in Table~\ref{table:rvt}.

Because the Galactic RV~Tauri stars with circumstellar dust are mainly associated with circumstellar discs \citep[e.g.,][]{lloydevans95, lloydevans97, vanwinckel99, gielen07}, we expected something similar to occur in the LMC. When we took a closer look at the subsamples these OGLE objects belong to, we indeed discovered that they all fulfil the selection criteria we used to select the objects with a disc, with four of them also living up to the criteria we imposed for sources with a shell. Detailed SED modelling and, if possible, IR spectroscopic data \citep[e.g.,][]{gielen09b} is needed to clarify the nature of the circumstellar material around these objects.

\subsubsection{R~Coronae Borealis stars}

R~Coronae Borealis (R\,CrB) stars are also linked to the post-AGB evolutionary stage. These are rare objects, of which only about 50 are known in the Galaxy \citep[e.g.,][]{tisserand08}. They are \element[][]{H}-poor supergiants that display sudden deep drops of up to eight magnitudes in their light curve, which are caused by dust obscuration in the line of sight. The star only returns to its normal brightness when the optical depth of the dust in the line of sight has decreased. While at maximum light, R\,CrB stars undergo small-scale, Cepheid-like variability.

Surprisingly many R\,CrB stars were found in the LMC as a byproduct of
the MACHO, EROS-2, and OGLE surveys \citep[][and references therein]{soszynski09a}. Of the 23 known R\,CrB stars and candidates, only one is a member of our final post-AGB sample: J052147.95-700957.0 or MACHO 6.6696.60. This object is a spectroscopically confirmed R\,CrB star that was observed by MACHO, but we obtained no spectrum for it.

One additional object that did not survive the SED-based luminosity-cut -- J051648.08-692222.0 (see Table~\ref{list_delobj}) -- is also listed as an R\,CrB candidate by \citet{soszynski09a} based on the variability of its light curve. This is not supported by its O-B type spectrum, however, in which no carbon chemistry is detected and which displays a strong \element[][]{H}$\beta$ emission peak.

\subsubsection{Other types of objects} \label{sssec:croscollothertypes}

The SEDs of luminous YSOs can be quite similar to those of post-AGB candidates. To estimate the number of YSOs among our post-AGB candidates, we cross-correlated our list with the 270 spectroscopically confirmed YSOs from \citet{seale09} (groups S, SE, P, PE, E and F), the 989 candidate YSOs selected by \citet{whitney08}, and the 1172 candidate YSOs from the catalogue of \citet{gruendl09} that they list as either 'definite' or 'probable'. The latter two catalogues are based on colour-magnitude criteria only. Neither of these catalogues imposed a luminosity criterion or obtained spectra, which renders an overlap with genuine post-AGB stars very likely. A comparison of both catalogues can be found in \citet{gruendl09}. Note that \citet{whitney08} made use of an earlier version of our catalogue to remove some post-AGB candidates from their sample.

\citet{seale09} list 270 embedded YSOs that were spectroscopically confirmed in the IR and selected from the catalogue of \citet{gruendl09} of which 54 are also part of our final sample. J052630.65-674036.6, an object that was discarded because its high final luminosity exceeded our constraints, was also observed by \citet{seale09}, but our optical spectrum indicated that it is a B1e star and it does not lie in an H$\alpha$-rich region of the LMC.

We found out that our sample contains 202 of the 989 objects listed by \citet{whitney08} as probable YSOs of which 119 have a high probability. For 14 of these, we have a low-resolution, optical spectrum, but none of the spectra are indicative of a YSO rather than a post-AGB star. Seven additional objects for which a spectrum was obtained but which have a final luminosity of more than 35\,000~$L_{\odot}$ are also found in this catalogue. Two of these show [\ion{O}{III}] in absorption owing to excess sky-subtraction from a surrounding emission nebula, but lie in an \element[][]{H}$\alpha$ poor region of the LMC. 
We also noted that J053557.86-671333.5 and J051654.04-672005.1, which we identified as a galaxy and a PN-like object respectively (see Sect.~\ref{app:spectra}), both occur in the list of high-probability YSOs.  

Of the objects listed by \citet{gruendl09}, 30 of the 317 probable and 134 of the 855 definite YSO candidates are part of our final sample. For three of these, we obtained a spectrum that is indicative of a post-AGB star rather than a YSO. One additional object -- J052630.65-674036.6 -- whose high final luminosity exceeded our constraints and is therefore no longer in our final sample, shows [\ion{O}{III}] in absorption, but does not lie in an \element[][]{H}$\alpha$-rich region. They also regard the PN-like object J051654.04-672005.1 (Sect.~\ref{sssec:pnlikeobj}) as a definite YSO candidate. 
\newline

We checked the presence of possible red supergiants in our sample by cross-correlating our list with the catalogue of candidate red supergiants of \citet{oestreicher97}. None of their 175 objects appear in our final list of post-AGB candidates and we lost the 24 sources that fulfil the colour criteria thanks to the luminosity-cut. We can therefore conclude that there are probably no red supergiants remaining in our sample of post-AGB candidates thanks to this cut. \newline

To estimate the number of AGB stars remaining in our sample, we used the 29 AGB stars classified by \citet{trams99} without listed spectral type. Four objects -- J044918.46-695314.5, J045538.96-674910.6, J051110.44-675210.6 and J051138.64-665109.8 -- are also part of our final list, but for none of
these was a spectrum obtained. All four objects have an SED that is indicative of a disc. 

\citet{cioni06} listed some colour criteria to identify evolved giant stars, which we used to obtain an extra, but rougher estimate of the number of remaining \element[][]{O} and \element[][]{C}-rich AGB stars. Based on their findings, our sample might contain 68 \element[][]{C}-rich and 446 \element[][]{O}-rich AGB stars, making a total of 514 objects. We have a spectrum for ten of these, none of which shows any indication of molecular bands. The central stars are clearly visible at shorter wavelengths and the estimated temperatures are often too high for AGB stars. The SEDs of the majority of these 514 objects, most of which show an indication of a disc, are therefore more indicative of a post-AGB star than of an AGB object. \newline 

We already discovered two PN-like objects among our post-AGB sample. To assess the total number of PNe that still remain, we cross-correlated our list with the catalogue of \citet{reid06}. We discovered that 59 of the 629 PNe listed by \citet{reid06} are still part of the sample. For one of those -- J051123.63-700157.4 -- and one additional object that was discarded because of its too high final luminosity -- J045433.85-692036.2 -- we already obtained a low-resolution, optical spectrum. These objects display a very peculiar spectrum, however, that is not indicative of a PN: the former is a peculiar emission-line object with a very weak continuum and the latter a B[e] star. Details of both spectra can be found in Table~\ref{table:em}. \newline

Low-resolution, optical spectra of all the sources are required to remove galaxies from our final sample. Nine galaxies that we identified from our spectra are listed in Table~\ref{table:galaxies}. To estimate the fraction of galaxies remaining on our list, we cross-correlated it with the NASA/IPAC Extragalactic Database (NED)\footnote{The NASA/IPAC Extragalactic Database (NED) is operated by the Jet  Propulsion Laboratory, California Institute of Technology, under contract with the National Aeronautics and Space Administration.}. 
We have spectra of eleven of the 172 objects both lists have in common, but none of these show a radial velocity that is incompatible with membership of the LMC. \newline

Finally, to take into account matches with other already studied objects,
we ran a search on SIMBAD. No other polluting types of objects in addition to the ones we already investigated were revealed.

\subsection{Variability}\label{ssec:var}

The variability of our full list of post-AGB candidates was examined by cross-correlating it with the catalogue of long-period variables (LPVs) from MACHO listed by \citet{fraser08}.  Among the 56\,453 LPVs listed in this catalogue, 439 are post-AGB candidates in our catalogue. The most common variable type (245 objects) is the sequence-D variable which is an LPV with a long secondary period \citep{wood99}.  These stars were shown by \citet{wood09} to exhibit a mid-IR excess due to circumstellar dust in a clumpy or disc-like configuration.  Indeed, all but one of our 245 candidates with a sequence-D variability have an SED that is possibly indicative of a disc (or other non-spherical dust distribution), and 13 of these also meet the criteria for a shell. Given the mid-IR excess associated with sequence-D variables, it is thus not surprising that they frequently appear in our catalogue of post-AGB star candidates.  Although the origin of the sequence-D variability is currently unknown \citep{wood04, nicholls09}, it is unlikely that all these objects are true post-AGB stars.  They are far too common, with their numbers being about 30~\% of the number of AGB stars, and their K, M, and C-type spectral types put them still on the AGB. This is consistent with the spectral types of the 3 post-AGB candidates that are listed in sequence~D and for which a spectrum was obtained.

Of the remaining 194 post-AGB candidates in the LPV catalog of \citet{fraser08}, 2 objects belong to sequence 4, 15 to sequence 3, 37 to sequence 2, 33 to sequence 1 and none to sequence E with the LMC period-luminosity sequences mentioned from shortest to longest period in the manner of \citet{fraser08}. All other objects are either part of the one-year artifact, which is caused by the annual observing schedule of the MACHO project (14 objects), fall outside the boundaries of any period-luminosity classification (39 objects), or were unable to be classified (54 objects).\newline

For those objects for which we have an optical spectrum that enabled us to determine their spectral type, we made a detailed examination of the light curves because catalogues of variable stars \citep[e.g.,][]{fraser08} often miss low-amplitude variables. We downloaded light curves from the MACHO website for all post-AGB candidates with a spectral type and a spectrum (a few were missed, presumably because of close bright companions or bad positions on the MACHO CCDs) and we also obtained OGLE~II and OGLE~III light curves for the stars in the OGLE~II and OGLE~III variable star catalogues \citep{udalski08, soszynski08, soszynski09b}. Light curves were obtained for a total of 53 post-AGB candidates from at least one of the above sources. 

\begin{figure*}
\centering
\includegraphics[width=17cm]{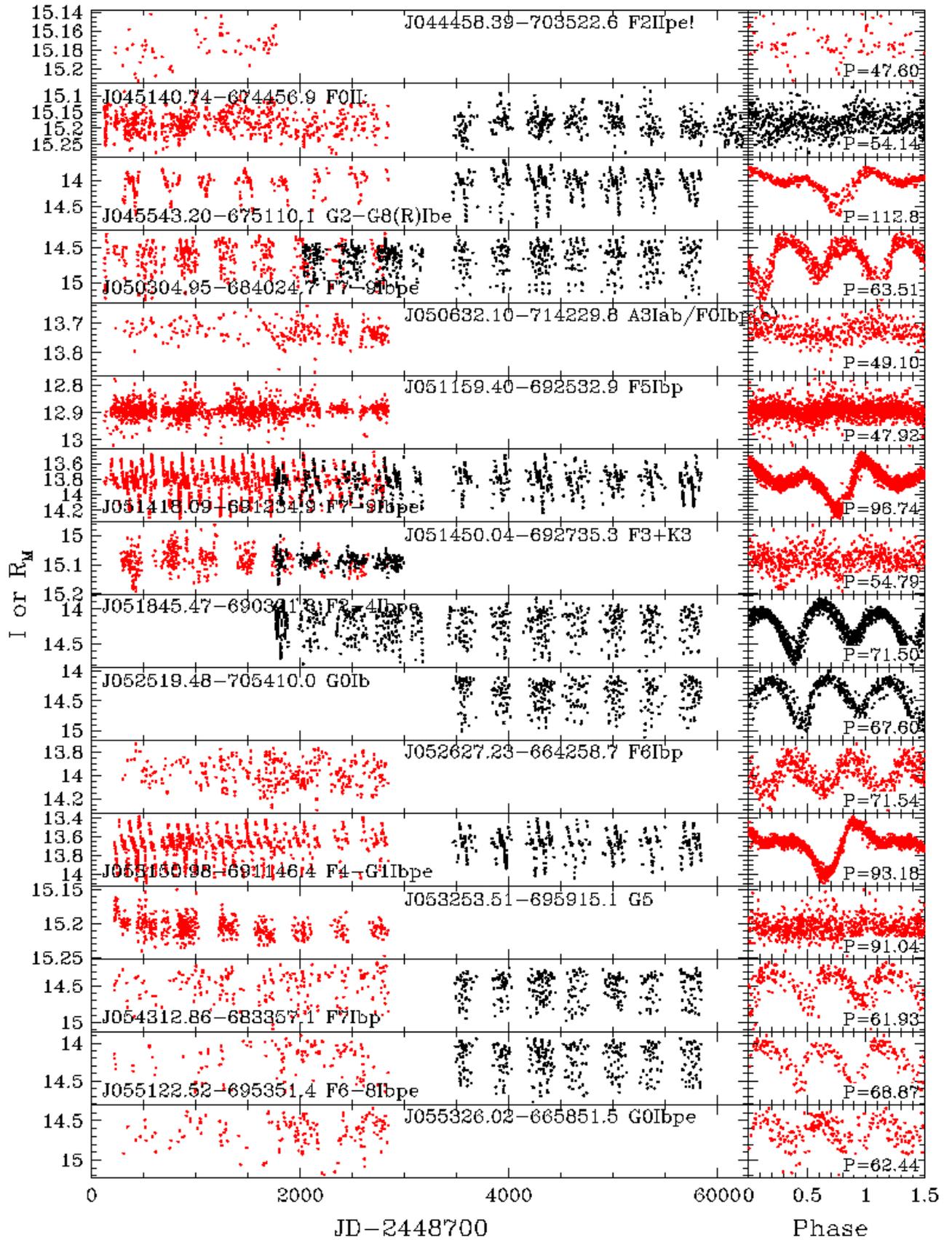}
\caption{Raw and phased light curves of post-AGB candidates that are Population~II Cepheids. Light curves coloured red are MACHO red magnitudes from the MACHO database, while the black curves are I magnitudes from OGLE II and/or OGLE III.  When light curves are available from both MACHO and OGLE, the MACHO red magnitude has been shifted so that its mean agrees with that of OGLE I. The phased light curves were made from one source only, as indicated by the colour.  The name of each object is listed in the figure along with the spectral type and the period, assumed to be the time between deep minima (or two cycles of the light curve where the minima are of equal depth).
}
\label{fig:lc_popIIceph}
\end{figure*}

\begin{figure*}
\centering
\includegraphics[width=17cm]{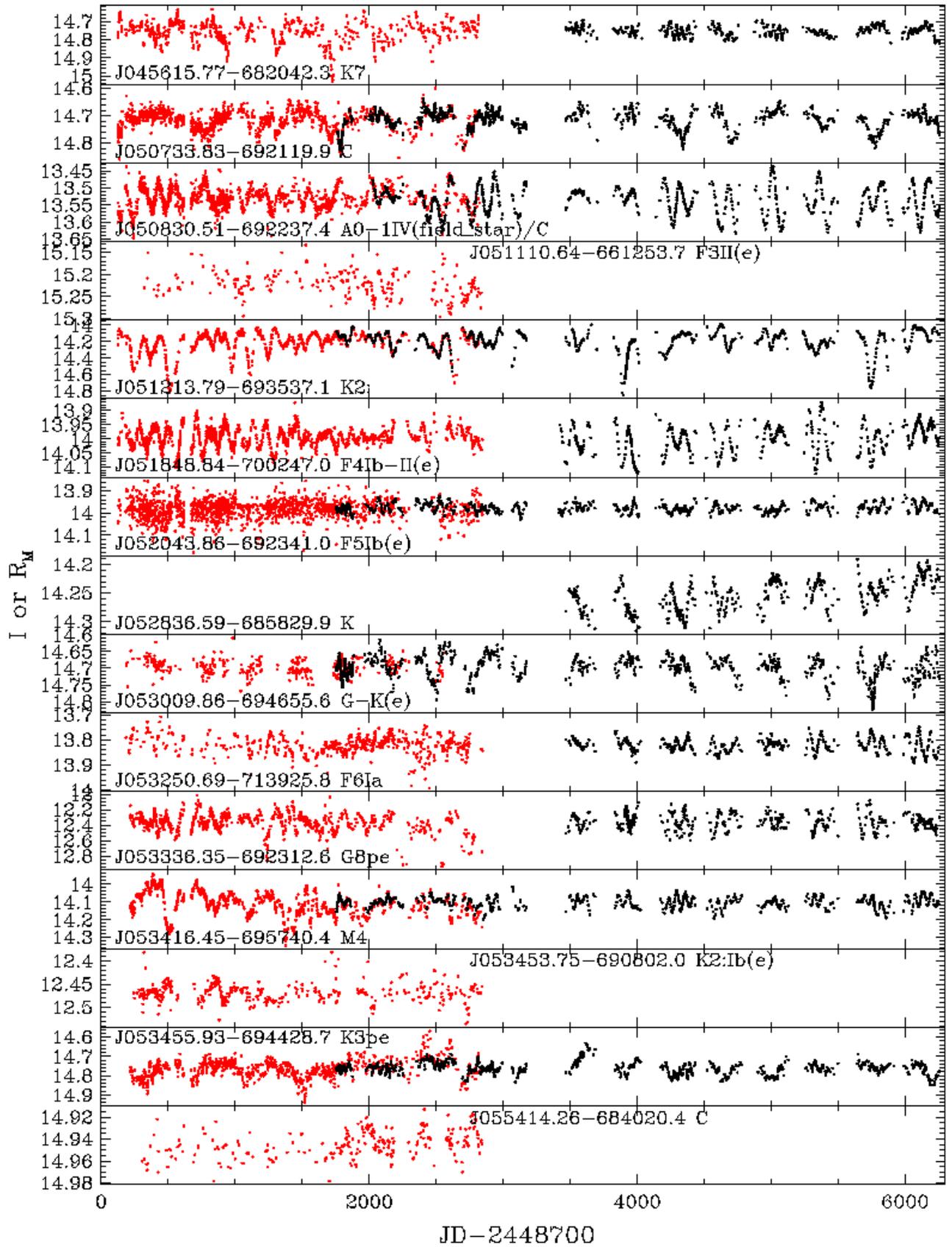}
\caption{Same as Fig.~\ref{fig:lc_popIIceph} but for the SRVs, and without the phased light curves.
}
\label{fig:lc_lpv}
\end{figure*}

\begin{figure*}
\centering
\includegraphics[width=17cm]{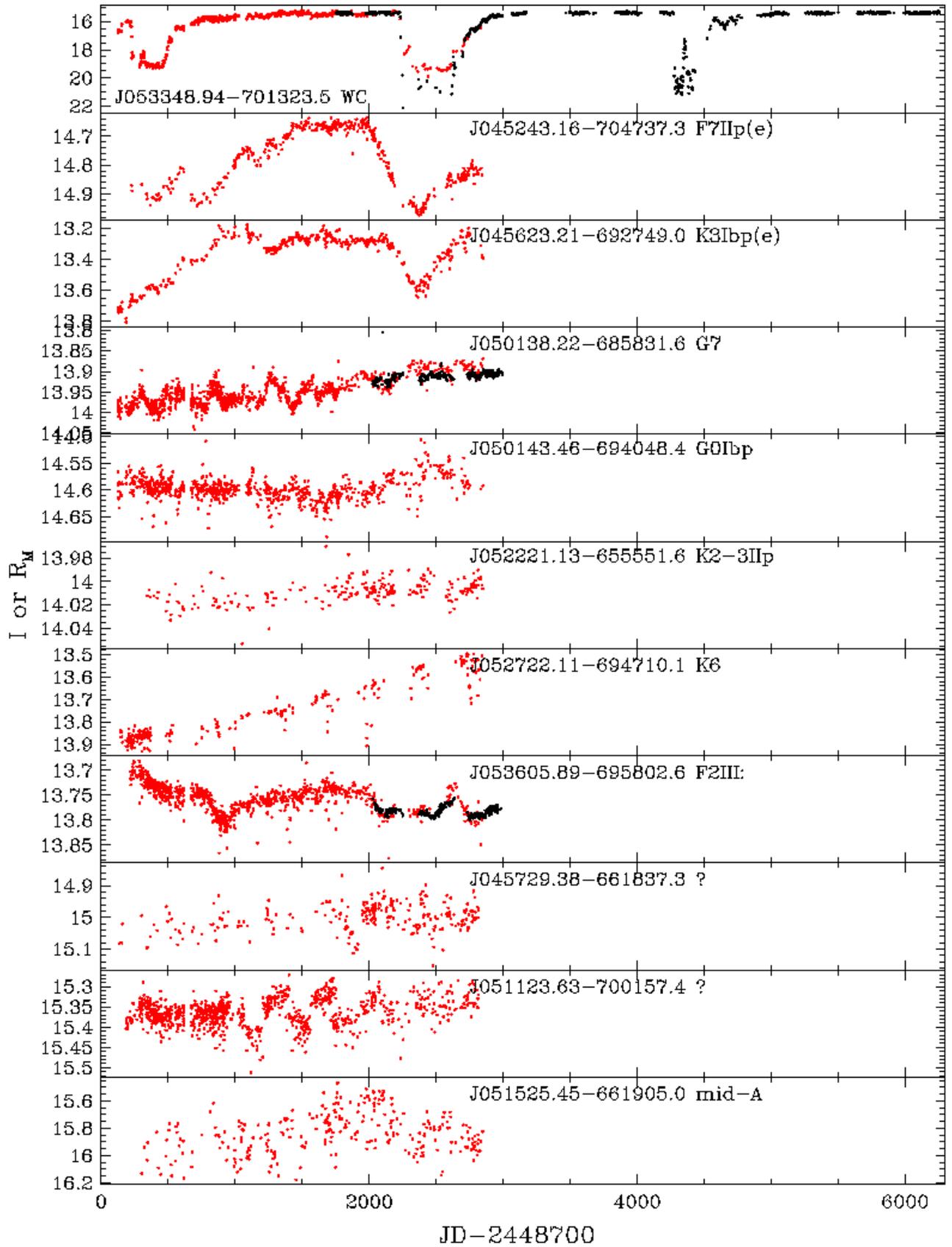}
\caption{Same as Fig.~\ref{fig:lc_lpv} but for post-AGB candidates showing slow variations, and for three objects in regions of strong nebular emission.
}
\label{fig:lc_other_slow}
\end{figure*}

\begin{figure*}
\centering
\includegraphics[width=17cm]{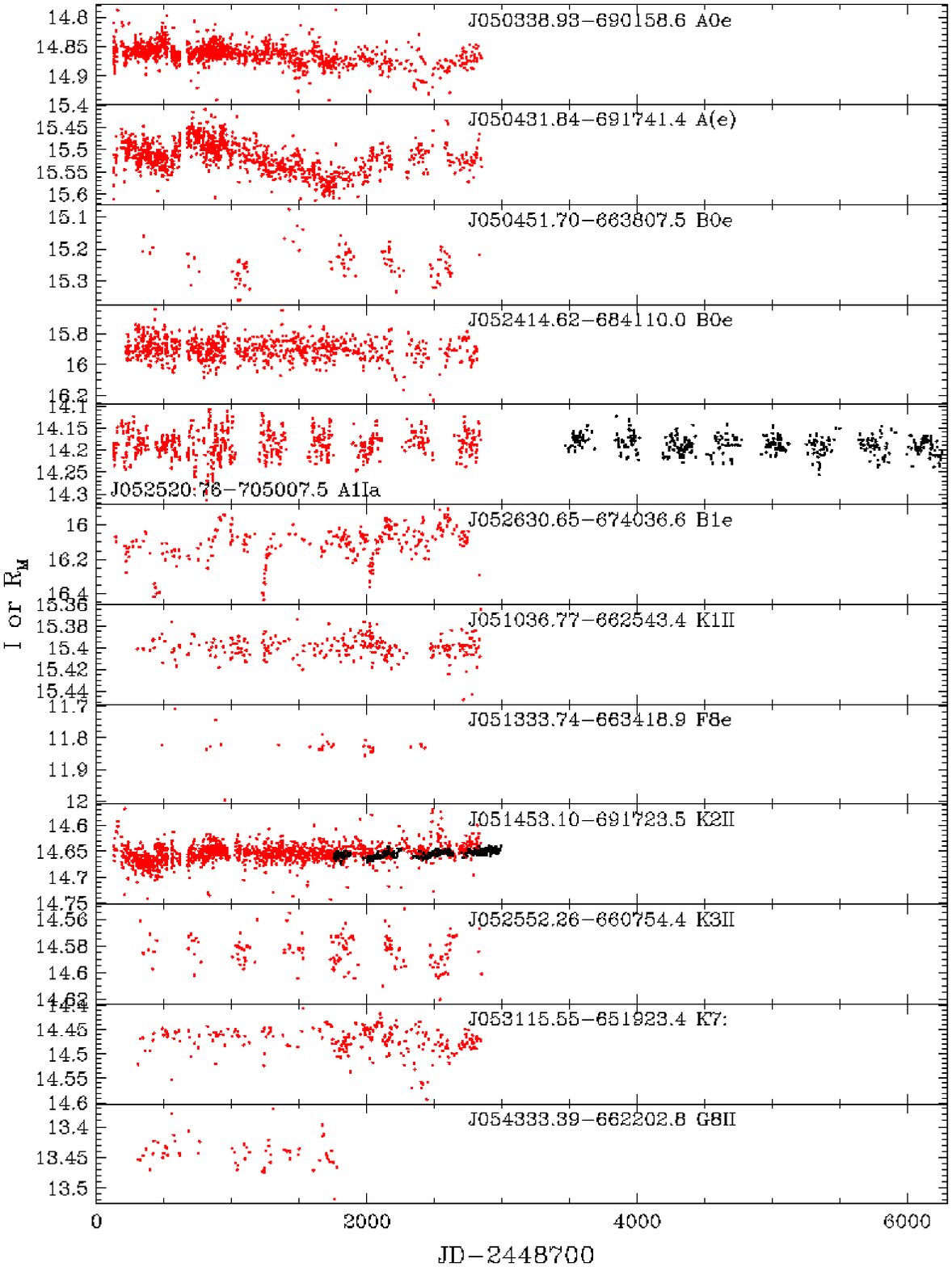}
\caption{Same as Fig.~\ref{fig:lc_lpv} but for post-AGB candidates with Ae or Be spectral types, or for post-AGB candidates without evidence of any significant light variation.
}
\label{fig:lc_other_be_nonvar}
\end{figure*}

The most common variability types among the post-AGB candidates are Population~II Cepheids (including RV~Tauri stars) and semi-regular variables (SRVs). Light curves of the 16 Population~II Cepheids we found are shown in Fig.~\ref{fig:lc_popIIceph}, their periods and MACHO numbers can be found in Table~\ref{table:ceph}. All periods were derived using the PDM routine in IRAF. The eight higher amplitude stars in this figure were previously-known variables, and they are the objects with a spectral type listed in Table~\ref{table:rvt} (see Sect.~\ref{sssec:rvt} for details of the other stars in Table~\ref{table:rvt}).  The other eight stars in Fig.~\ref{fig:lc_popIIceph} are new Population~II Cepheids. The objects J052627.23-664258.7 and J055326.02-665851.5 have moderate amplitudes of $\sim$ 0.3 magnitudes, but most of the new Population~II Cepheids are of very low amplitude ($<0.1$ magnitudes).  These are more examples of the ultra-low amplitude variables discussed by \citet{buchler05}, \citet{buchler09} and \citet{vanwinckel09}.  One of the objects found here (J050304.95-684024.7) was also discussed by \citet{buchler09}.  As expected, all Population~II Cepheids have F and G spectral types. The bulk of these objects (10 sources) have SEDs that are indicative of a disc, two are shell sources and the remaining four display the criteria for both groups. All these stars are presumably genuine post-AGB stars evolving away from the AGB across the HR~diagram.

\begin{table}
\caption{Periods, MACHO numbers and spectral types of the Population~II Cepheids in our sample.}             
\label{table:ceph}      
\centering                          
\begin{tabular}{lccl}        
\hline\hline                 
Object Name (IRAC) & P (days) & MACHO & SpT\\ 
\hline                        
J044458.39-703522.6   & 47.6004 & 46.761.549   & F2IIpe! \\ 
J045140.74-674456.9   & 54.1438 & 47.1893.34   & F0II: \\ 
J045543.20-675110.1   & 112.828 & 47.2496.8    & G2-8(R)Ibe:\\ 
J050304.95-684024.7   & 63.51   & 19.3694.19   & F7-9Ibpe \\ 
J050632.10-714229.8   & 49.1    & 38.4253.12   & A3Iab\\
                      &         &              & /F0Ibp(e) \\ 
J051159.40-692532.9   & 47.92   & 79.5135.14   & F5Ibp \\ 
J051418.09-691234.9   & 96.745  & 79.5501.13   & F7-9Ibpe \\ 
J051450.04-692735.3   & 54.785  &  5.5618.3940 & F3 + K3\\ 
J051845.47-690321.8   & 71.505  &              & F2-4Ibpe \\ 
J052519.48-705410.0   & 67.600  &              & G0Ib \\ 
J052627.23-664258.7   & 71.537  & 62.7474.1378 & F6Ibp \\ 
J053150.98-691146.4   & 93.180  & 82.8405.15   & F4-G1Ibpe\\ 
J053253.51-695915.1   & 91.04   & 81.8514.57   & G5 \\ 
J054312.86-683357.1   & 61.933  & 33.10229.18  & F7Ibp \\ 
J055122.52-695351.4   & 68.874  & 30.11540.16  & F6-8Ibpe \\ 
J055326.02-665851.5   & 62.444  & 69.11947.14  & G0Ibpe \\ 
\hline                                   
\end{tabular}
\end{table}

Light curves of the 15 SRVs are shown in Fig.~\ref{fig:lc_lpv}. Five of these SRVs -- J045615.77-682042.3, J050733.83-692119.9, J053009.86-694655.6, J053416.45-695740.4 and J053455.93-694428.7 -- show a long secondary period and hence they are sequence-D variables. Three were first recognised by \citet{fraser08} and the other two by us. They are of K, M, or C spectral type, just like another five of the SRVs. Most of these stars, including the two carbon stars J050830.51-692237.4 and J055414.26-684020.4, may still be on the AGB, but we identified the third carbon star (the sequence-D variable J050733.83-692119.9) as a candidate post-AGB star (Sect.~\ref{ssec:Cstar}). The other five SRVs have unusually warm spectral types of F or G. This is not expected for semi-regular variables.

Some of the post-AGB candidates with known spectral type showed evidence for long-term, slow variations and their light curves are shown in Fig.~\ref{fig:lc_other_slow}.  The first object, J053348.94-701323.5, is a previously-know R\,CrB star that is discarded from the final sample because of its WC-like spectrum. The next seven objects down the figure show a bizarre behaviour, including some resemblance to that of R\,CrB stars as well as long-term trends in brightness.  This behaviour is most likely associated with rapidly changing extinction, which might be expected for AGB stars that have just ejected their envelopes, or which have discs that have not settled into a steady state.  The final three objects in Fig.~\ref{fig:lc_other_slow} are in regions of very strong nebular emission.  All of them seem to show long-term brightness variations.

The final group of light curves (Fig.~\ref{fig:lc_other_be_nonvar}) consists of Ae and Be stars (the top six light curves) and post-AGB candidates that do not show strong evidence of variability (the lower six light curves).  Most of the Ae and Be stars show evidence for short period oscillations, as well as longer-term trends.  This behaviour is common in Ae and Be systems and the variability is thought to be caused by binarity along with possible nonradial pulsation \citep{mennickent05,mennickent06,mennickent10}.

In summary, the light curves of the post-AGB candidates provide useful information to support or reject post-AGB status.  Stars with Population~II Cepheid variations must be in the instability strip, and they can only be in this part of the HR~diagram if they have evolved away from the AGB along their post-AGB evolutionary tracks. Some of the stars show unusual long-term variations that are suggestive of variable extinction, which is likely the result of the clearing and/or accretion of recently ejected matter: these stars are also likely to be post-AGB stars.

\subsection{Efficiency of our selection criteria}

\begin{figure*}
\centering
\includegraphics[width=17cm]{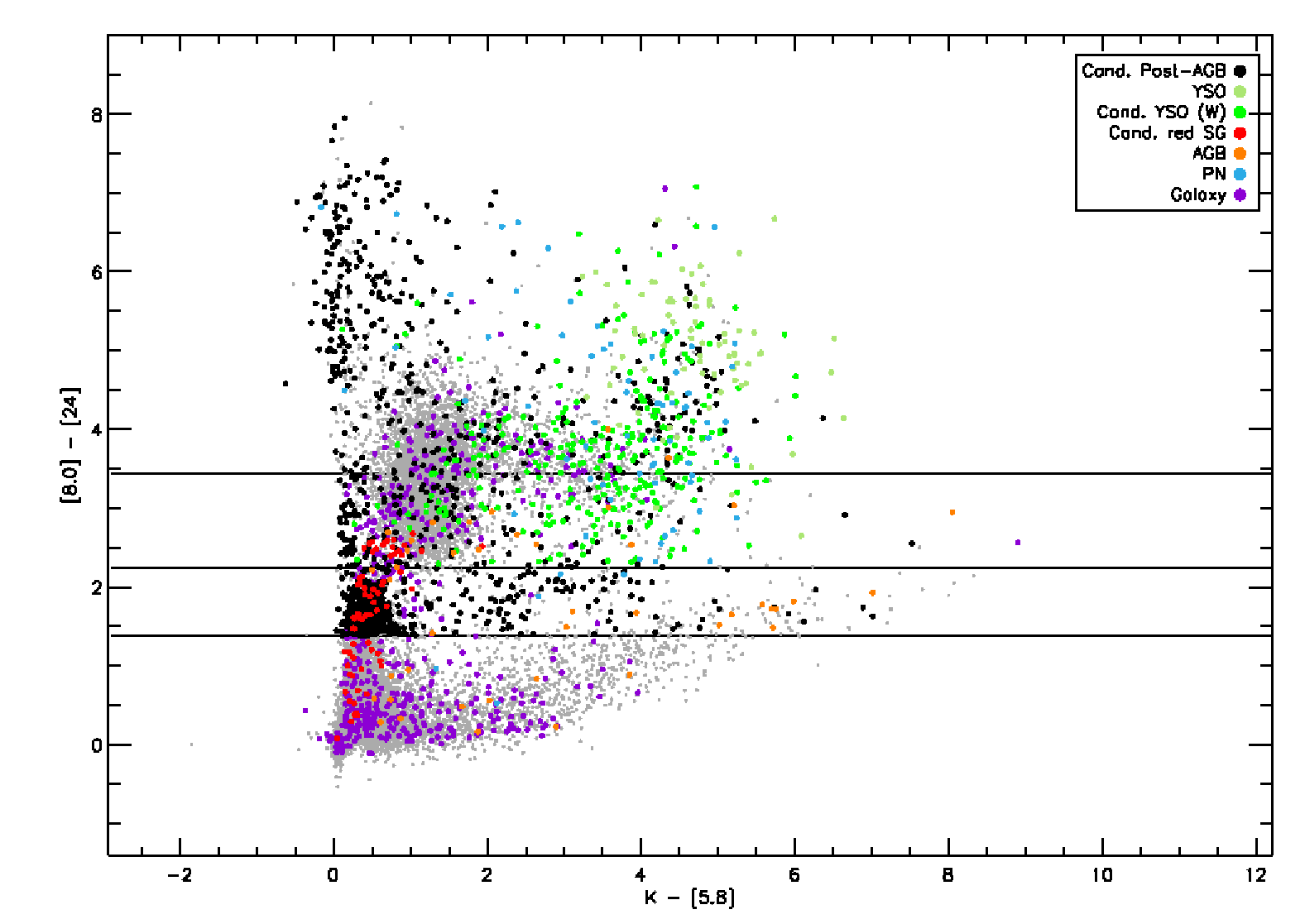}
\caption{Colour-colour plot of $K - [5.8]$ vs $[8] - [24]$. The displayed objects are the final sample of post-AGB candidates (black dots) together with the YSOs from \citet{seale09} (khaki dots), the high-probability candidate YSOs from \citet{whitney08} (green dots), the red supergiants from \citet{oestreicher97} (red dots), the AGB stars from \citet{trams99} (orange dots), the PNe from \citet{reid06} (blue dots) and all galaxies from NED that are in the line of sight of the LMC (purple dots). The grey dots in the background represent the initial SAGE catalogue and the black horizontal lines the boundaries of the different subtypes of our post-AGB sample based on the shape of the SED. Objects with an SED that is indicative of a disc can be found between the bottom and the top line, while those with a shell are above the one in the middle. The bottom line also indicates the colour-cut we used.}
 \label{fig:effcolcrit}
\end{figure*}

To evaluate our selection procedure, we reviewed the different selection criteria one by one. In Fig.~\ref{fig:effcolcrit} we display $K - [5.8]$ versus $[8.0] - [24]$ for all objects in our final sample of post-AGB candidates together with the YSOs from \citet{seale09}, the high-probability candidate YSOs from \citet{whitney08}, the red supergiants from \citet{oestreicher97}, the AGB stars from \citet{trams99}, the PNe from \citet{reid06} and all galaxies from NED that are in the line of sight of the LMC. The grey dots in the background represent the initial SAGE catalogue. All objects in this plot with $[8.0] - [24]$ larger than 1.384 are initially part of our sample of post-AGB candidates. Obviously our initial criterion already removes some galaxies, red supergiants, and AGB stars from the sample, but the other types have colours in strong overlap with the post-AGB candidates. \newline

\begin{figure*}
\centering
\includegraphics[width=17cm]{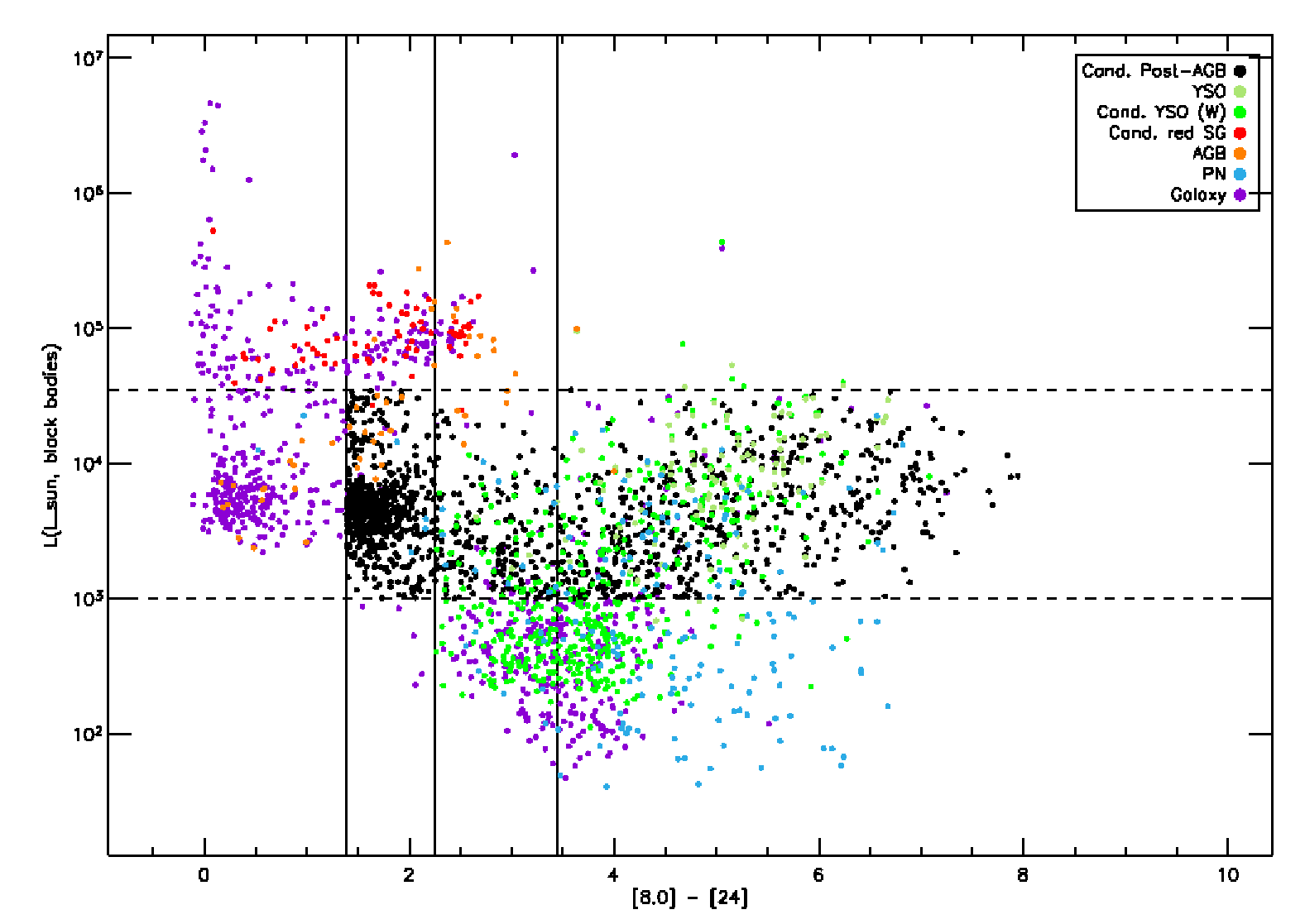}
\caption{Plot of the estimated luminosities from Sect.~\ref{ssec:lumcut} versus the $[8] - [24]$ colour. The displayed objects are the same as in Fig.~\ref{fig:effcolcrit}. The vertical lines are the boundaries of the different subtypes of our post-AGB sample based on the shape of the SED, where the objects with a disc can be found between the two outer lines and those with a shell are to the right of the middle one. The dashed horizontal lines indicate the luminosity-cut we used.}
 \label{fig:efflumcrit}
\end{figure*}

We computed the luminosity estimates of the objects from the different catalogues by fitting up to three black bodies to their photometry and integrating, as was previously described for our sample of post-AGB candidates (see Sect.~\ref{ssec:lumcut}). Fig.~\ref{fig:efflumcrit} shows that the luminosity-cut we performed to remove the bulk of the supergiants from the sample was indeed effective. All remaining supergiants and some extra galaxies and AGB stars were discarded from our sample. Distinguishing between YSOs and post-AGB stars proved to be a bit more difficult: the lower limit we imposed for the luminosity discards the bulk of the candidate YSOs from \citet{whitney08} but leaves almost all more luminous and spectroscopically confirmed YSOs of \citet{seale09} in the sample. Because of the similarity of the SEDs of YSOs and post-AGB stars, no other colour-colour plane can help us solve this issue. Spectra are therefore needed to discard the YSOs from our final catalogue. Some dim PNe and galaxies are also left out of the sample by this lower limit.

\subsection{Completeness of the catalogue}

\begin{figure}
\resizebox{\hsize}{!}{ 
\includegraphics{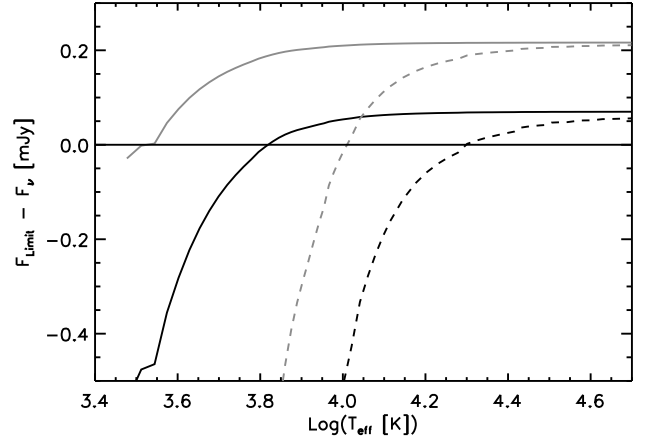}}
\caption{Extra amount of flux above the naked atmosphere needed in the system of a post-AGB star to be detected by Spitzer. We display the difference between the faint limit at 8 (black) and 24~$\mu$m (grey) and the minimal amount of flux that should be present in the system of a naked post-AGB candidate in those filters for a range of effective temperatures of the central star. The solid and dashed lines correspond to a luminosity of 1000 and 35\,000~$L_{\odot}$ respectively.}
 \label{fig:completeness}
\end{figure}

The completeness of the catalogue of optically bright post-AGB candidates is limited by the sensitivity of the IRAC 8.0~$\mu$m and MIPS 24~$\mu$m filters, which have a faint limit of respectively 14.9 and 11.3 in magnitude
\footnote{\url{http://data.spitzer.caltech.edu/popular/sage/20090922_enhanced/documents/SAGEDataProductsDescription_Sep09.pdf}}. For the naked stellar atmosphere models introduced in Sect.~\ref{subsection:SEDs}, this means that an atmosphere model placed at the distance of the LMC and with a luminosity of 1000~$L_{\odot}$ is no longer detectable at 8~$\mu$m if it has a temperature above 6750~K. For a luminosity of 35\,000~$L_{\odot}$ this only happens at 20\,000~K. 

The initial selection criterion we required for post-AGB candidates, $[8]-[24]>1.384$, implies the presence of dust in the system at 24~$\mu$m and consequently a lower limit for the flux emitted at this wavelength for an object to become part of our final sample. The above mentioned models with no excess at 8~$\mu$m, a flux at 24~$\mu$m that obeys the criterion, and luminosities of 1000 and 35\,000~$L_{\odot}$ will only be detected for effective temperatures below 3500 and 10\,500~K. 

Accordingly we detect all post-AGB candidates with an effective temperature less than 3500~K, while objects with a temperature above 20\,000~K need a dust excess at 8 and 24~$\mu$m to be found. For all temperatures in between, we lack mainly the hotter objects with minimal dust excess. In Fig.~\ref{fig:completeness} we show the minimal amount of flux needed in each filter for a post-AGB candidate to be detected for a range of effective temperatures. \newline

\citet{wood01} found 25 possible post-AGB candidates in the LMC. The candidates had a detection in the 8.28~$\mu$m band of the Midcourse Space Experiment (MSX) satellite, together with luminosities and $J-K$ colours that suggested these candidates were transiting from the AGB to higher temperatures in the HR~diagram. These objects should have been detected in the IRAC 8~$\mu$m band, but only 17 can be found in the SAGE catalogue constructed by us (Sect.~\ref{ssec:colcrit}). One object did not fulfill the colour criteria we imposed, four did not have optical magnitudes in the requested filters and two additional MSX-selected post-AGB candidates were detected at 70 and/or 160~$\mu$m in the MIPS filters and had an SED that monotonically increased towards wavelengths redder than 70~$\mu$m.
The final object from the MSX PAGB candidate list that did not make it to the present survey is J053348.94-701323.5: it is an R\,CrB star \citep{wood01, alcock01} and was discarded from our final sample because it is also a WC star (see Sect.~\ref{sssec:wc_spec}).

Nine post-AGB candidates are common to our current candidate list and to the MSX post-AGB candidate list of Wood \& Cohen (2001, and unpublished). The object J050718.33-690742.9 in Table~\ref{catalogue_specundet} with a spectral type A7II/A9I is an MSX post-AGB candidate: the spectra obtained by Wood \& Cohen give a similar spectral type of F0e:, it has emission in the core of \element[][]{H}$\alpha$.  The objects J051845.23-700534.5, J052147.95-700957.0, J052337.91-693854.1, J052804.80-685947.2, J053010.30-690933.8, J053306.86-703033.9, J053327.17-700951.5 and J053431.48-683513.7 are also MSX post-AGB candidates, but we did not obtain a spectrum for any of them.  Of these stars, J052147.95-700957.0 is an R\,CrB star \citep{wood01, alcock01}, J051845.23-700534.5, J052804.80-685947.2 and J053431.48-683513.7 are stars with \element[][]{H}$\alpha$ emission, while J053010.30-690933.8 and J053306.86-703033.9 are M stars.

There are two MSX post-AGB candidates that appear to be genuine stellar candidate post-AGB stars but which are not in our current catalogue. J052048.20-701212.6 is a star with an R\,CrB-like light curve \citep{wood01, alcock01} and J052551.84-684634.2 is an M star. Both objects had no counterpart in the optical catalogues we used. The other MSX post-AGB candidates (making up the total of 25 in \citet{wood01}) are in regions of strong nebular emission (perhaps because of the MSX source) or do not have optical counterparts and so their post-AGB status is uncertain.  

Ignoring the M stars (which are probably still on the AGB) and the objects in regions of strong nebular emission, we find that of the eight MSX post-AGB candidates for which there is spectral or light curve information that supports the initial post-AGB selection, seven are in our current catalogue. In addition to this, we found in Sect.~\ref{sssec:rvt} that all 11 known RV~Tauri stars that are detected by SAGE at both 8 and 24~$\mu$m are members of our sample. Our selection criteria may thus have rejected roughly 1 in 20 (5~\%) of the potential optically bright post-AGB candidates.

\section{Conclusions}

We present our catalogue of post-AGB candidates based on the second data release of the point-source catalogue of the SAGE-Spitzer LMC survey \citep{meixner06}. Our method combines photometry, low-resolution spectroscopy, and detailed SED fitting and includes constraints on the luminosity. Our conclusions and future research lines can be summarised as follows: 

\begin{itemize}

\item The catalogue of 70 spectroscopically confirmed and 1337 good post-AGB candidates we constructed is the first step towards a homogeneous and complete list of post-AGB stars in the LMC. Spectroscopic confirmation is still needed for the majority of the candidates. About 25 post-AGB stars were known in the LMC before our endeavours.

\item Our separation between sources with a disc and a shell is based on the SED characteristics of thermal emission from the circumstellar dust. This yields 624 discs, 568 shells and 215 undetermined sources. 

\begin{itemize}
\item As in the Galaxy, the disc sources form a very significant fraction of the optically bright post-AGB stars. The evolution rate of post-AGB stars depends on the rate at which the remaining \element[][][][]{H}-rich envelope is consumed. The consumption can be from within by \element[][][][]{H}-shell burning, which just depends on the luminosity and is the same for disc and shell sources, or by mass-loss from the exterior. As in the Galaxy, some disc sources in the LMC show the effect of accretion \citep{reyniers07b, gielen09b}, which increases the mass of the \element[][][][]{H}-rich envelope and slows down the evolutionary timescale of the central star. Because the discs are stable, the IR lifetime of these objects is expected to be much longer than for shell objects \citep{vanwinckel03}, which allows strong grain processing to occur. In a limited study of \citet{gielen09b}, this is also detected in the few LMC disc sources that have been published until now. Disc sources in the Galaxy are associated with binary central stars \citep[e.g.,][]{vanwinckel09}. In our future research, we will focus on a detailed full sample study of the disc sources, but the direct proof of the high binary rate in this sample will be an observational monitoring challenge.

\item The objects that we characterise as shell sources cover a wide range in luminosity and hence initial mass.
Detailed spectroscopic analyses of the photospheres will allow testing of the chemical enrichment predictions of AGB theoretical models in full detail with a well constrained sample of objects with known luminosities (and hence final and initial masses). Systematic IR spectroscopic surveys were defined before the construction of our catalogue, but the available Spitzer spectra \citep{kemper10, woods11} form an ideal complement that is needed to fully characterise the post-AGB stars.

\end{itemize}

\item We have shown that light curves provide useful information on whether a post-AGB candidate is indeed a genuine post-AGB star.  Looking at light curves doubled the number of known Population~II Cepheids in our sample of 53 stars with spectra from 8 to 16 (these are almost certainly post-AGB stars).  The light curves also revealed long-term changes in magnitude in eight other objects that strongly hint at genuine post-AGB status. 

\end{itemize}

The catalogue offers strong potential in the study of mass-dependant properties of post-AGB stars such as AGB nucleosynthetic yields via photospheric abundances studies, and mass-loss rates by the distribution of post-AGB stars in the HR~diagram. The disc sources that are likely associated with binary central stars will allow testing of the poorly understood binary evolution channels. Clearly, our extensive catalogue with well-constrained luminosities is just the first step in the exploitation of the SAGE project for extragalactic post-AGB research.

\begin{acknowledgements}
E. van Aarle acknowledges support from the Fund for Scientific Research of
Flanders (FWO) under  grant number G.0470.07. 

We would like to thank the referee for instructive remarks.

This paper utilizes public domain
data obtained by the MACHO Project, jointly funded by the US
Department of Energy through the University of California, Lawrence
Livermore National Laboratory under contract No. W-7405-Eng-48, by the
National Science Foundation through the Center for Particle
Astrophysics of the University of California under cooperative
agreement AST-8809616, and by the Mount Stromlo and Siding Spring
Observatory, part of the Australian National University. 
\end{acknowledgements}

\bibliographystyle{aa} 
\bibliography{15834} 

\longtab{3}{
\begin{longtable}{cccccccccclc}
\caption{\label{table:prinfeat_em} Principal features in the spectra of J045433.85-692036.2, J045729.38-661837.3, J054034.00-692509.8, J051525.45-661905.0, and J060323.20-680437.5 and their identification. Some extra information on certain features is given with a preceeding letter. Here, 'A' stands for an absorption feature, in some cases associated with P~Cygni emission, 'P' for P~Cygni emission, 'b' for broad line, 'c' for crowded line or composite feature, 'd' for a line that appears to be an unresolved double feature, 'w' for a weak feature and '?' notes a doubtful feature. 
}\\
\hline\hline
\multicolumn{10}{c}{Observed Features} & \multicolumn{2}{c}{Feature Identification} \\
\multicolumn{10}{c}{(\AA{})} & \multicolumn{2}{c}{(\AA{})} \\
\multicolumn{2}{c}{J045433.85} & \multicolumn{2}{c}{J045729.38\tablefootmark{a}} & \multicolumn{2}{c}{J054034.00} & \multicolumn{2}{c}{J051525.45} & \multicolumn{2}{c}{J060323.20} &                     &         \\
\hline
\endfirsthead
\caption{continued.}\\
\hline\hline
\multicolumn{10}{c}{Observed Features} & \multicolumn{2}{c}{Feature Identification} \\
\multicolumn{10}{c}{(\AA{})} & \multicolumn{2}{c}{(\AA{})} \\
\multicolumn{2}{c}{J045433.85} & \multicolumn{2}{c}{J045729.38\tablefootmark{a}} & \multicolumn{2}{c}{J054034.00} & \multicolumn{2}{c}{J051525.45} & \multicolumn{2}{c}{J060323.20} &                     &         \\
\hline
\endhead
\hline
\endfoot
  &                 &   &                 & b &  3651.93           &   &                   &   &                   &                      &         \\
  &  3680.68        &   &                 &   &                    &   &                   &   &                   &                      &         \\
  &  3688.25        &   &                 &   &                    &   &                   &   &                   &                      &         \\
  &                 &   &                 &   &  3730.12           &   &                   &   &                   & [OII](1F)            & 3726.16 \\ 
  &                 &   &                 &   &                    &   &                   &   &                   &                      & 3728.91 \\
  &  3762.91        &   &                 &   &                    &   &                   &   &                   &                      &         \\       
  &                 &   &                 &   &                    & A & 3800.26           &   &                   & HI(2)                & 3797.90 \\
  &                 &   &                 &   &                    & A & 3837.59           & A & 3835.07           & HI(2)                & 3835.39 \\
\hline                  
\end{longtable}
 \tablefoot{
\tablefoottext{a}{It seems anomalous that an apparent P~Cygni structure is seen at H$\gamma$ and H$\delta$, whereas the emission appears symmetrical with no sign of absorption at H$\alpha$ and H$\beta$. The alternative, that the emission and absorption appear in different objects, is unattractive unless we see a binary with a high velocity amplitude as the absorption components are blueshifted. Higher resolution spectra would be useful.}\\
\tablefoottext{b}{The triplet at 7774~\AA{} is normally seen in absorption and is a useful luminosity criterion from late B to early G type stars, becoming  very strong in the most luminous supergiants. It may appear in emission when the line at 8446~\AA{} is strongly in emission. A caveat is that, given our lower resolution, the emission line near 8446~\AA{} must be a blend of the OI triplet with the stronger unidentified line found  by \citet{thackeray62}.}
}
}

\appendix

\section{Tables of the subsample of post-AGB candidates with a low-resolution, optical spectrum} \label{app:catsample}

We display all data for the post-AGB candidates in our final sample that have a spectrum in three tables subdivided into the three subclasses based on the shape of the SED. 

\begin{landscape}
\begin{table}
\caption{\label{catalogue_specdisc} Catalogue of the objects in our final sample that have a low-resolution, optical spectrum and an SED indicative of a disc. We list for each object the name, the position, the calculated values of the luminosity (both based on the BB fit and the SED), the effective temperature, the interstellar reddening, the spectral type, some variability remarks and the catalogues of other types of objects in the LMC in which it is listed.}
\centering                                      
\begin{tabular}{lcccccclll}          
\hline\hline                        
Object & Right Ascension & Declination & $L_{\mathrm{BB}}$ & $L_{\mathrm{SED}}$ & $T_\mathrm{eff,\ SED}$ & $E(B-V)_{\mathrm{SED}}$ & Spectral Type\tablefootmark{a} & Variability\tablefootmark{b} & Remarks\tablefootmark{c} \\
& (h:m:s) & (d:m:s) & ($L_{\odot}$) & ($L_{\odot}$) & (K) & & & &  \\
\hline
J044458.39-703522.6 & 04:44:58.39 & -70:35:22.6&   2000 & $   2000 \pm      50 $ & $   7000 \pm     250$ & $ 0.04 \pm  0.01$ &                                                     F2IIpe! &                         Ceph. (47.60004) &                          \\
J045243.16-704737.3 & 04:52:43.17 & -70:47:37.4&   2000 & $   2170 \pm      37 $ & $   5500 \pm     250$ & $ 0.17 \pm  0.01$ &                                                    F7IIp(e) &                                       SV &                          \\
J045623.21-692749.0 & 04:56:23.21 & -69:27:49.0&   6000 & $   5780 \pm      90 $ & $   4250 \pm     250$ & $ 0.07 \pm  0.01$ &                                                    K3Ibp(e) &                  SV, Fr$_{\mathrm{-99}}$ &                          \\
J050143.46-694048.4 & 05:01:43.47 & -69:40:48.5&   2000 & $   2900 \pm      60 $ & $   5000 \pm     250$ & $ 0.23 \pm  0.01$ &                                                       G0Ibp &                                       SV &                          \\
J050304.95-684024.7 & 05:03:04.96 & -68:40:24.8&   3000 & $   2700 \pm      40 $ & $   5750 \pm     250$ & $ 0.05 \pm  0.01$ &                                                    F7-9Ibpe &                       Ceph. (63.51), RVT &                        AG\\
J051159.40-692532.9 & 05:11:59.40 & -69:25:33.0&  11000 & $  12200 \pm     200 $ & $   6250 \pm     250$ & $ 0.12 \pm  0.01$ &                                                       F5Ibp &                            Ceph. (47.92) &                          \\
J051333.74-663418.9\tablefootmark{d} & 05:13:33.74 & -66:34:19.0&  20000 & $  32500 \pm     270 $ & $   6250 \pm     250$ & $ 0.00 \pm  0.00$ &                                                    F8-G0Ipe &                                       NV &                          \\
J051418.09-691234.9 & 05:14:18.10 & -69:12:35.0&   7000 & $   8300 \pm     800 $ & $   6250 \pm     250$ & $ 0.25 \pm  0.04$ &                                                    F7-9Ibpe &                      Ceph. (96.745), RVT &                        AG\\
J051845.47-690321.8 & 05:18:45.47 & -69:03:21.8&   4000 & $   6300 \pm     190 $ & $   6750 \pm     250$ & $ 0.29 \pm  0.01$ &                                                    F2-4Ibpe &                      Ceph. (71.505), RVT &                        AG\\
J052221.13-655551.6 & 05:22:21.14 & -65:55:51.6&   4000 & $   5200 \pm      80 $ & $   4250 \pm     250$ & $ 0.26 \pm  0.01$ &                                                     K2-3IIp &                                       SV &                          \\
J052519.48-705410.0 & 05:25:19.49 & -70:54:10.0&   3000 & $   2140 \pm      27 $ & $   5500 \pm     250$ & $ 0.00 \pm  0.00$ &                                                        G0Ib &                      Ceph. (67.600), RVT &                        AG\\
J052627.23-664258.7 & 05:26:27.23 & -66:42:58.7&   4000 & $   4590 \pm     100 $ & $   6250 \pm     250$ & $ 0.14 \pm  0.01$ &                                                       F6Ibp &                           Ceph. (71.537) &                         G\\
J053253.51-695915.1 & 05:32:53.51 & -69:59:15.1&   1200 & $   1440 \pm      38 $ & $   4750 \pm     250$ & $ 0.11 \pm  0.01$ &                                                          G5 &                            Ceph. (91.04) &                          \\
J053336.35-692312.6 & 05:33:36.35 & -69:23:12.7&  19000 & $  11640 \pm     180 $ & $   4750 \pm     250$ & $ 0.02 \pm  0.01$ &                                                        G8pe &                 SRV, Fr$_{\mathrm{-99}}$ &                         G\\
J053453.75-690802.0 & 05:34:53.76 & -69:08:02.1&  30000 & $  32400 \pm    1400 $ & $   4250 \pm     250$ & $ 0.26 \pm  0.03$ &                                                    K2:Ib(e) &                                      SRV &                          \\
J053605.89-695802.6 & 05:36:05.90 & -69:58:02.7&   6000 & $   9210 \pm     180 $ & $   6750 \pm     250$ & $ 0.23 \pm  0.01$ &                                                      F2III: &                                       SV &                         G\\
J054312.86-683357.1 & 05:43:12.86 & -68:33:57.2&   3000 & $   3900 \pm     210 $ & $   6250 \pm     250$ & $ 0.33 \pm  0.02$ &                                                       F7Ibp &                      Ceph. (61.933), RVT &                         A\\
J054333.39-662202.8 & 05:43:33.39 & -66:22:02.9&   5000 & $   5330 \pm      80 $ & $   4500 \pm     250$ & $ 0.08 \pm  0.01$ &                                                        G8II &                    NV, Fr$_{\mathrm{9}}$ &                          \\
J055326.02-665851.5 & 05:53:26.03 & -66:58:51.6&   3000 & $   2560 \pm      50 $ & $   5500 \pm     250$ & $ 0.01 \pm  0.01$ &                                                      G0Ibpe &                           Ceph. (62.444) &                          \\
\hline                                             
\end{tabular}
\tablefoot{
\tablefoottext{a}{All spectral types indicated with a superscript \textit{a} were found in the catalogue of massive stars of \citet{bonanos09}, the others were determined based on our low-resolution, optical spectra.}\\
\tablefoottext{b}{The variability remarks are based on both the luminosity curves as discussed in Sect.~\ref{ssec:var} and cross-correlation with the catalogue of \citet{fraser08}. The latter is indicated with 'Fr' with the correct sequence in subscript. Sequences 9, 0 and -99 indicate subsequently that the star is identified with the One-Year Artifact which is caused by the annual observing schedule of the MACHO project, that the star is outside the boundaries of any period-luminosity classification, or that \citet{fraser08} were unable to classify this object. The other abbreviations used to characterise the variability of the different objects are Population~II Cepheids (Ceph.) which include the subclass of RV~Tauri stars (RVT), semi-regular variables (SRV) with the subclass of sequence-D variables (Seq. D) that were recognized by us, objects with lightcurves in Fig.~\ref{fig:lc_other_slow} that show long-term, slow variations (SV), R\,CrB stars (R\,CrB), objects in regions of very strong nebular emission (em), Ae and Be stars (AB) and objects that do not show strong evidence of variability (NV). For all Population~II Cepheids, the periods we computed between deep minima or two cycles of the light curve if the minima are of equal depth, are listed in days.}\\
\tablefoottext{c}{The final column contains the empirically confirmed cross-matches of the object in the catalogues of Sect.~\ref{sssec:croscollothertypes} with 'A' standing for the RV\,T stars listed in the recent OGLE-III Catalogue of Variable Stars \citep{soszynski08}, 'B' for the R\,CrB stars in \citet{soszynski09a}, 'C' for the MSX post-AGB stars of \citet{wood01}, 'D' are the YSOs from \citet{seale09}, 'E' the AGB stars from \citet{trams99}, 'F' the PNe in \citet{reid06} and 'G' the galaxies in the line of sight of the LMC from NED.}\\
\tablefoottext{d}{This star is unique in our sample in combining a solar type, with unusually strong broad \ion{Ca}{II} absorption, and emission of \ion{He}{I} (Table~\ref{table:em}).}}
\end{table}
\end{landscape}

\begin{landscape}
\begin{table}
\caption{\label{catalogue_specoutflow} The same as Table~\ref{catalogue_specdisc} but for objects with a shell.}
\centering                                      
\begin{tabular}{lcccccclll}          
\hline\hline                        
Object & Right Ascension & Declination & $L_{\mathrm{BB}}$ & $L_{\mathrm{SED}}$ & $T_\mathrm{eff,\ SED}$ & $E(B-V)_{\mathrm{SED}}$ & Spectral Type\tablefootmark{a} & Variability\tablefootmark{b} & Remarks\tablefootmark{c} \\
& (h:m:s) & (d:m:s) & ($L_{\odot}$) & ($L_{\odot}$) & (K) & & & &  \\
\hline
J045140.74-674456.9 & 04:51:40.75 & -67:44:57.0&   4000 & $   6100 \pm     330 $ & $   7750 \pm     250$ & $ 0.86 \pm  0.07$ &                                                       F0II: &       Ceph. (54.1438), Fr$_{\mathrm{0}}$ &                          \\
J045330.05-691749.5 & 04:53:30.06 & -69:17:49.5&   30000 & & & &                                      similar to Of/Of?p$^a$ &                                          &                         G\\
J050138.22-685831.6 & 05:01:38.23 & -68:58:31.6&  11000 & $  21800 \pm     500 $ & $   4500 \pm     250$ & $ 1.30 \pm  0.02$ &                                                          G7 &                  SV, Fr$_{\mathrm{-99}}$ &                          \\
J050629.89-682409.0 & 05:06:29.89 & -68:24:09.0&    1300 & & & &                                                           ? &                                          &                          \\
J050632.10-714229.8 & 05:06:32.11 & -71:42:29.9&   6000 & $  25900 \pm     400 $ & $  10500 \pm     500$ & $ 0.42 \pm  0.01$ &                                            A3Iab / F0Ibp(e) &                             Ceph. (49.1) &                          \\
J051110.64-661253.7 & 05:11:10.64 & -66:12:53.8&   8000 & $  15500 \pm     300 $ & $   7000 \pm     250$ & $ 1.27 \pm  0.01$ &                                                     F3II(e) &                 SRV, Fr$_{\mathrm{-99}}$ &                          \\
\hline
\end{tabular}
\end{table}
\end{landscape}

\begin{landscape}
\begin{table}
\caption{\label{catalogue_specundet} Same as Table~\ref{catalogue_specdisc}, but for objects for which the shape of the SED did not permit to distinguish between objects with a shell and those with a disc.}
\begin{tabular}{lcccccclll}
\hline\hline
Object & Right Ascension & Declination & $L_{\mathrm{BB}}$ & $L_{\mathrm{SED}}$ & $T_\mathrm{eff,\ SED}$ & $E(B-V)_{\mathrm{SED}}$ & Spectral Type\tablefootmark{a} & Variability\tablefootmark{b} & Remarks\tablefootmark{c} \\
& (h:m:s) & (d:m:s) & ($L_{\odot}$) & ($L_{\odot}$) & (K) & & & &  \\
\hline
J044555.34-705041.7 & 04:45:55.34 & -70:50:41.8&    4000 & & & &                                                       ? (e) &                                          &                          \\
J045444.13-701916.1 & 04:54:44.14 & -70:19:16.1&  20000 & $  22000 \pm    2800 $ & $   4000 \pm     250$ & $ 0.22 \pm  0.13$ &                                                    M0 (III) &                                          &                          \\
J045515.73-701742.1 & 04:55:15.74 & -70:17:42.2&   4000 & $  22600 \pm     600 $ & $   9000 \pm     250$ & $ 0.77 \pm  0.01$ &                                                        A3II &                                          &                          \\
J045543.20-675110.1 & 04:55:43.20 & -67:51:10.1&   3000 & $   3080 \pm     150 $ & $   4750 \pm     250$ & $ 0.00 \pm  0.03$ &                                                 G2-8(R)Ibe: &                     Ceph. (112.828), RVT &                         A\\
J045615.77-682042.3 & 04:56:15.77 & -68:20:42.3&   1900 & $   2600 \pm     220 $ & $   4250 \pm     250$ & $ 0.25 \pm  0.08$ &                                                          K7 &           SRV, Seq. D, Fr$_{\mathrm{D}}$ &                          \\
J045655.23-682732.9 & 04:56:55.23 & -68:27:33.0&   4000 & $   6780 \pm     110 $ & $   5000 \pm     250$ & $ 0.55 \pm  0.01$ &                                                       G1Ibp &                                          &                          \\
\hline
\end{tabular}
\end{table}
\end{landscape}

\section{Table of all objects that have a low-resolution, optical spectrum but are no longer part of the post-AGB sample}\label{app:delobj}

\longtab{1}{
\begin{landscape}
\begin{longtable}{lcclccccll}
\caption{\label{list_delobj} List of objects for which a low-resolution, optical spectrum or a spectral type was obtained but that are no longer part of the final sample. We list their names, positions, spectral type and the reason why they are no longer in our sample for all objects. These reasons are: A: Not in the current version of the SAGE catalogue, B: Does not obey the colour criterion, C: Does not survive the black-body luminosity-cut (Sect.~\ref{ssec:lumcut}), D: Is listed as a cool carbon star by \citet{kontizas01} or an AGB star with spectral type by \citet{trams99}, E: Is not a post-AGB star according to the spectral type, F: Does not survive the SED luminosity-cut (Sect.~\ref{subsection:SEDs}). The subsample they belong to, the black-body-based luminosity and the SED-based luminosity, effective temperature and reddening are also listed if the available photometry and/or the spectral type allows to deduce them. A more extended version of this table that also lists the magnitudes of these objects in the different filters and some variability information can be found at the CDS. This catalogue contains the same information as the final catalogue of post-AGB stars (see Sect.~\ref{ssec:finlum}), but with an extra final column listing the reason why these objects are no longer in our sample.}\\
\hline\hline
Object & Right Ascension & Declination & Subsample & $L_{\mathrm{BB}}$ & $L_{\mathrm{SED}}$ & $T_\mathrm{eff,\ SED}$ & $E(B-V)_{\mathrm{SED}}$ & Spectral Type\tablefootmark{a} & Reason \\
& (h:m:s) & (d:m:s) & & ($L_{\odot}$) & ($L_{\odot}$) & (K) & & & \\
\hline
\endfirsthead
\caption{continued.}\\
\hline\hline
Object & Right Ascension & Declination & Subsample & $L_{\mathrm{BB}}$ & $L_{\mathrm{SED}}$ & $T_\mathrm{eff,\ SED}$ & $E(B-V)_{\mathrm{SED}}$ & Spectral Type\tablefootmark{a} & Reason \\
& (h:m:s) & (d:m:s) & & ($L_{\odot}$) & ($L_{\odot}$) & (K) & & & \\
\hline
\endhead
\hline
\endfoot
J044904.66-690455.0 & 04:49:04.67 & -69:04:55.0 & disc or shell &    3000 & $ 240000 \pm   40000$ & $  47500 \pm    6000 $ & $ 0.51 \pm  0.02$ &                                                          O-B &                                           F \\
J044922.46-692434.5 & 04:49:22.46 & -69:24:34.5 & disc or shell &   90000 & $ 118000 \pm    2300$ & $   3500 \pm     250 $ & $ 0.38 \pm  0.01$ &                                                            M &                                           C \\
J044943.46-691955.3 & 04:49:43.47 & -69:19:55.4 &         shell &    1500 & $ 186000 \pm    6000$ & $  47500 \pm    2500 $ & $ 0.66 \pm  0.01$ &                                                          O-B &                                           F \\
J044957.99-702102.6 & 04:49:57.99 & -70:21:02.6& & & & & &                                                     B5II-III                                           & A \\
J045433.85-692036.2 & 04:54:33.85 & -69:20:36.3 & disc or shell &   10000 & $ 124000 \pm   17000$ & $  17000 \pm    1000 $ & $ 1.03 \pm  0.02$ &                                                         B[e] &                                           F \\
J045503.04-692912.7 & 04:55:03.05 & -69:29:12.8 & disc or shell &  170000 & $ 230000 \pm    4000$ & $   3500 \pm     250 $ & $ 0.78 \pm  0.01$ &                                                           M2 &                                           C \\
J045521.62-694716.8 & 04:55:21.62 & -69:47:16.8 & disc or shell &  100000 & $ 121000 \pm    2600$ & $   3500 \pm     250 $ & $ 0.49 \pm  0.01$ &                                                            M &                                           C \\
J045617.53-661818.9 & 04:56:17.54 & -66:18:19.0 &         shell &   19000 & $ 170000 \pm   40000$ & $  18000 \pm    1700 $ & $ 0.27 \pm  0.03$ &                                                  B2.5Iab$^a$ &                                          F \\
\end{longtable}
\tablefoot{
\tablefoottext{a}{All spectral types indicated with a superscript \textit{a} were found in the catalogue of massive stars of \citet{bonanos09}, the others were determined based on our low-resolution, optical spectra.}
}
\end{landscape}
}

\Online

\section{Spectral analysis of the intruders}\label{app:spectra}

Fourteen of the low-resolution, optical spectra we took, do not belong to post-AGB stars. We were able to identify nine galaxies, two PN-like objects and two WC-like objects. 

\subsection{Galaxies}\label{ssec:specgal}

\begin{table}
\caption{Redshifts and types of the galaxies in our sample for which we obtained low-resolution optical spectra.}             
\label{table:galaxies}      
\centering                          
\begin{tabular}{lll}        
\hline\hline                 
Object Name (IRAC) & Redshift & Type \\    
\hline                        
J050303.97-663345.9 & 0.063 & Active galaxy \\
J050524.35-673435.4 & 0.047 & Normal spiral or irregular galaxy \\
J051424.45-720705.5 & 0.027 & Normal spiral or irregular galaxy \\
J051852.75-683848.9 & 0.059 & Starburst galaxy \\
J052026.68-695125.4 & 0.004 & Normal spiral or irregular galaxy \\
J052229.16-710814.9 & 0.036 & Starburst galaxy \\
J053308.25-722644.9 & 0.046 & Normal spiral or irregular galaxy \\
J053556.06-714135.4 & 0.041 & Starburst galaxy \\
J053557.86-671333.5 & 0.078 & Normal spiral or irregular galaxy \\
\hline                                   
\end{tabular}
\end{table}

We observed nine objects for which the spectrum does not correspond to a source in the LMC. They all have too high a redshift and are therefore galaxies. Some details on these sources can be found in Table~\ref{table:galaxies}. All these objects are listed as galaxies on NASA/IPAC Extragalactic Database (NED)\footnote{The NASA/IPAC Extragalactic Database (NED) is operated by the Jet Propulsion Laboratory, California Institute of Technology, under contract with the National Aeronautics and Space Administration.}. A redshift is noted only for J050303.97-663345.9 and J053308.25-722644.9 and both values agree with the ones we determined.

\subsection{PN-like objects}\label{sssec:pnlikeobj}

\begin{figure*}
\centering
\includegraphics[width=17cm]{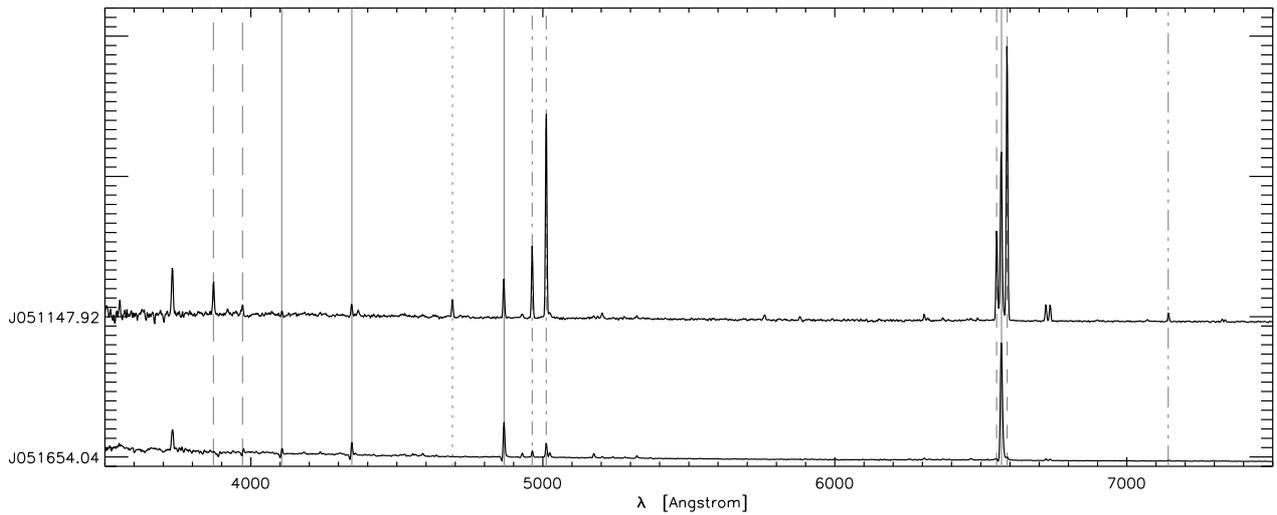}
\caption{Spectra of the two PN-like objects -- J051654.04-672005.1 and J051147.92-692342.3 -- that occur in our sample. All spectra were normalised and displaced vertically by 15 units to increase the visibility of the plot. The vertical grey lines indicate the wavelengths at which spectral features of certain elements are expected, corrected for the average radial velocity of the LMC. These are: Balmer lines (solid), [\ion{O}{III}] (dash dotted), [\ion{N}{II}] (dashed), \ion{He}{II} (dotted), [\ion{Ar}{III}] (dash dot dotted) and [\ion{Ne}{III}] (long dashes).}
\label{fig:PNspectra_all}
\end{figure*}

Two objects in our sample -- J051654.04-672005.1 and J051147.92-692342.3 -- show emission features of \element[][]{H}$\alpha$, \element[][]{H}$\beta$, [\ion{O}{II}], [\ion{O}{III}], [\ion{N}{II}], [\ion{S}{II}], [\ion{Ar}{III}] and [\ion{Ne}{III}] (see Fig.~\ref{fig:PNspectra_all}). In addition, \ion{He}{II} emission can be seen from J051147.92-692342.3. Both spectra bear resemblance to spectra of \ion{H}{II} regions as well as PNe.\newline

The spectrum of J051147.92-692342.3 contains emission features of elements like [\ion{O}{III}], [\ion{Ar}{III}] and [\ion{Ne}{III}] at higher levels of ionisation than can be found in \ion{H}{II} regions. The line intensity ratio of [\ion{O}{III}] 5007~\AA{} to \element[][]{H}$\beta$ is about 5, which is somewhat lower than the theoretically expected value of 9 \citep{reid06}, but still compatible with a PN. Since [\ion{N}{II}] 6583~\AA{} $\geq$ \element[][]{H}$\alpha$ and \ion{He}{II} 4686~\AA{} is also clearly visible, this object is probably a high excitation Type I PN.\newline

The most significant differences in the spectrum of J051654.04-672005.1 with respect to J051147.92-692342.3 are the less prominent features of [\ion{O}{III}] and [\ion{N}{II}] when compared to the Balmer lines. The line intensity ratio of [\ion{O}{III}] 5007~\AA{} to \element[][]{H}$\beta$ for this object is 0.4, which is incompatible with both classical PNe and \ion{H}{II} regions. A very low excitation PN or an extremely compact, low-excitation \ion{H}{II} can explain these ratios, however.  The latter possibility is supported by the [\ion{N}{II}]/\element[][]{H}$\alpha$ ratio which is well below 0.7 \citep{kennicutt00}.

Some of the most prominent features in J051654.04-672005.1 show a P~Cygni profile which is indicative of a fast stellar wind. The low-resolution of the spectrum only allows for a rough estimation of the velocity of this wind which turns out to be between 650 and 1050~km/s. \newline

Of the two objects we identified as PNe-like objects, only J051147.92-692342.3 can be found in the PNe catalogue of \citet{reid06}.

\subsection{Wolf-Rayet carbon star-like objects}\label{sssec:wc_spec}

\begin{figure*}
\centering
\includegraphics[width=17cm]{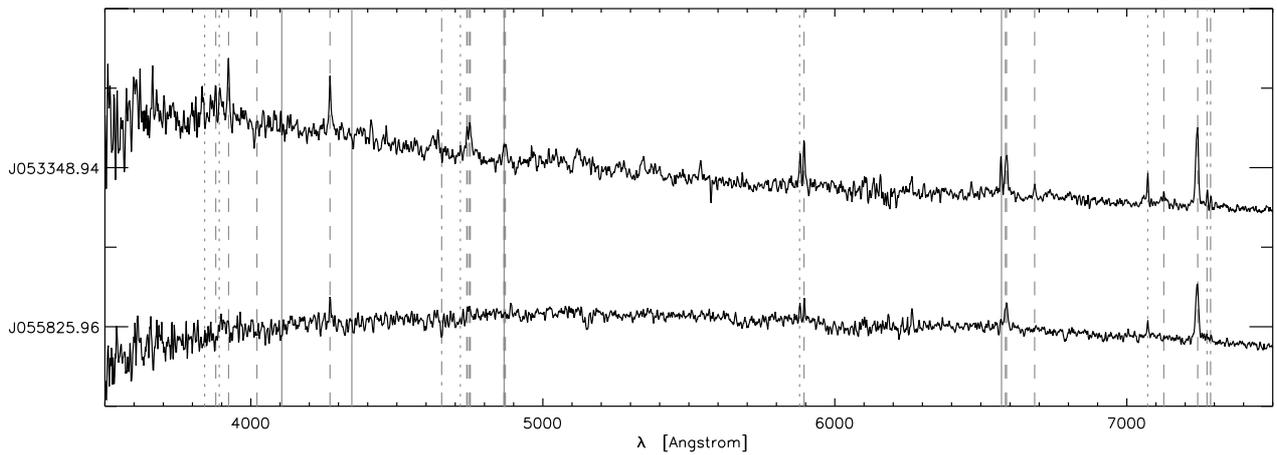}
\caption{Spectra of the two possible WC stars -- J053348.94-701323.5 and J055825.96-694425.8 -- that occur in our sample. All spectra were normalised and translated vertically in steps of four units to increase the visibility of the plot. The vertical grey lines indicate the wavelengths at which spectral features of \ion{H}{I} (solid lines), \ion{He}{I} (dotted lines), \ion{C}{II} (dashed lines) and \ion{O}{II} (dot-dashed lines) are expected, corrected for the average radial velocity of the LMC.}
\label{fig:wrspectra_all}
\end{figure*}

The spectra of J053348.94-701323.5 and J055825.96-694425.8 are clearly distinct from the other spectra, but resemble each other (see Fig.~\ref{fig:wrspectra_all}). Both spectra are similar to Wolf-Rayet carbon (WC) stars, but the emission lines are rather weak and they lack the \ion{C}{III} feature at 5696~\AA{} which is still present even in WC stars of spectral types as late as WC10 \citep{crowther98}. The resemblance of both spectra to that of V~348~Sagitarii, a Galactic R\,CrB star with a very peculiar spectrum \citep[e.g.,][]{leuenhagen94,houziaux68}, makes it plausible that our objects are also of spectral type [WC12]. One of these objects, J053348.94-701323.5, is also listed as an R\,CrB star by \citet{soszynski09a} even though it is not \element[][]{H} deficient.

\subsection{Some additional spectra}

In addition to the 105 objects from our sample of post-AGB candidates, we observed 31 bright sources with IR detections, which were originally identified using the first release of the SAGE catalogue of only Epoch~1 data. These objects were not recovered while creating our final list. Most of them were lost due to the black-body luminosity-cut, but some simply do not have the required measurements in the different SAGE catalogues we used to define our final sample or are listed as a cool carbon star by \citet{kontizas01} or an AGB star with spectral type by \citet{trams99}. For these objects, and some objects with a low-resolution, optical spectrum that were later on discarded from our final sample, we list the spectral types and the reason why they are no longer in our final catalogue in Table~\ref{list_delobj}.

\clearpage

\section{SEDs of all post-AGB candidates with a known spectral type}\label{app:SEDs}

\subsection{Post-AGB candidates with a disc}

\onlfig{1}{
\begin{figure*}
\resizebox{\hsize}{!}{
\includegraphics{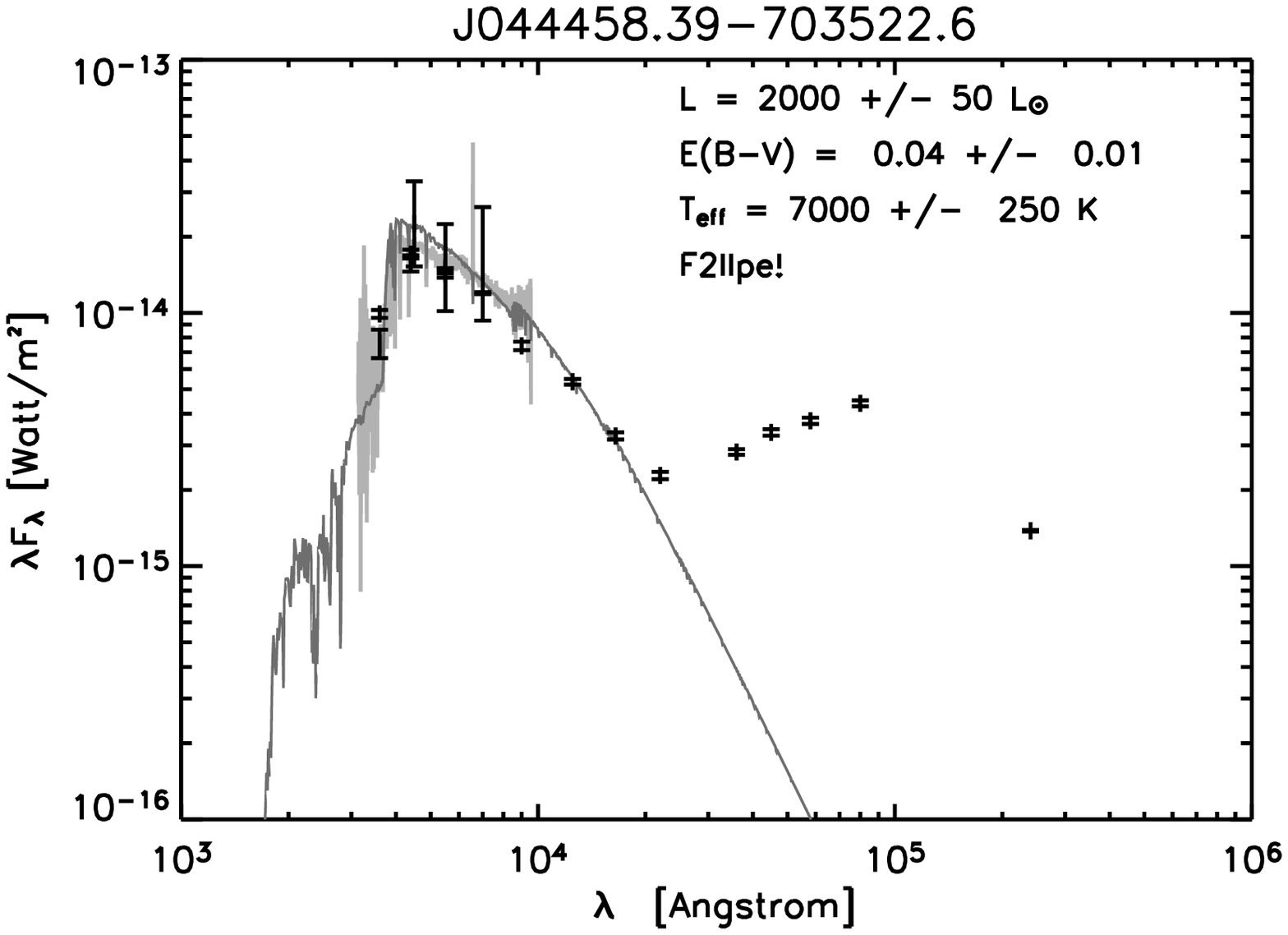}
\includegraphics{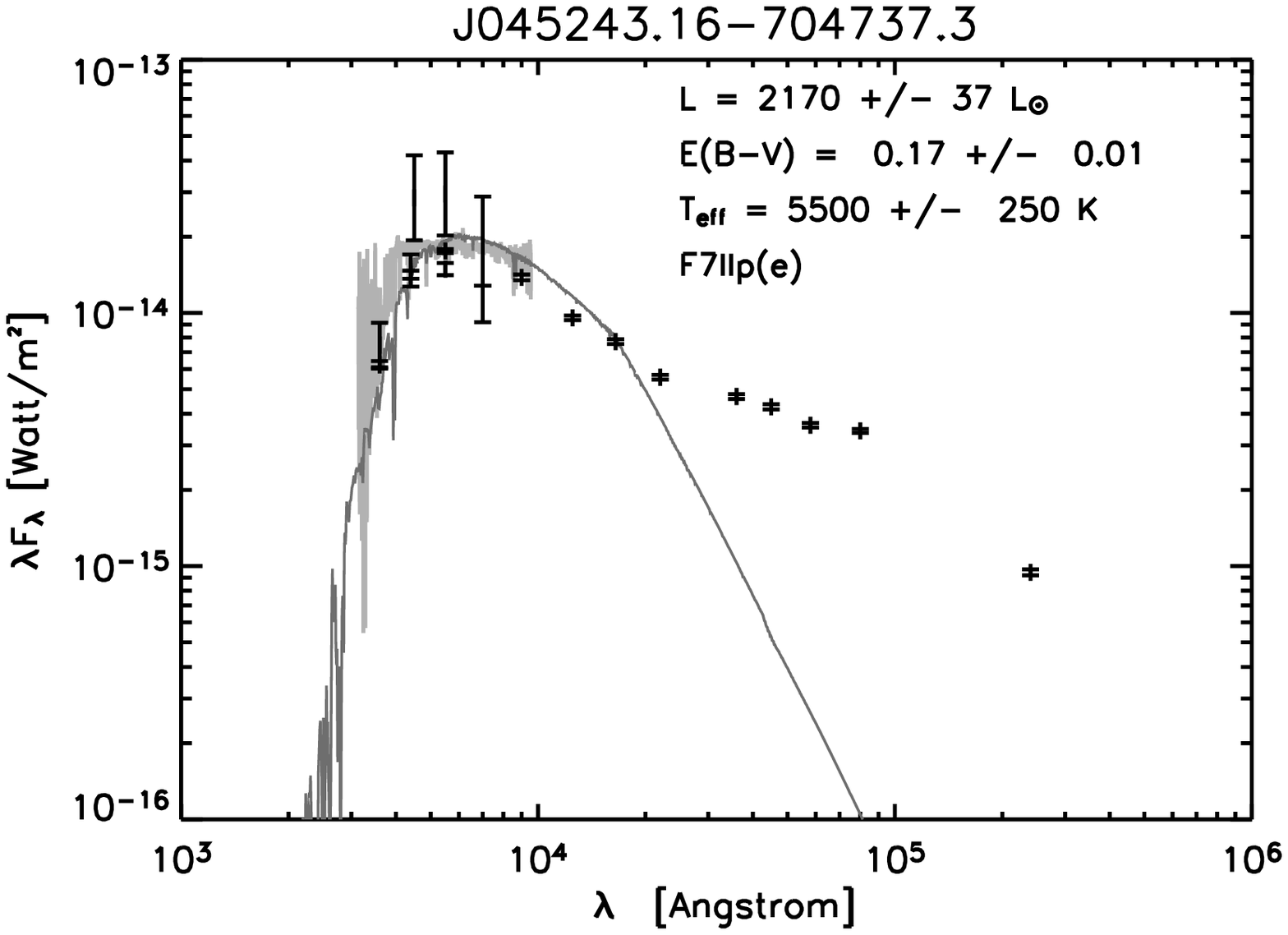}
\includegraphics{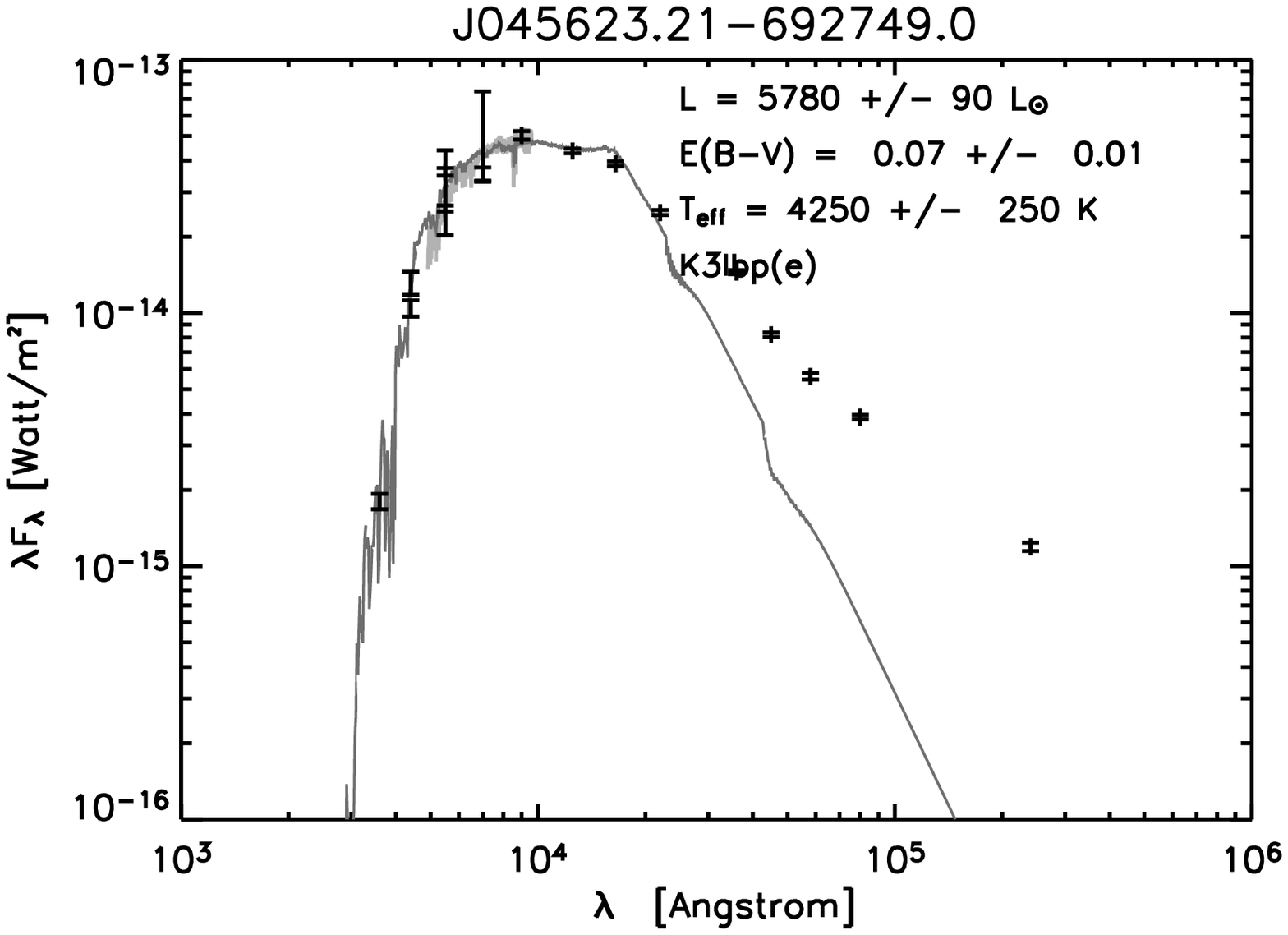}
}
\resizebox{\hsize}{!}{
\includegraphics{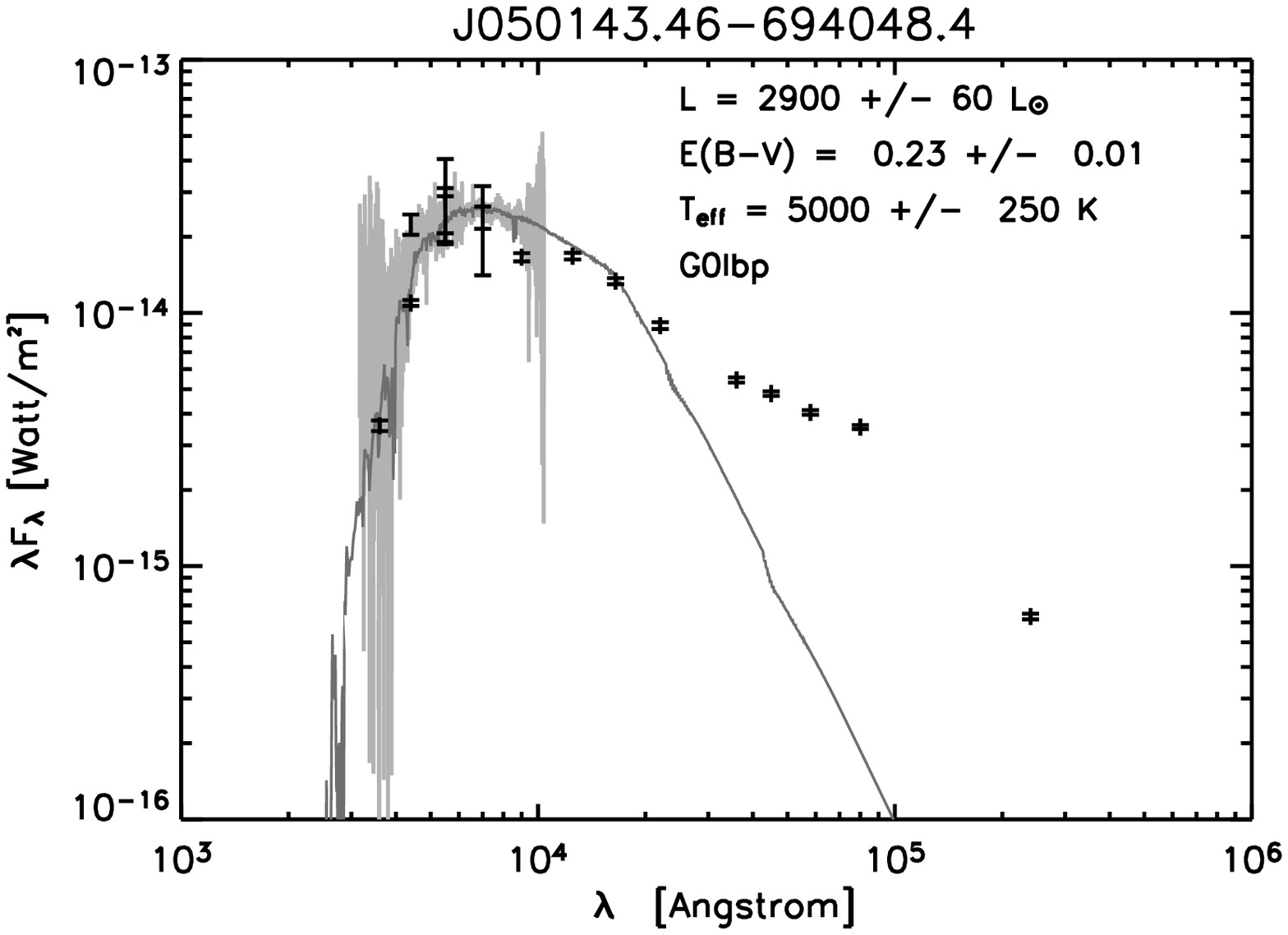}
\includegraphics{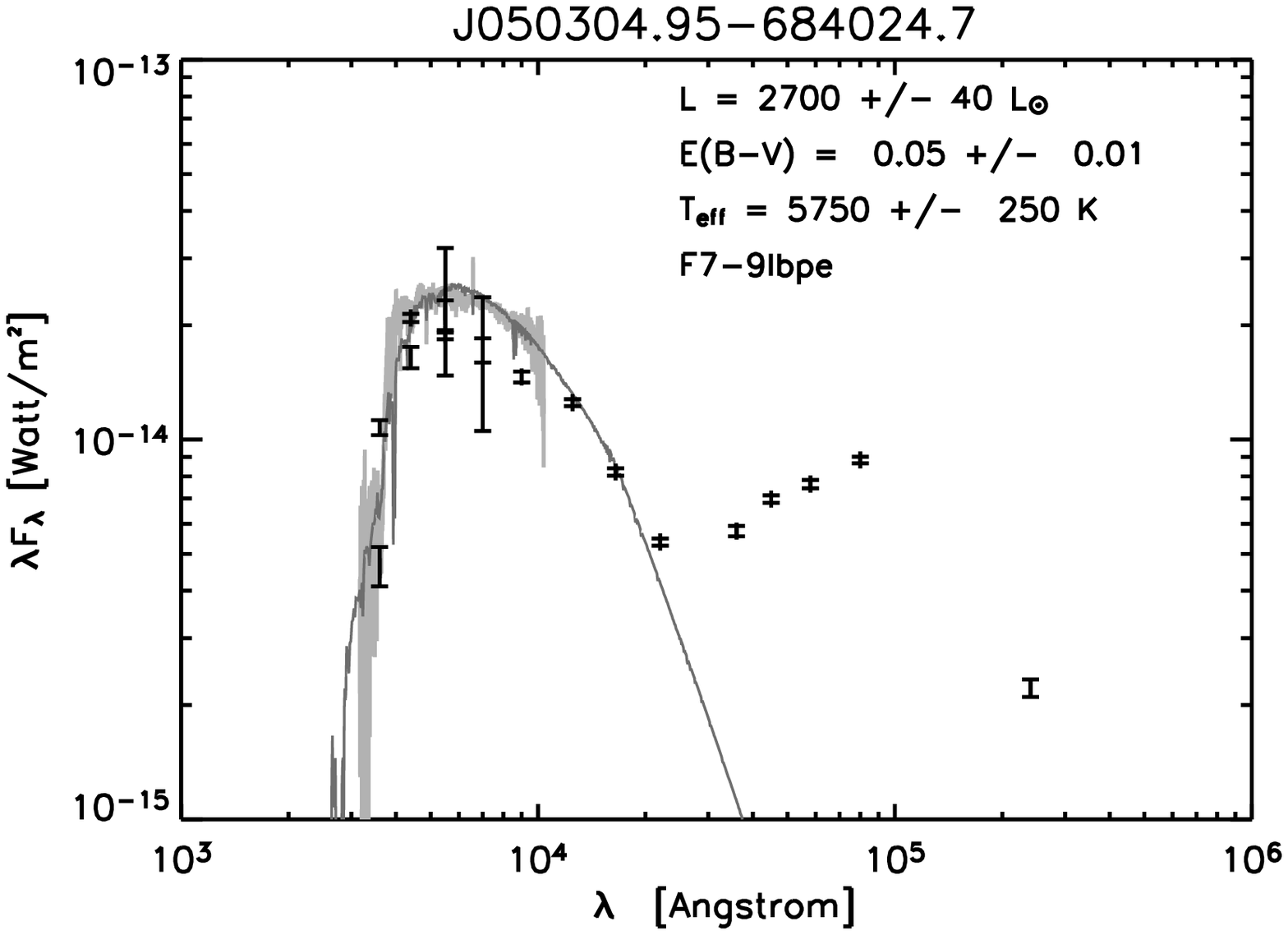}
\includegraphics{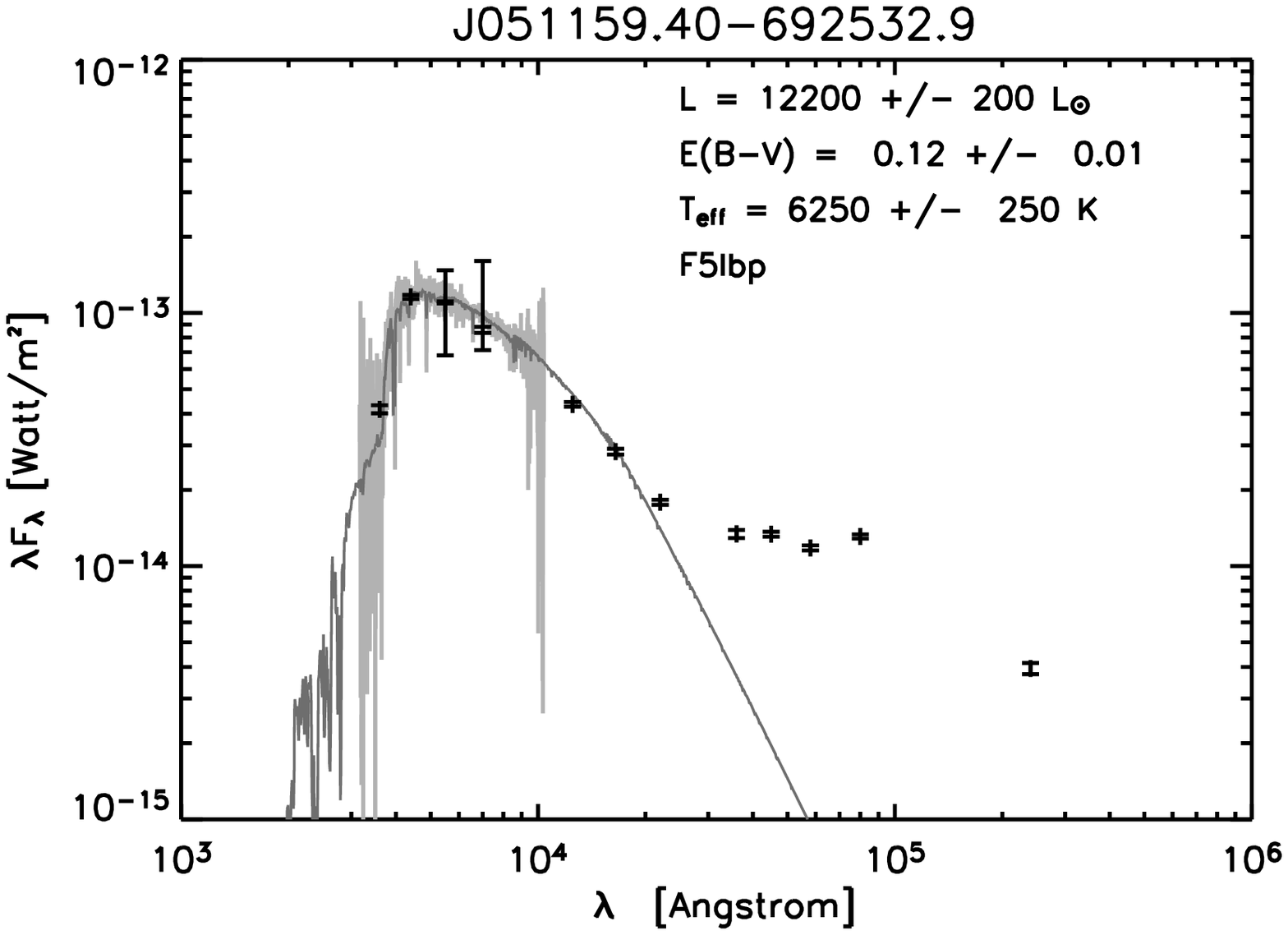}
}
\resizebox{\hsize}{!}{
\includegraphics{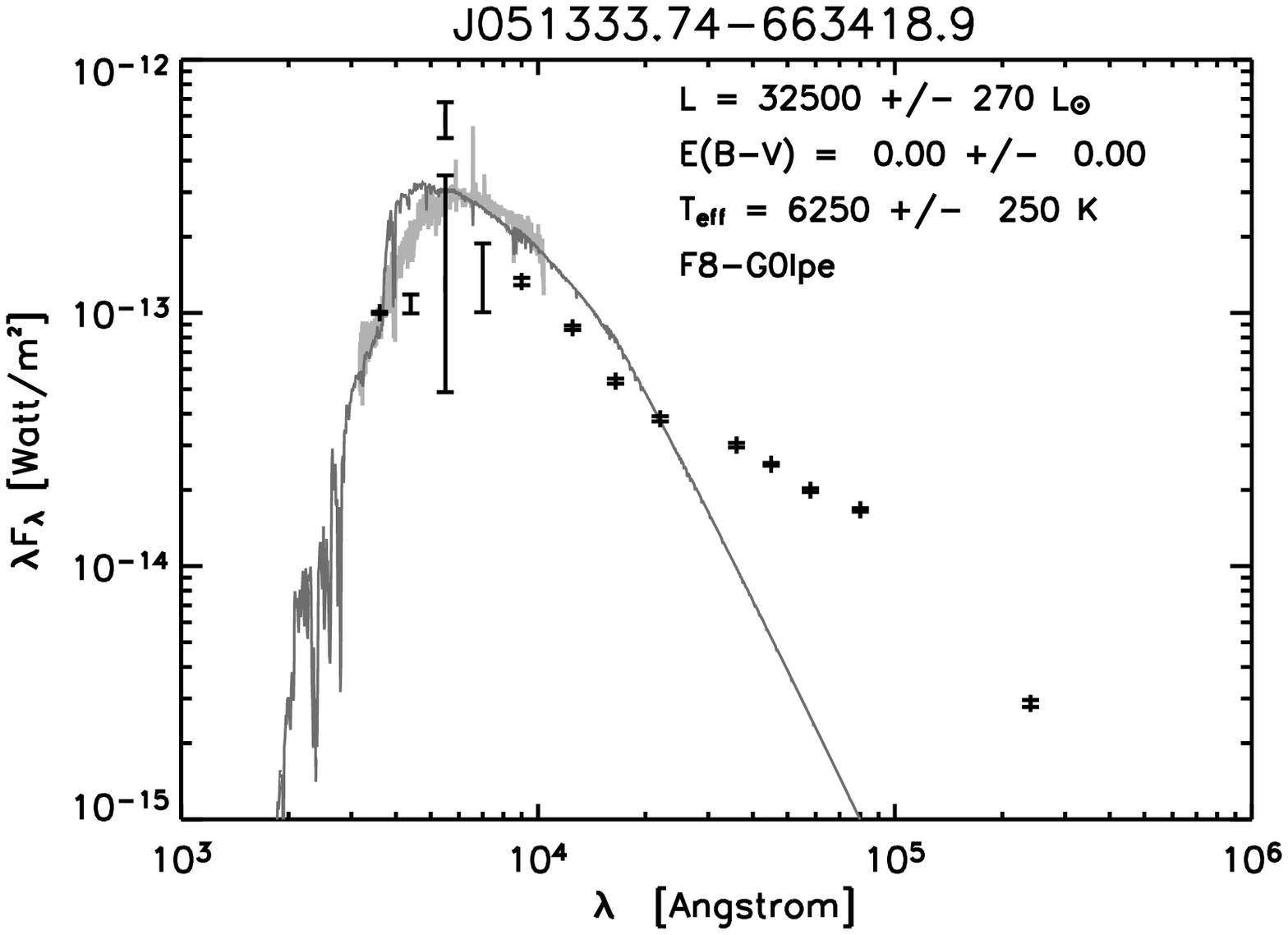}
\includegraphics{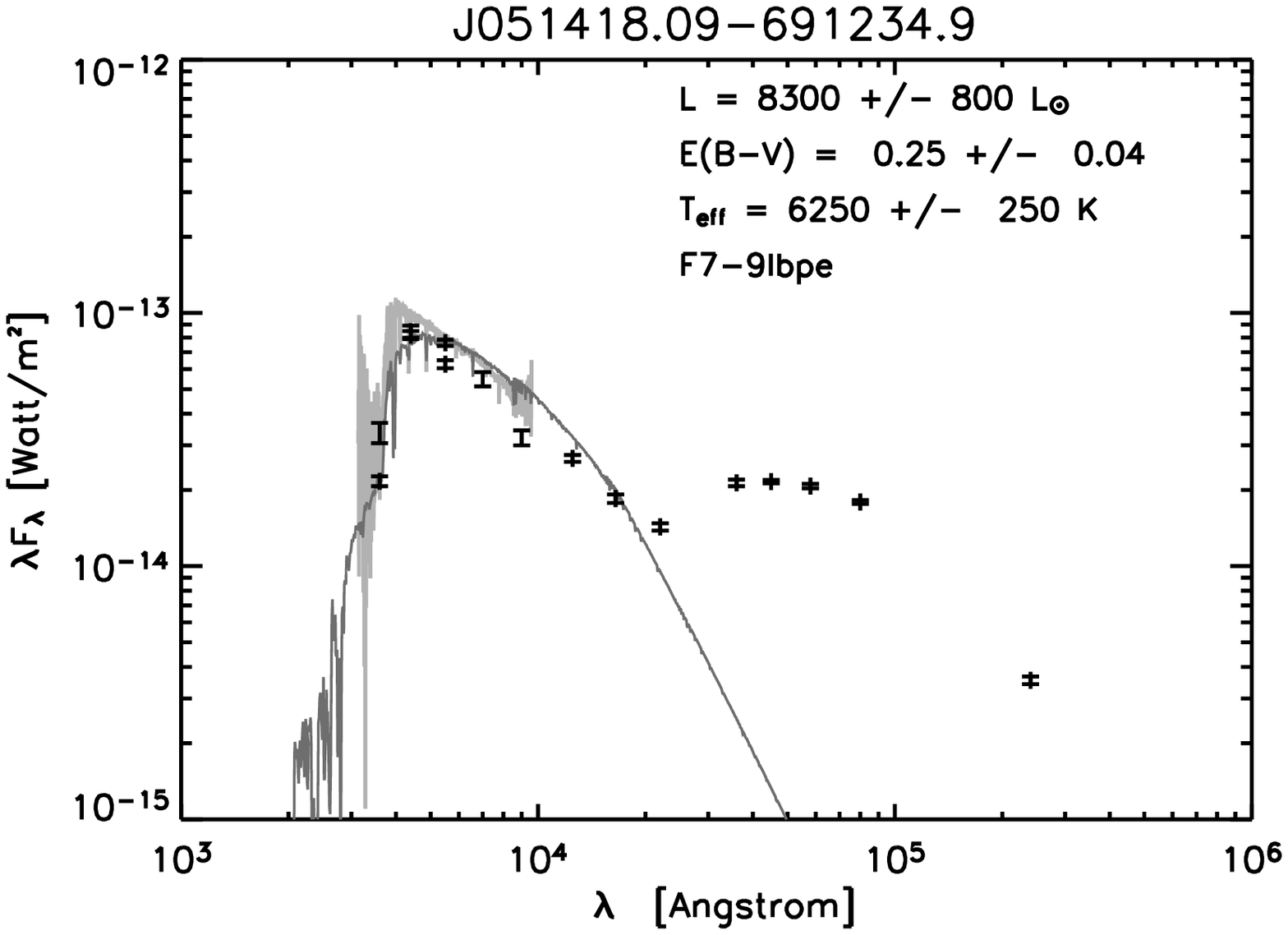}
\includegraphics{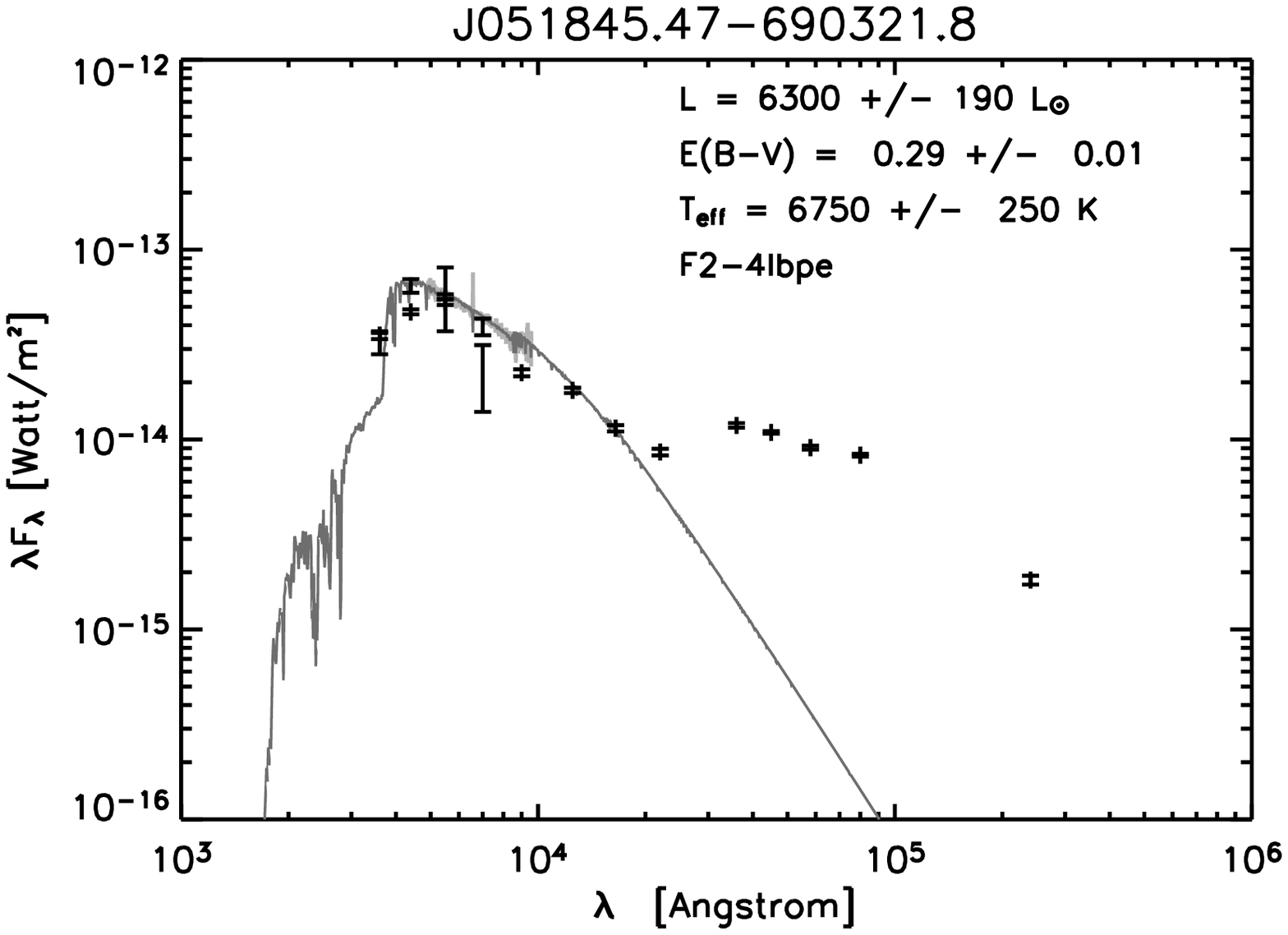}
}
\resizebox{\hsize}{!}{
\includegraphics{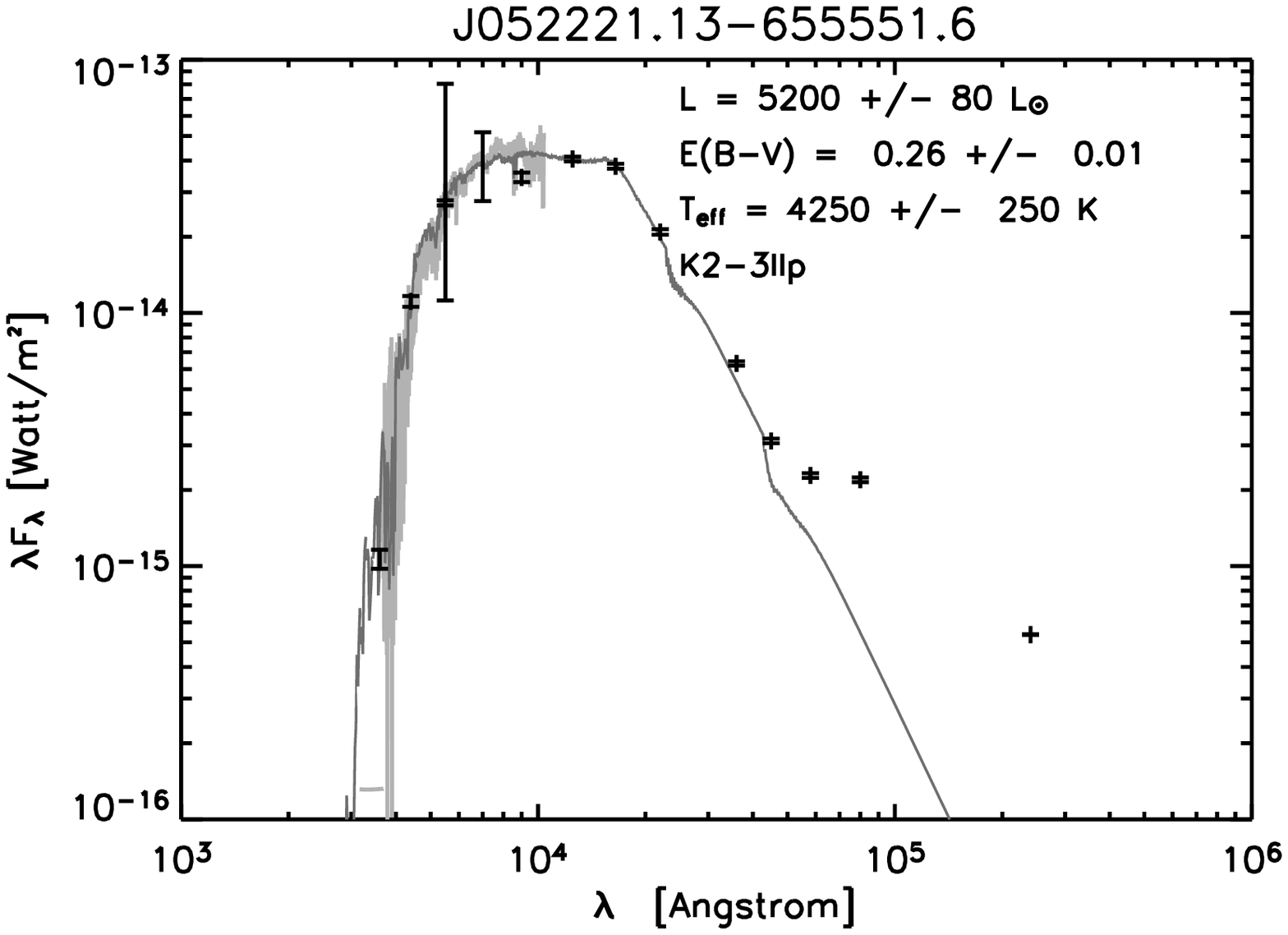}
\includegraphics{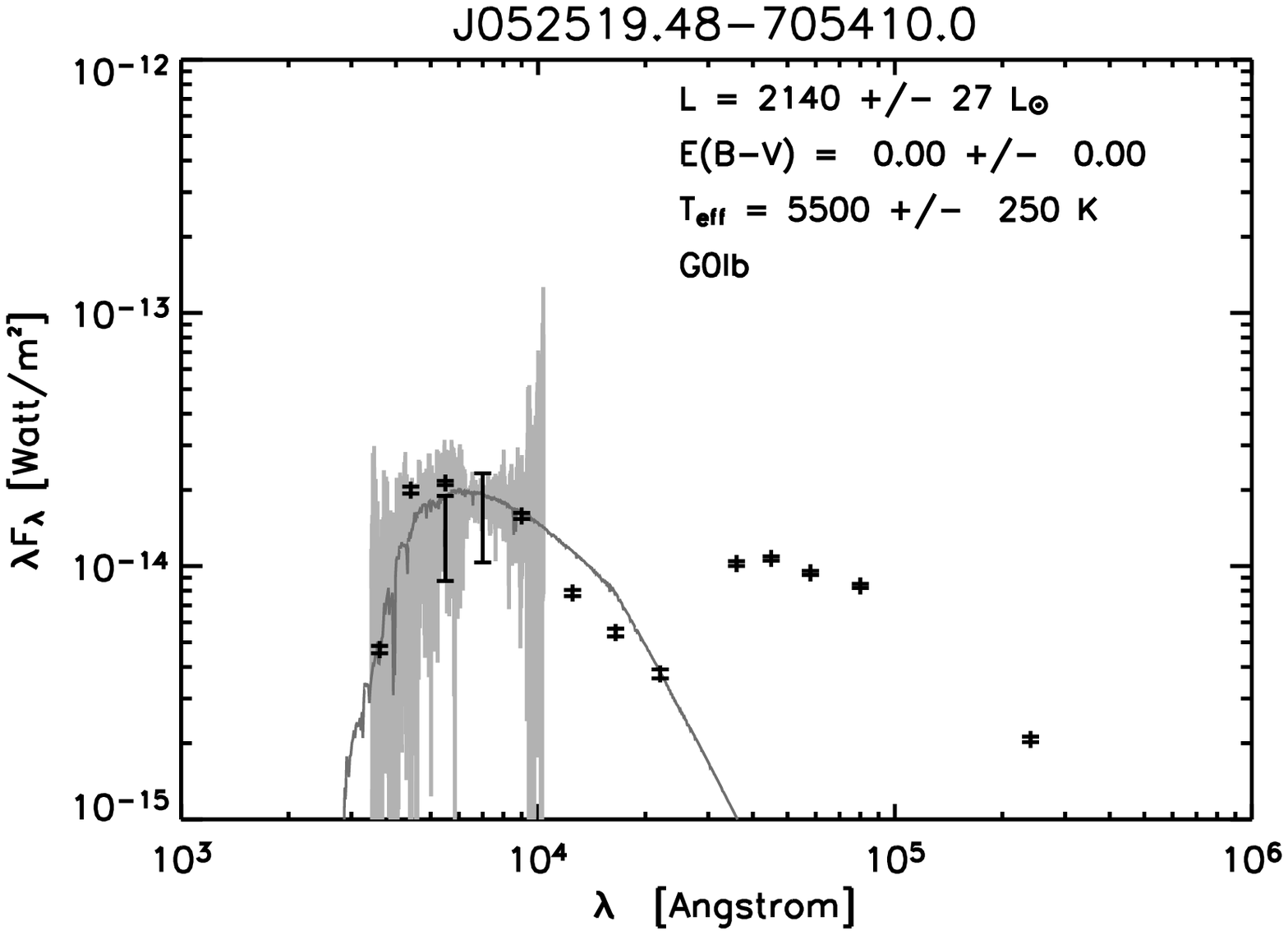}
\includegraphics{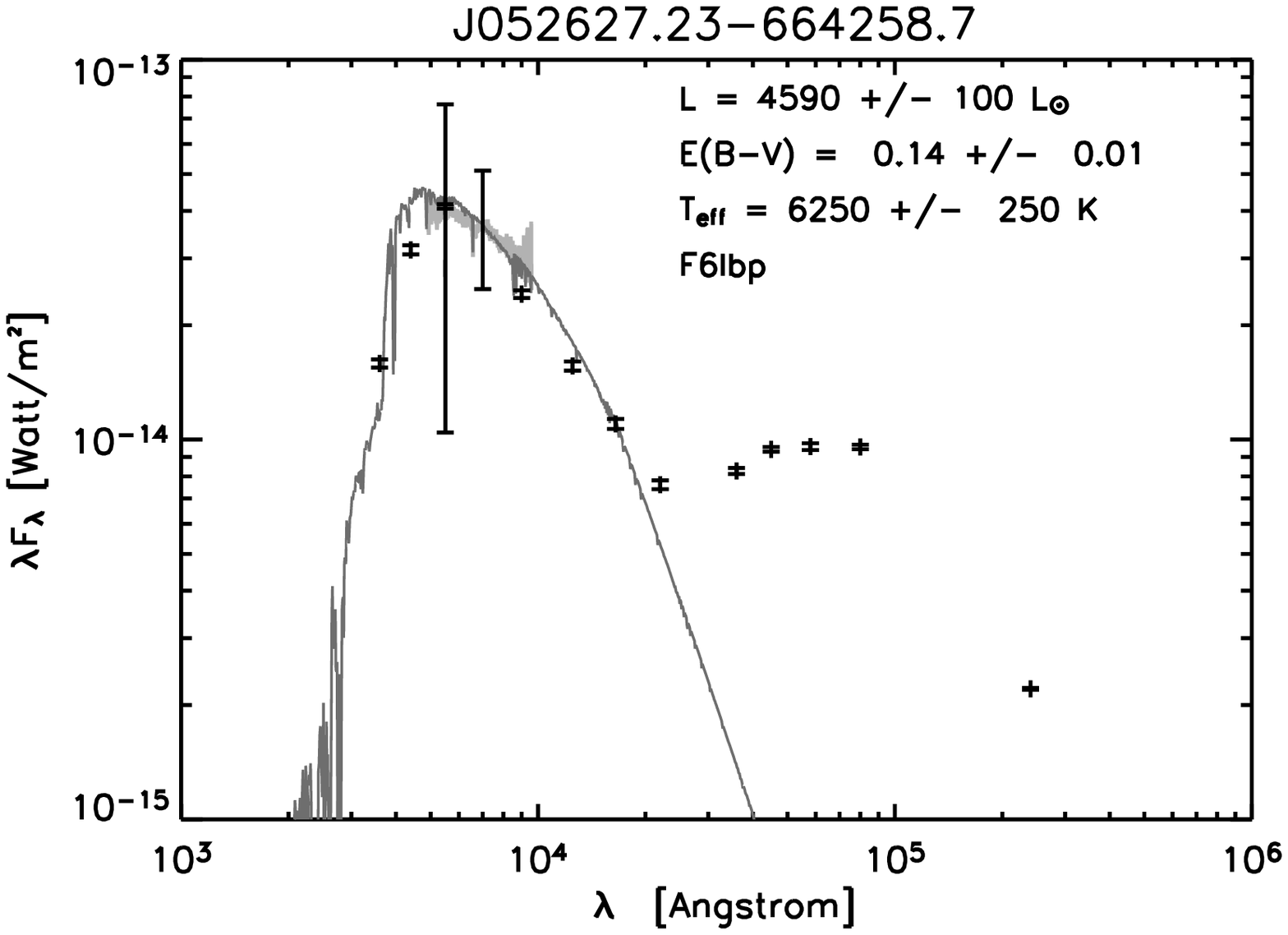}
}
\resizebox{\hsize}{!}{
\includegraphics{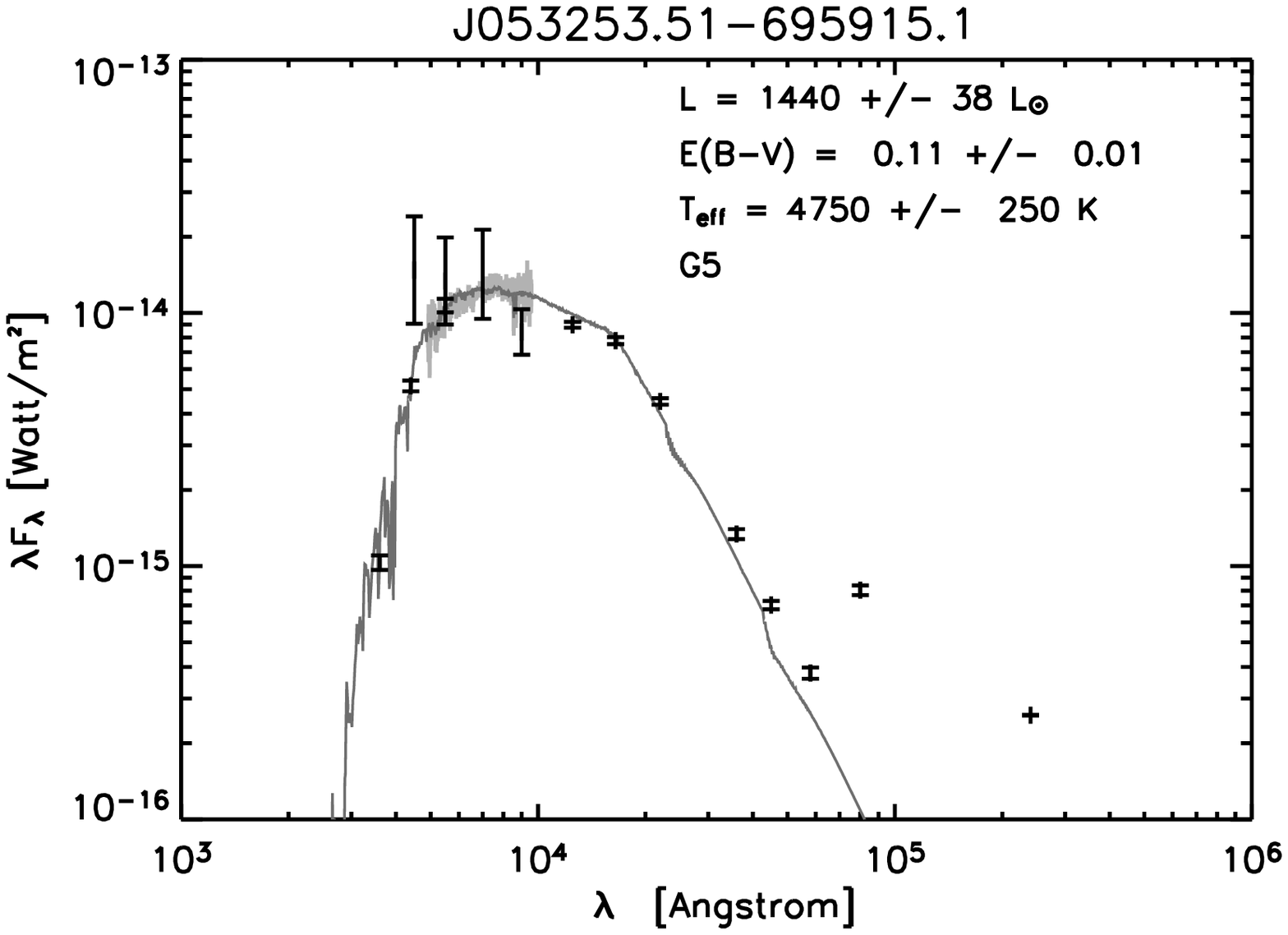}
\includegraphics{15834fg7a.ps}
\includegraphics{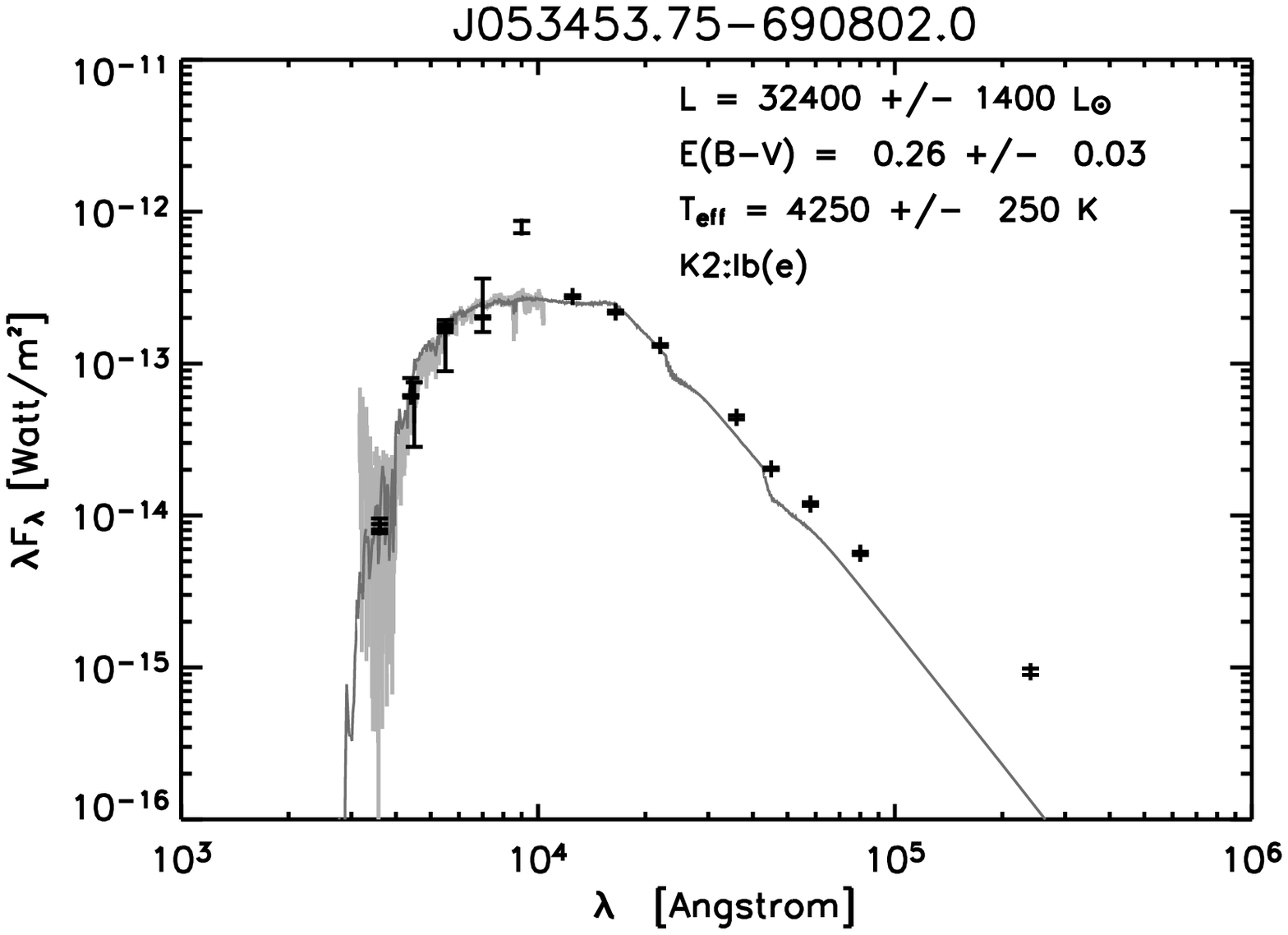}
}
\label{fig:discspec11}
\caption{SEDs of all post-AGB candidates with a known spectral type and an SED indicative of a disc. The dark and light grey line are the Kurucz atmosphere model used and the low-resolution, optical spectrum, respectively. All photometric datapoints are dereddened.}
\end{figure*}
}

\onlfig{2}{
\begin{figure*}
\resizebox{\hsize}{!}{
\includegraphics{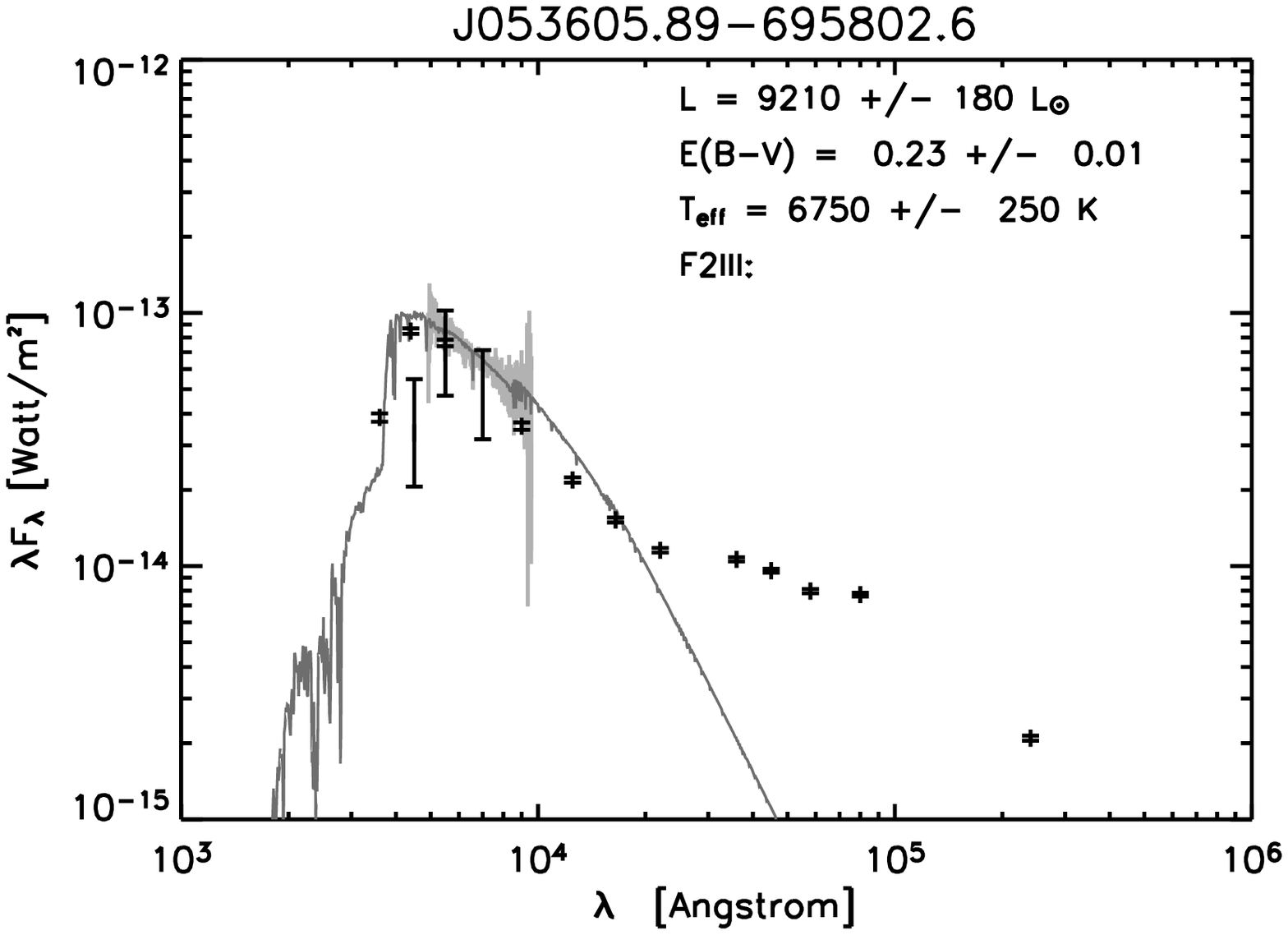}
\includegraphics{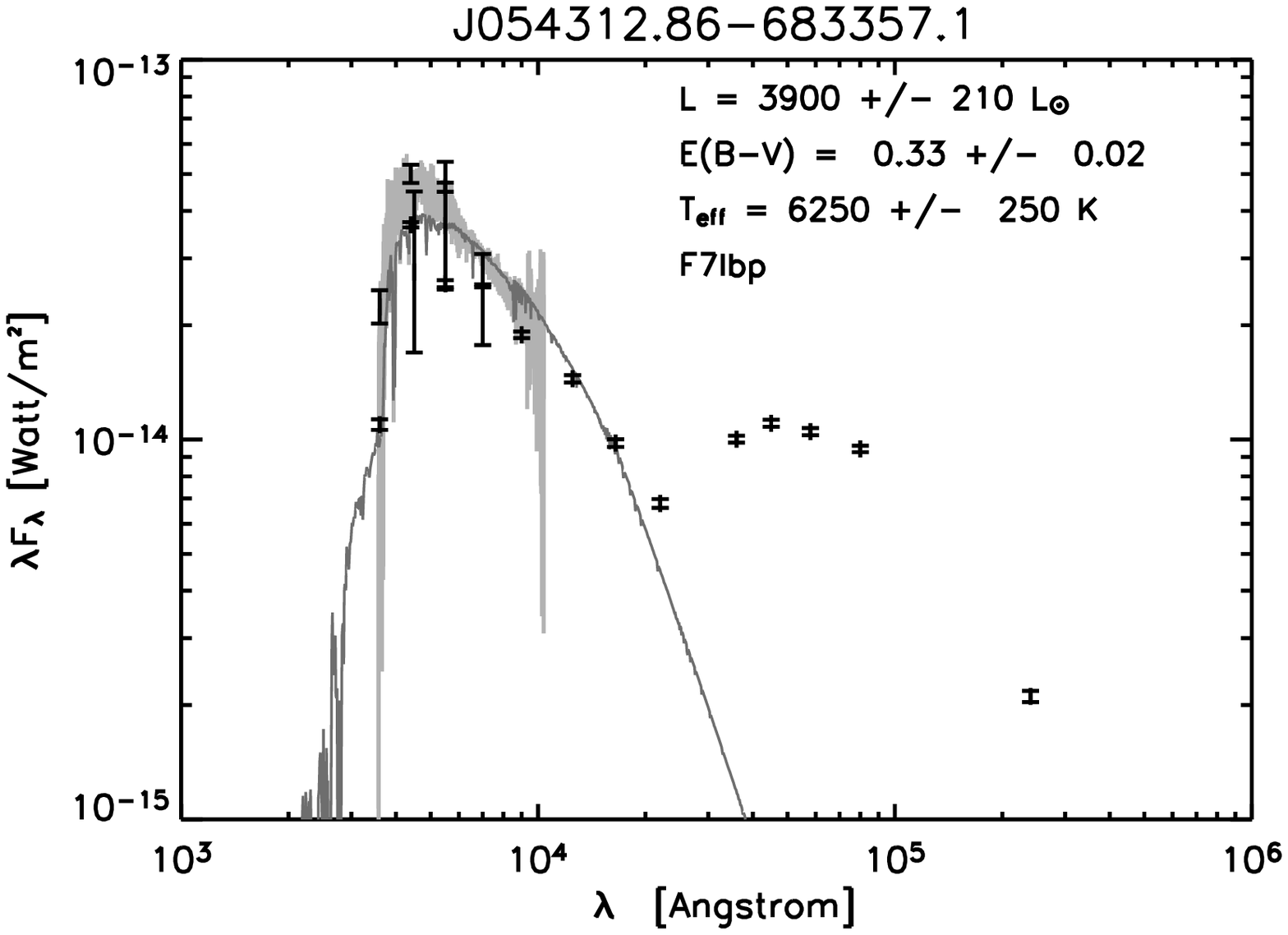}
\includegraphics{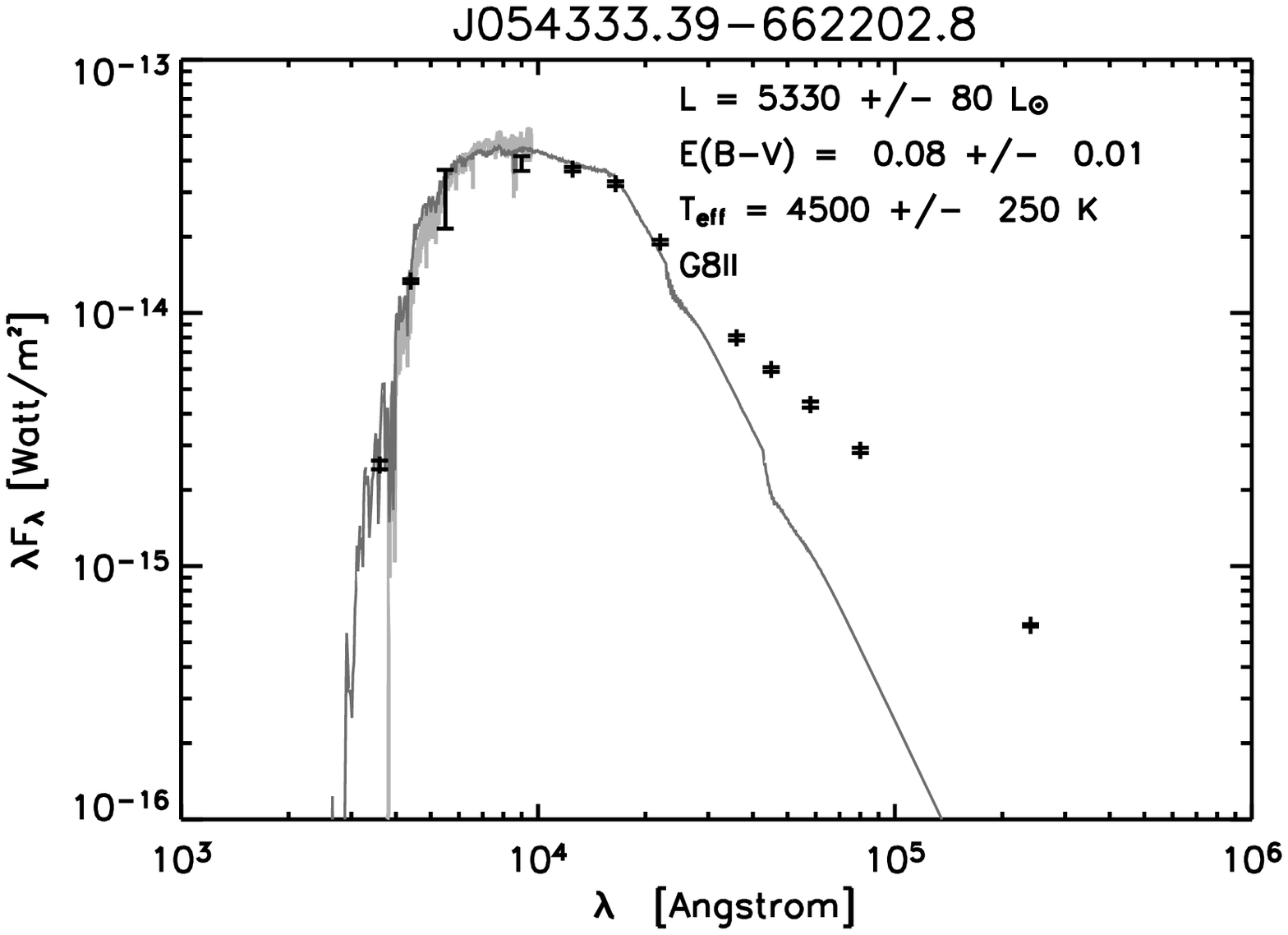}
}
\resizebox{\hsize}{!}{
\includegraphics{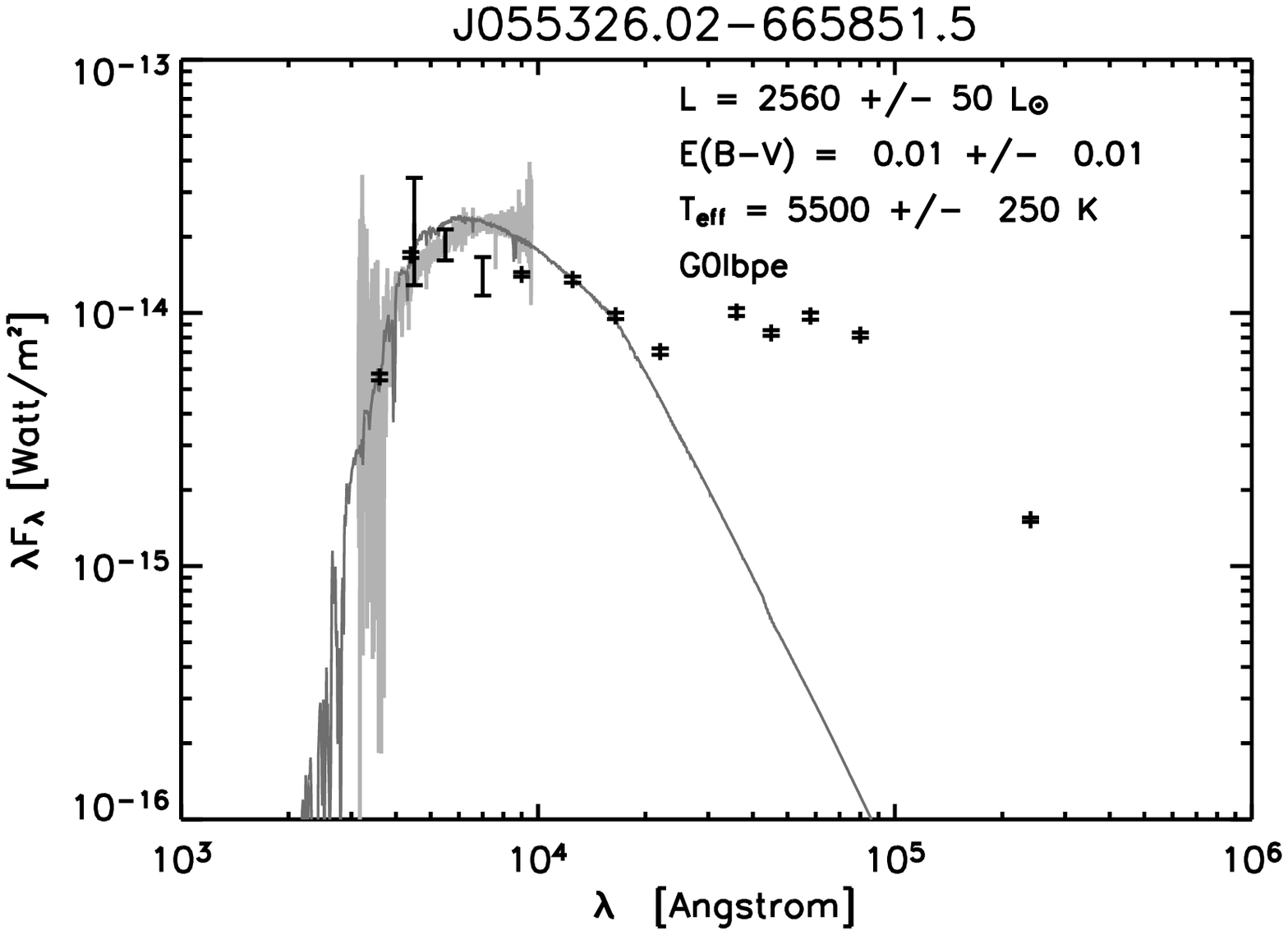}
\includegraphics{15834fgd2e.ps}
\includegraphics{15834fgd2e.ps}
}
\caption{Continuation of Fig.~D.1.}
\label{fig:discspec2}
\end{figure*}
}

\subsection{Post-AGB candidates with a shell}

\onlfig{3}{
\begin{figure*}
\resizebox{\hsize}{!}{
\includegraphics{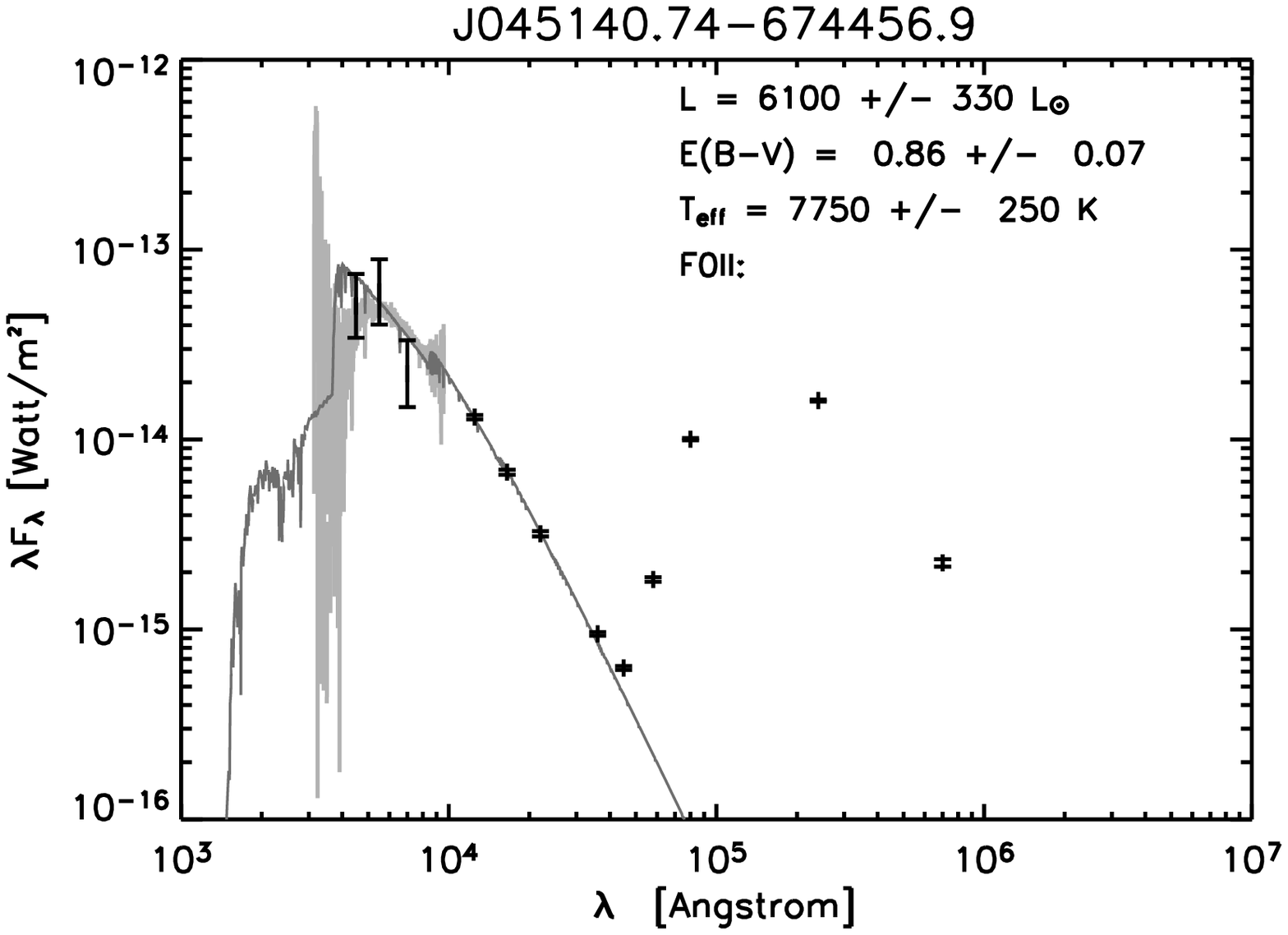}
\includegraphics{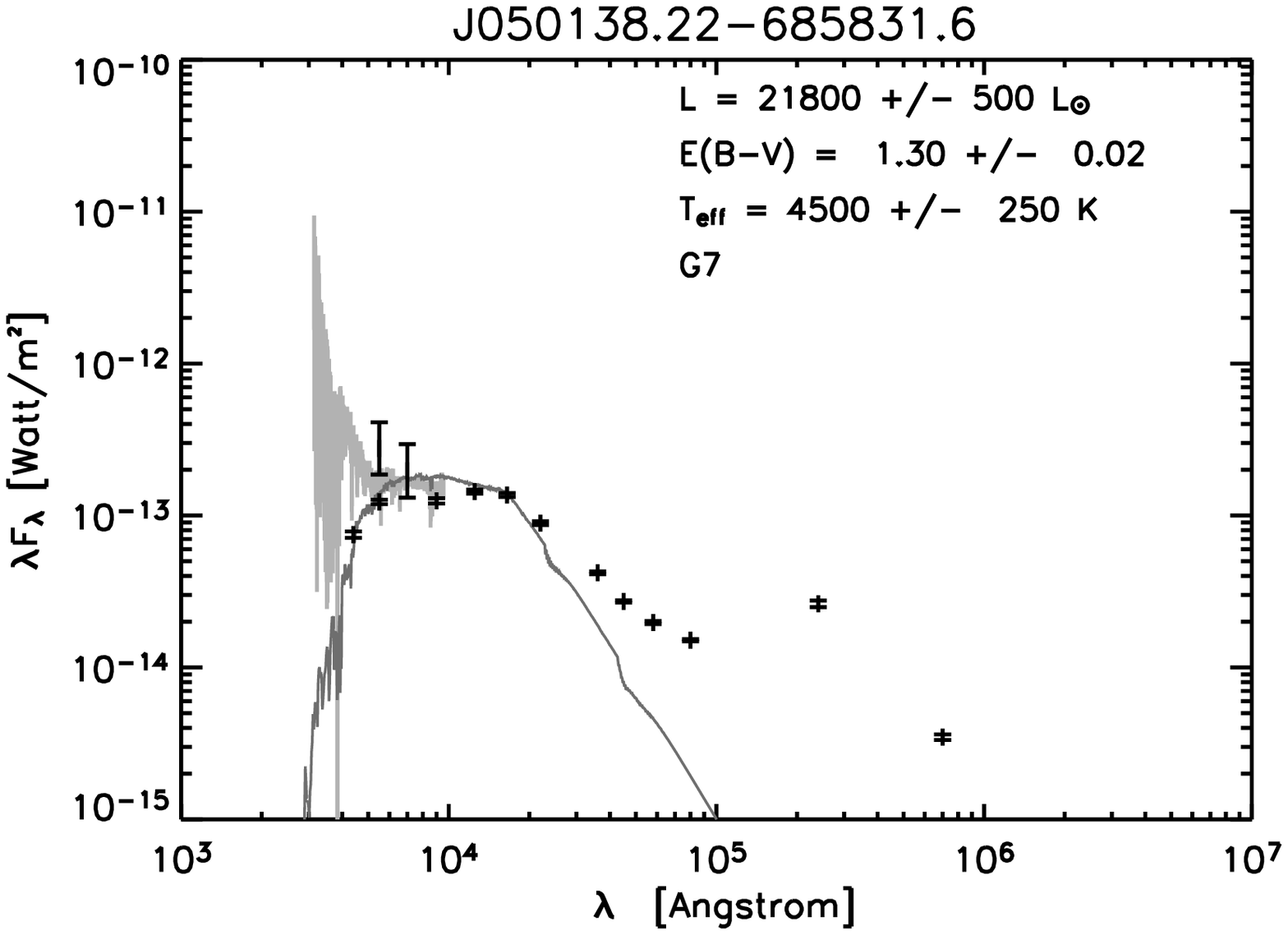}
\includegraphics{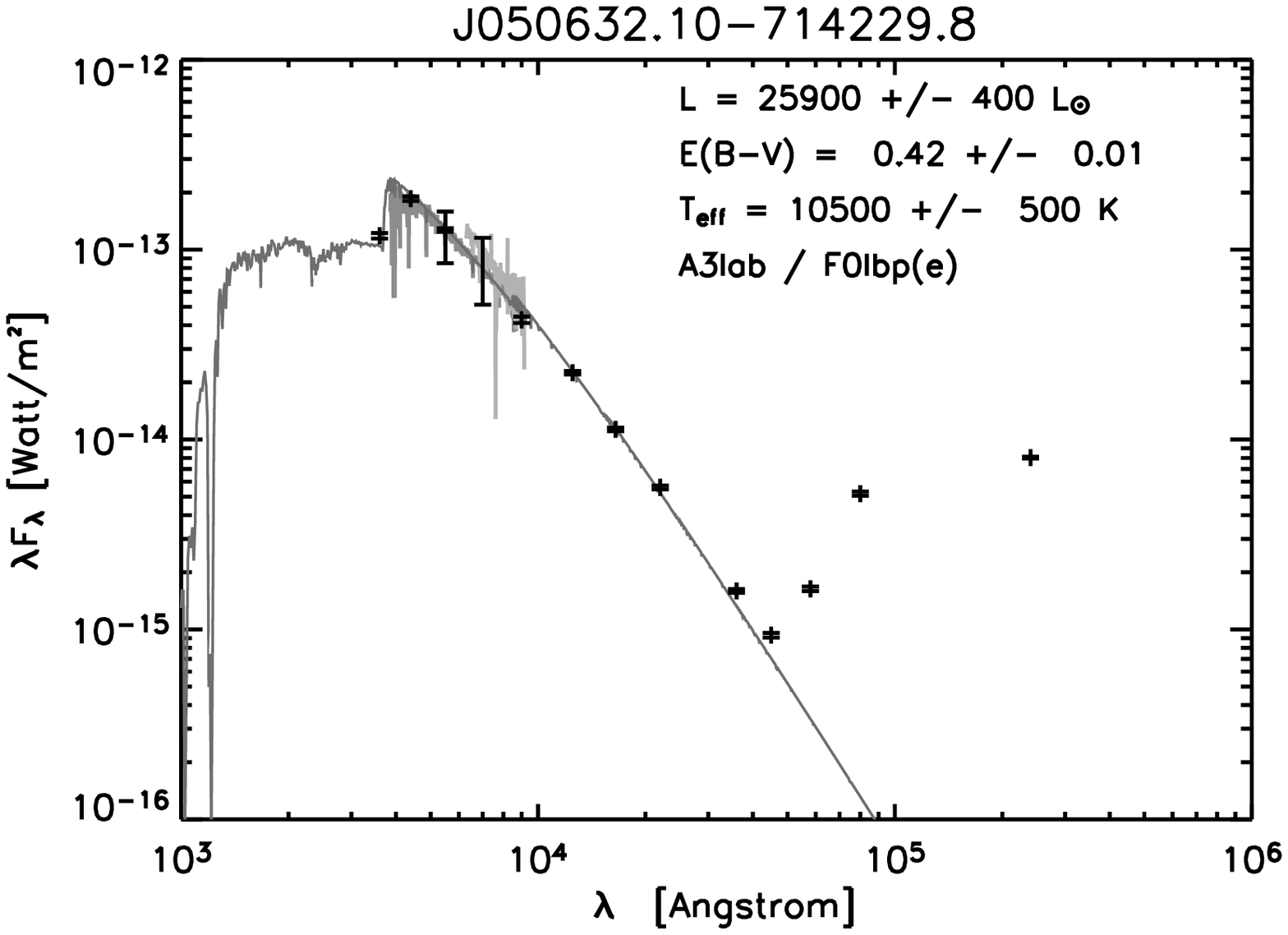}
}
\resizebox{\hsize}{!}{
\includegraphics{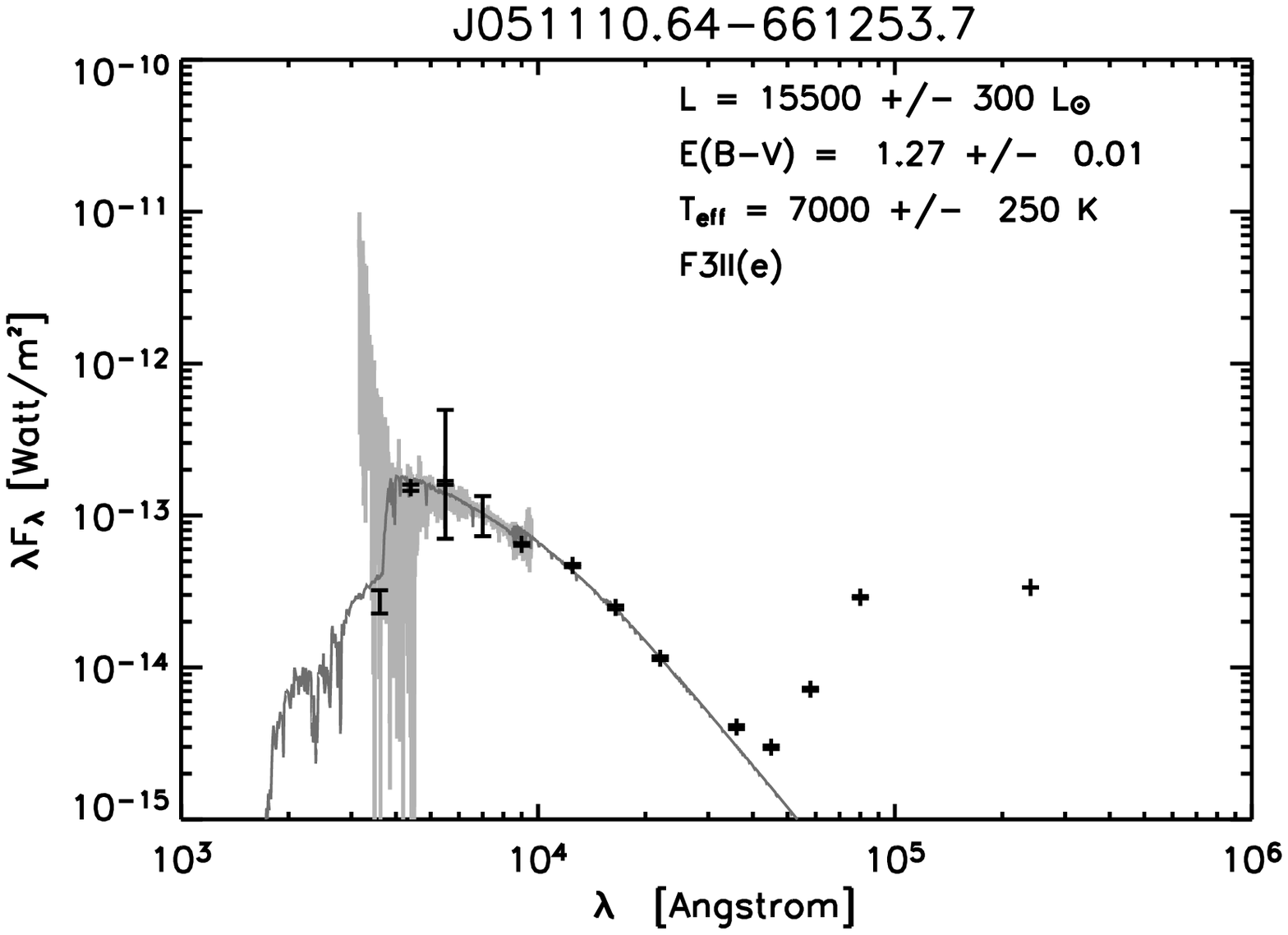}
\includegraphics{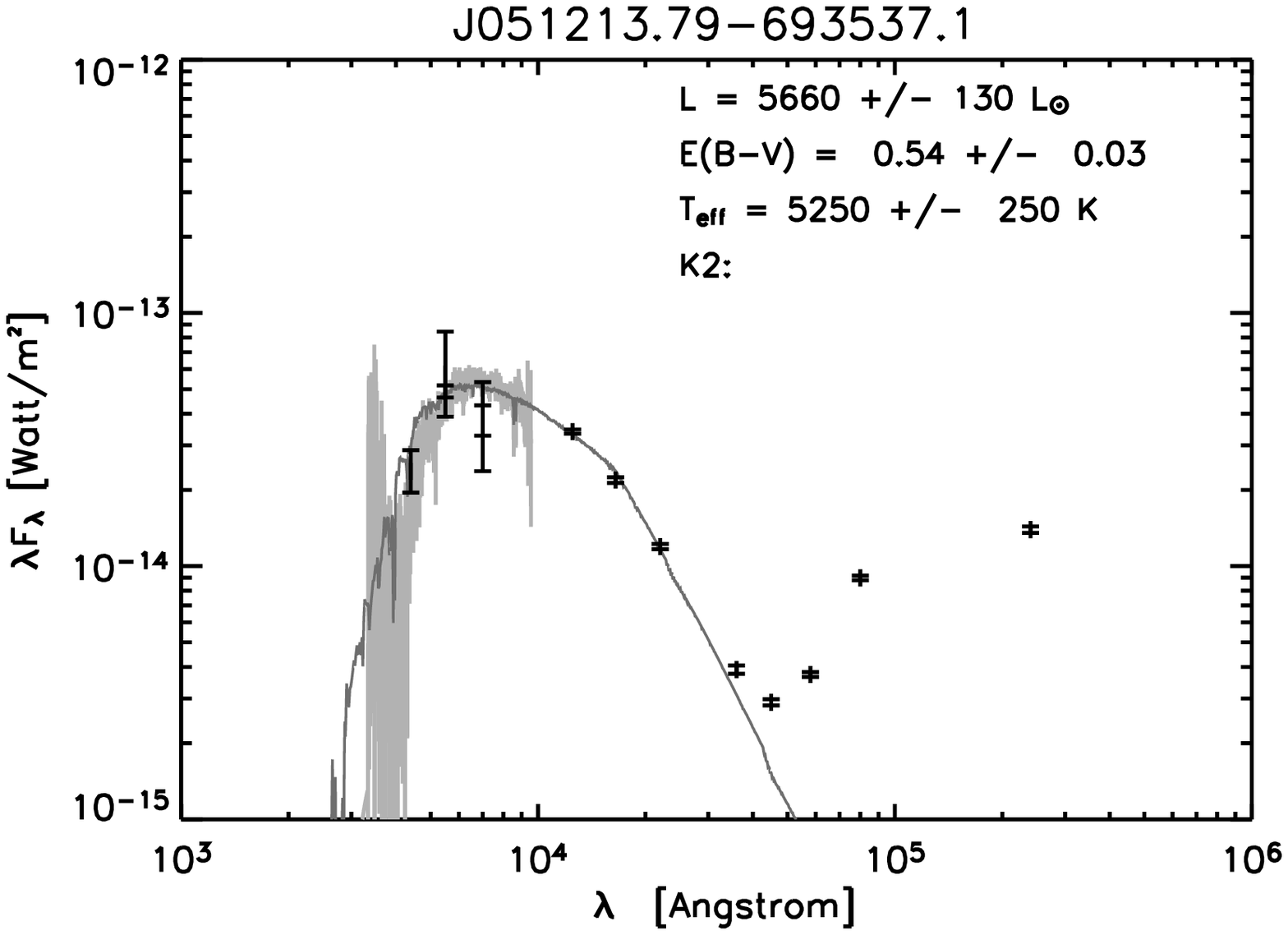}
\includegraphics{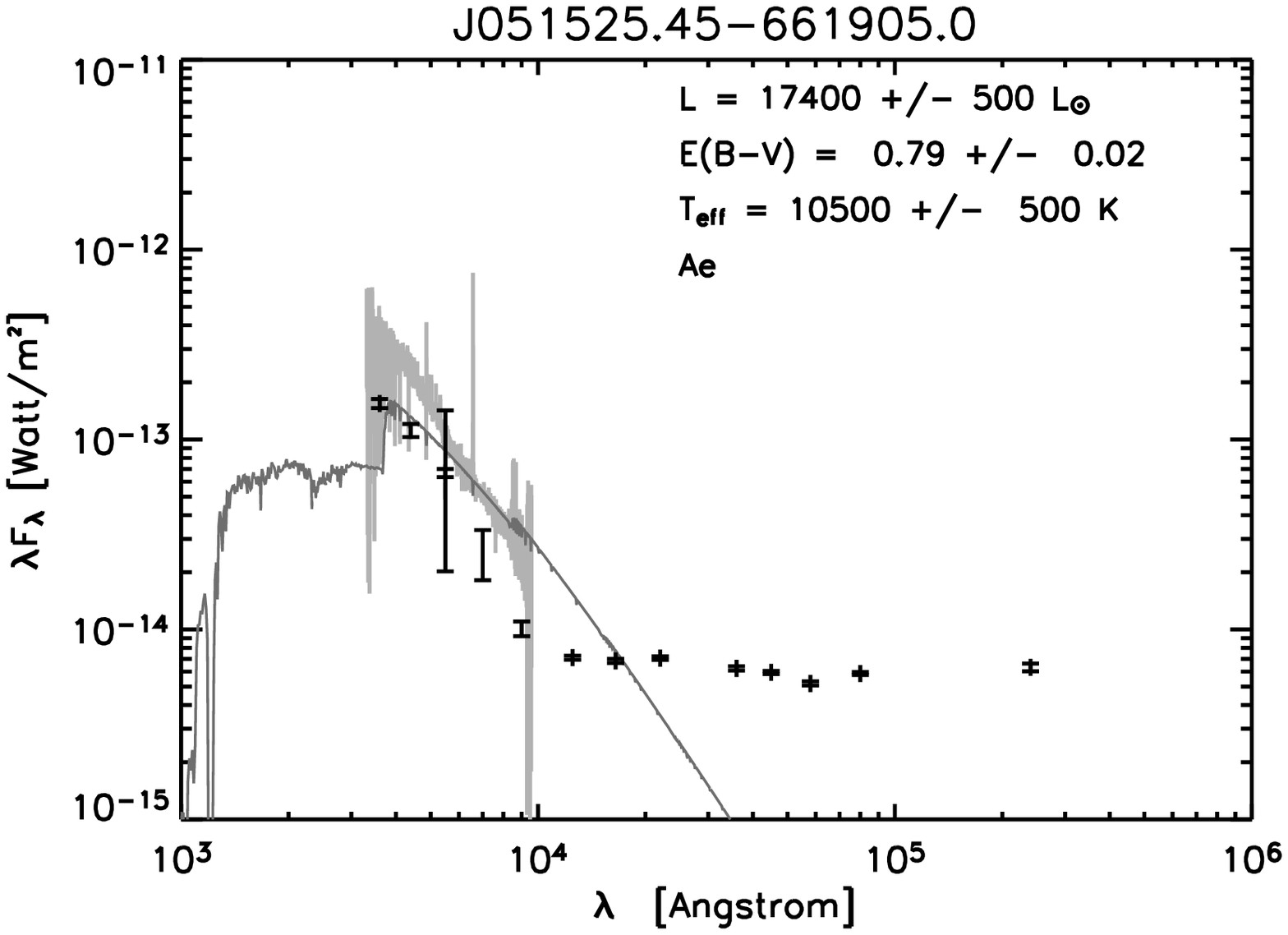}
}
\resizebox{\hsize}{!}{
\includegraphics{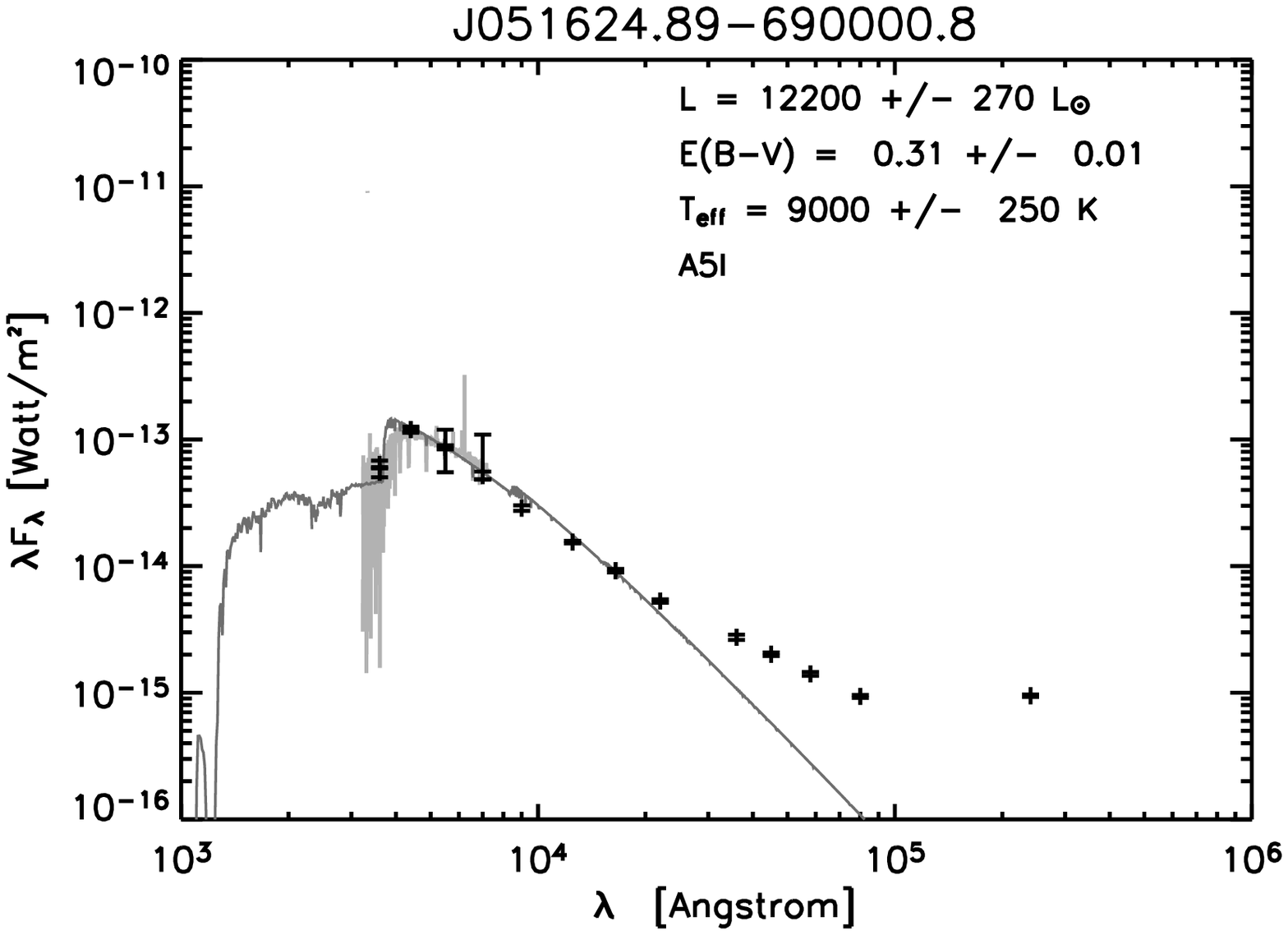}
\includegraphics{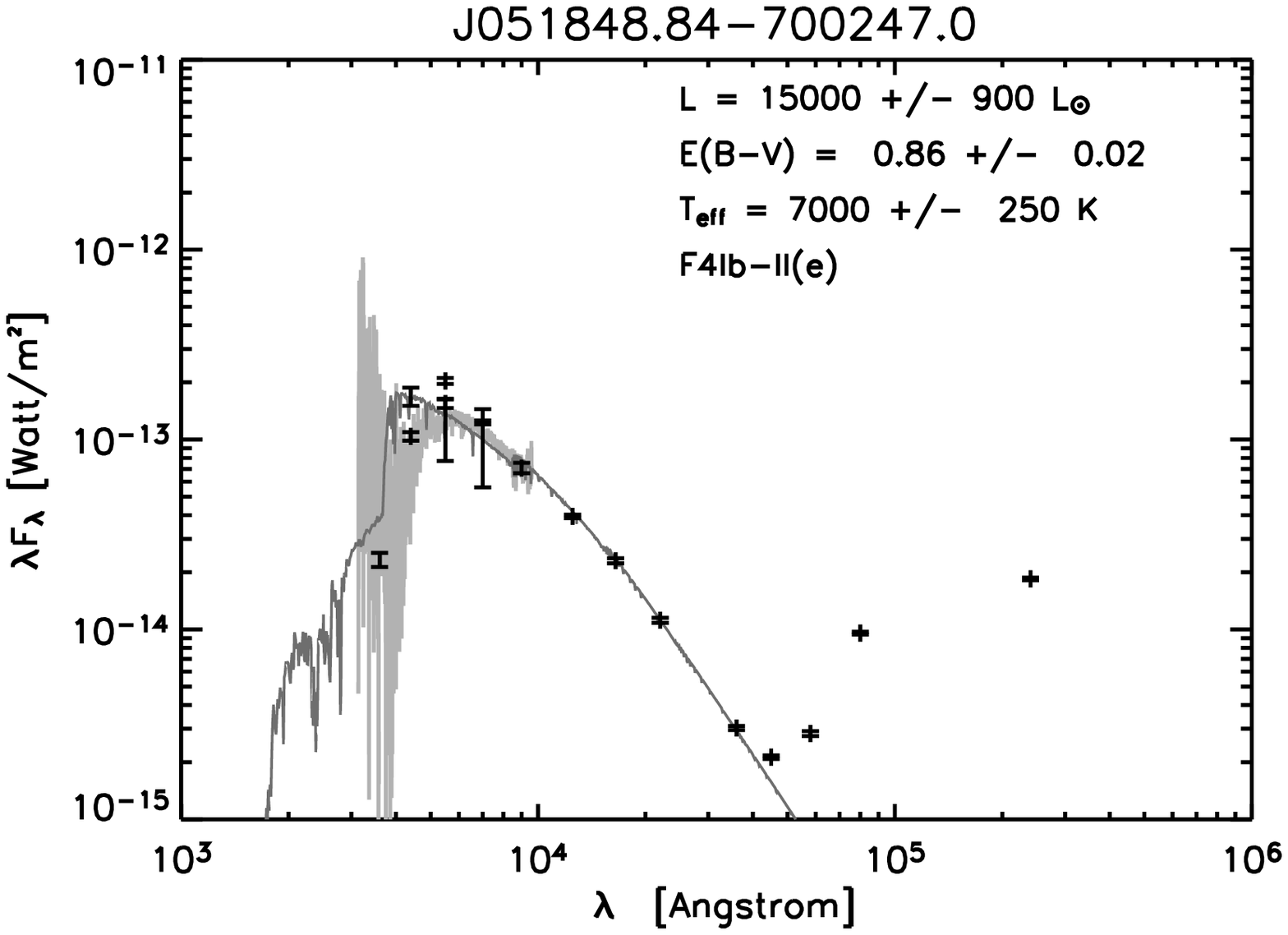}
\includegraphics{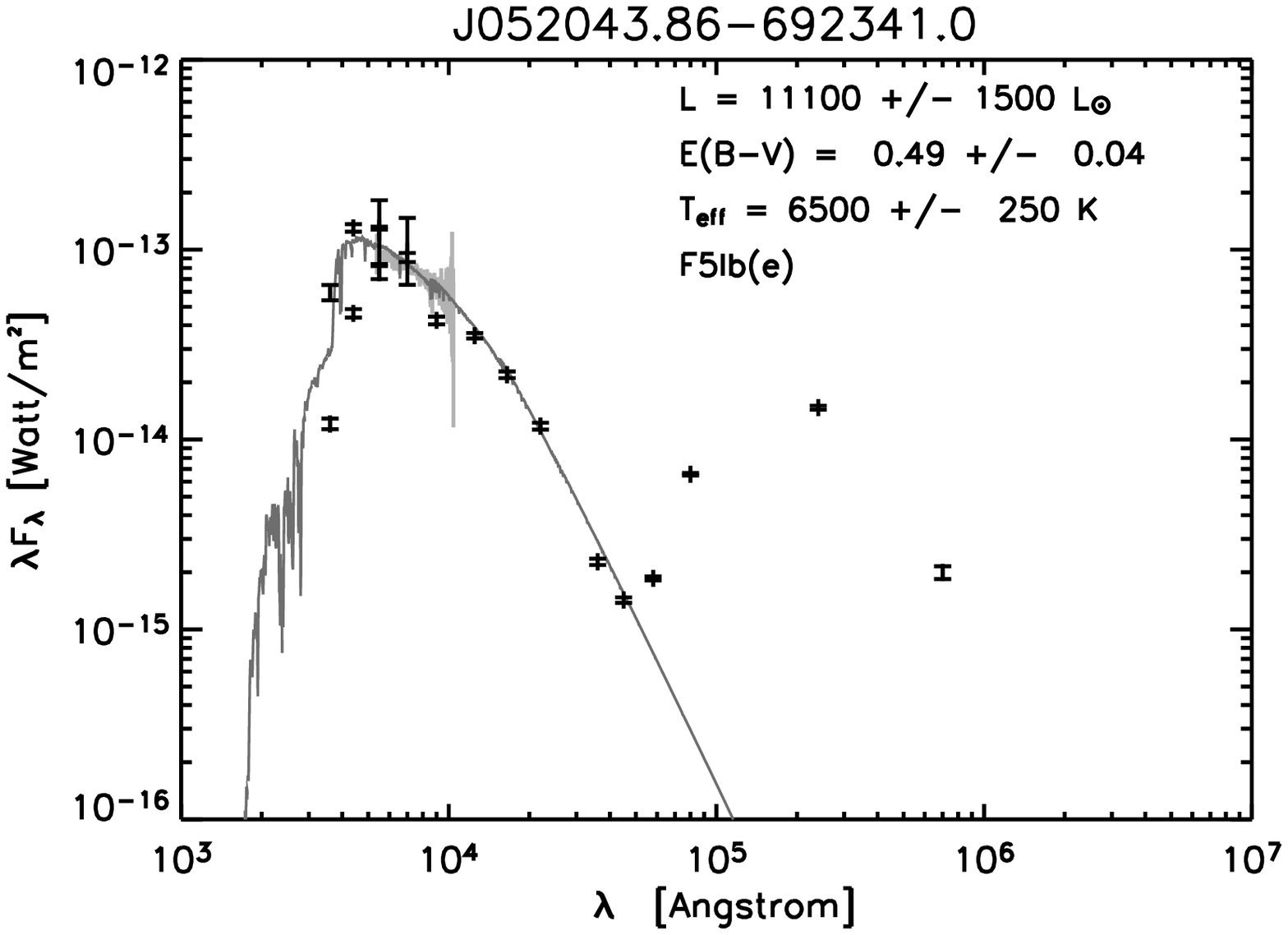}
}
\resizebox{\hsize}{!}{
\includegraphics{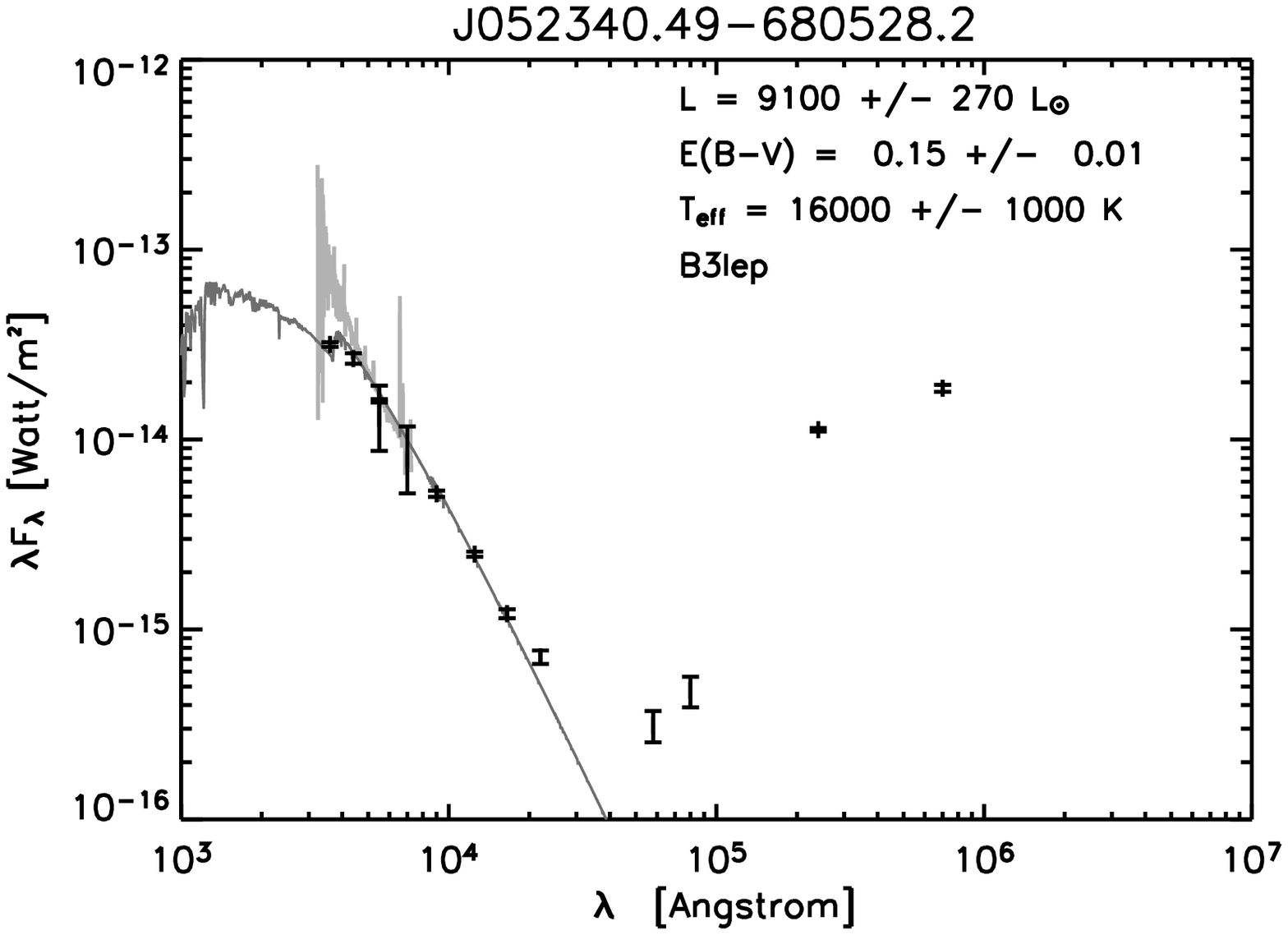}
\includegraphics{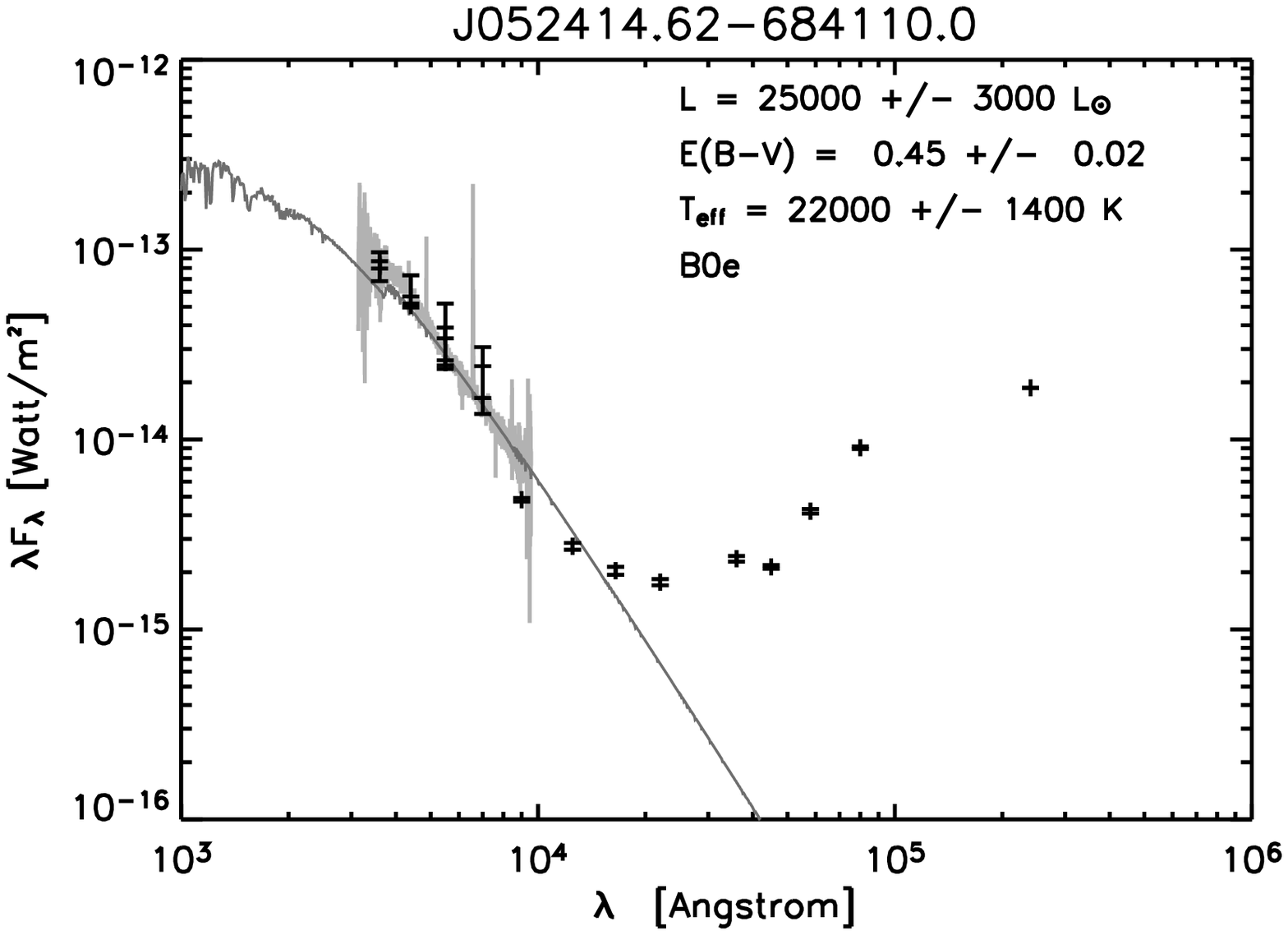}
\includegraphics{15834fg7b.ps}
}
\resizebox{\hsize}{!}{
\includegraphics{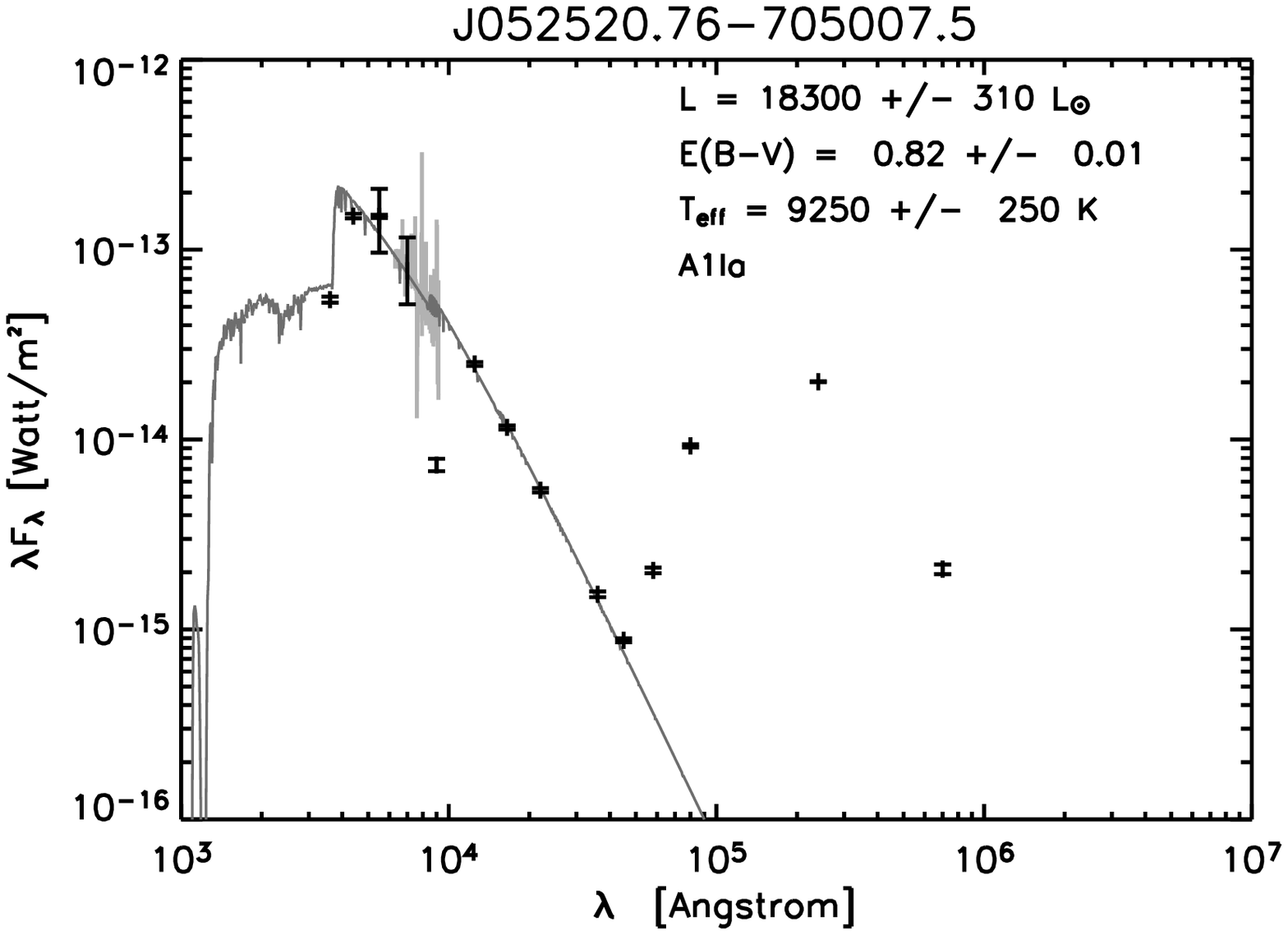}
\includegraphics{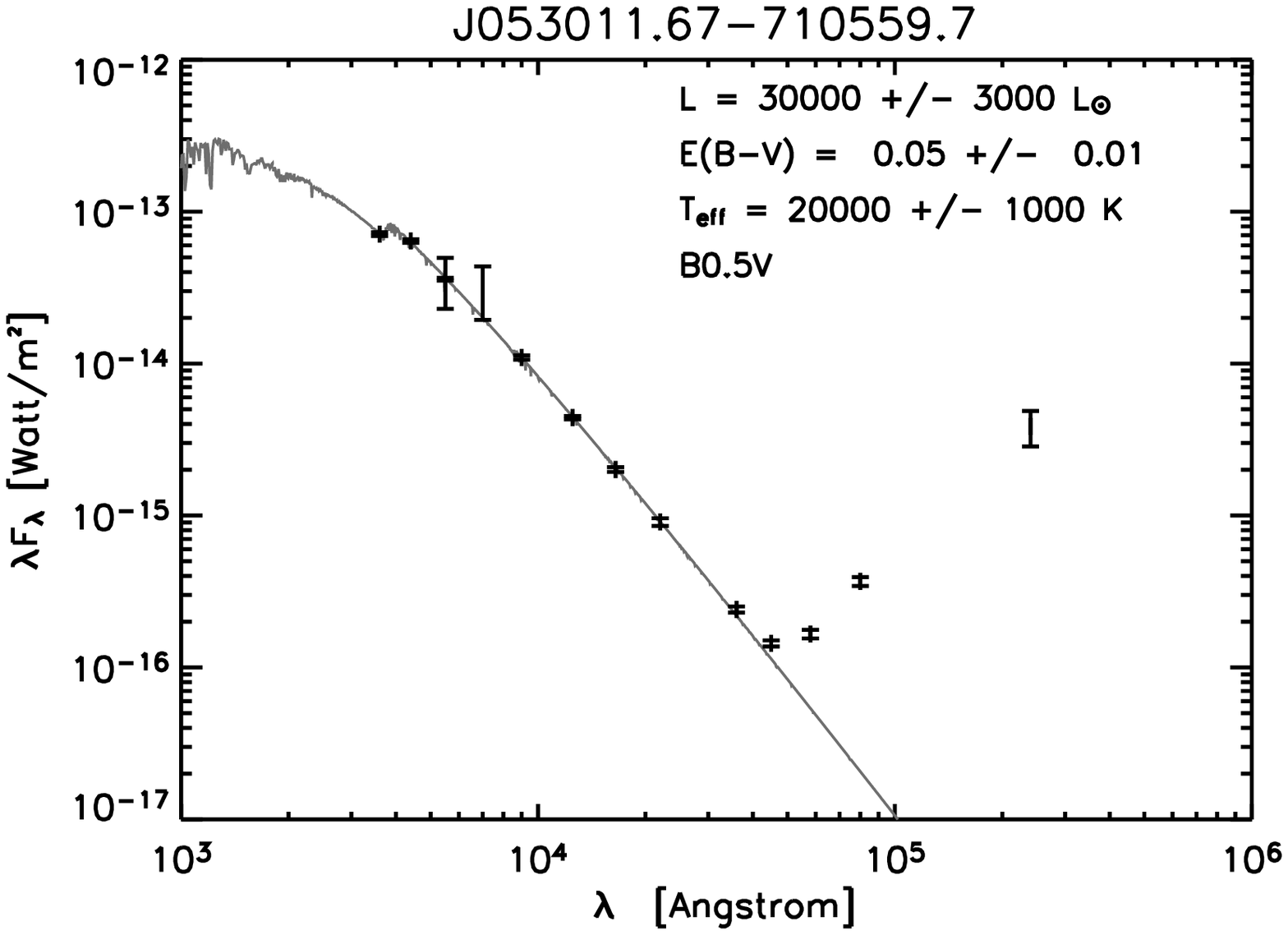}
\includegraphics{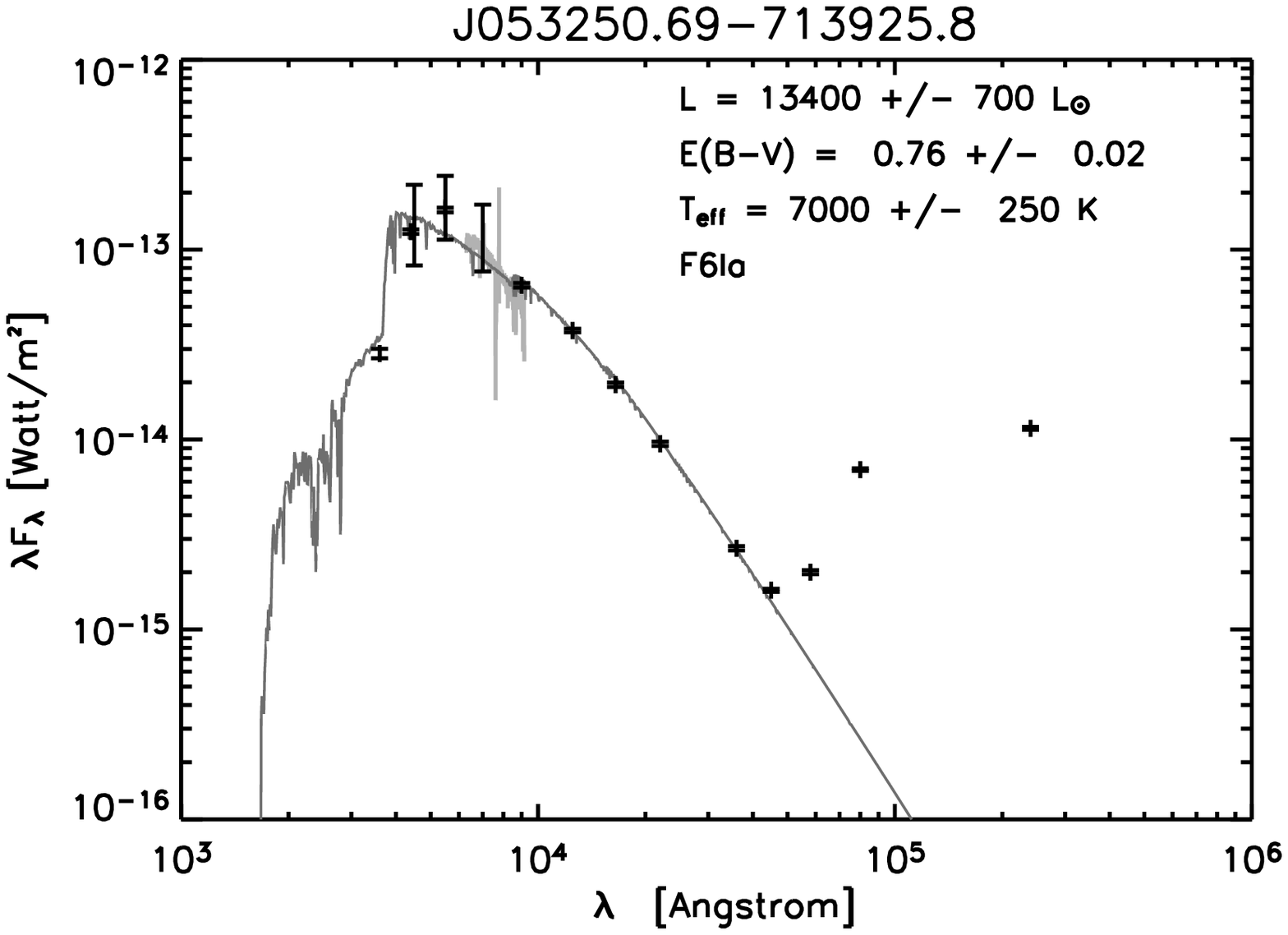}
}
\caption{Same as Fig.~D.1, but for objects with a shell.}\label{fig:shellspec44}
\end{figure*}
}

\onlfig{4}{
\begin{figure*}
\resizebox{\hsize}{!}{
\includegraphics{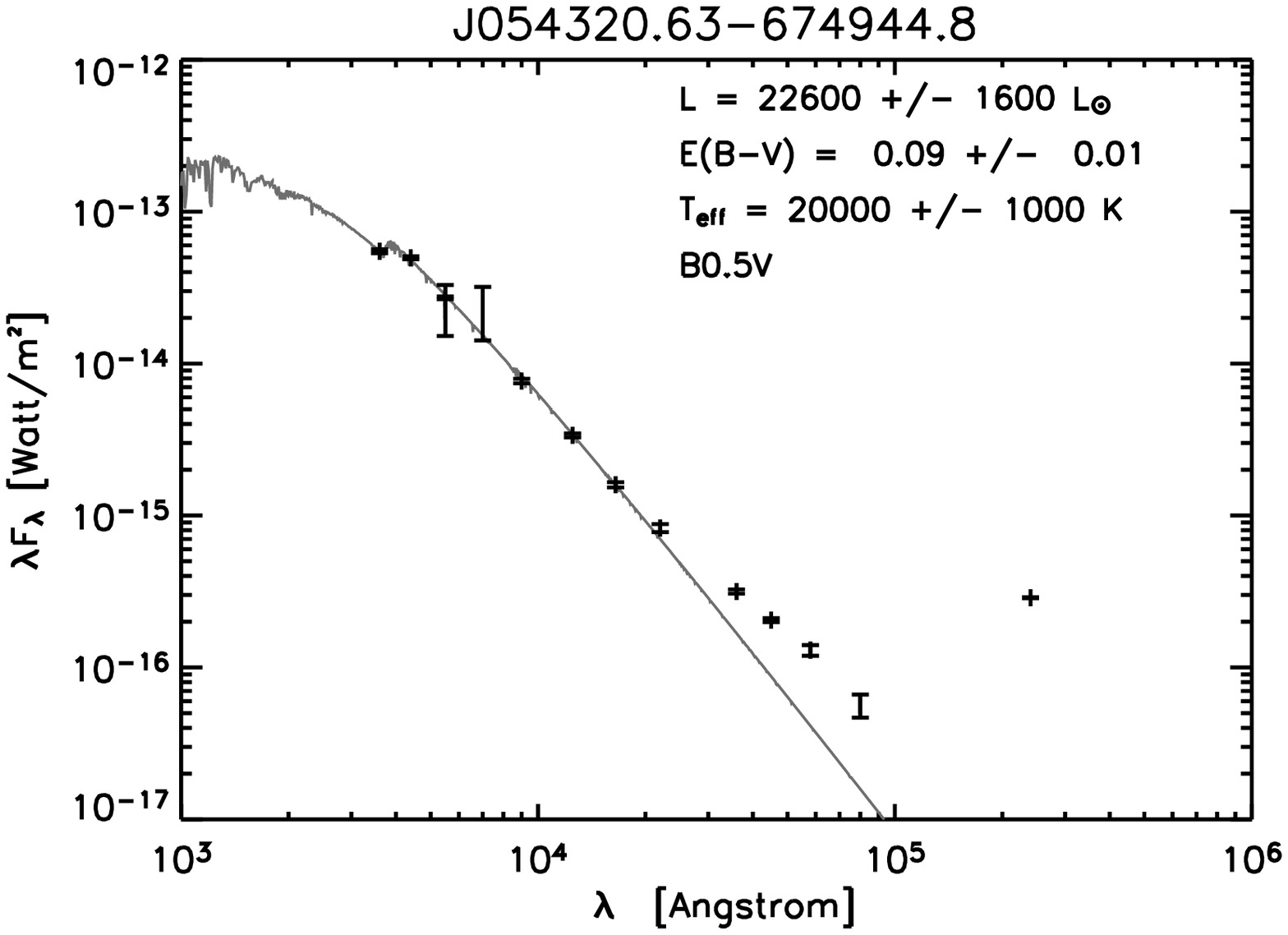}
\includegraphics{15834fgd2e.ps}
\includegraphics{15834fgd2e.ps}
}
\caption{Continuation of Fig.~\ref{fig:shellspec44}.}
\label{fig:shellspec45}
\end{figure*}
}

\subsection{Post-AGB candidates that fulfil all criteria}

\onlfig{5}{
\begin{figure*}
\resizebox{\hsize}{!}{
\includegraphics{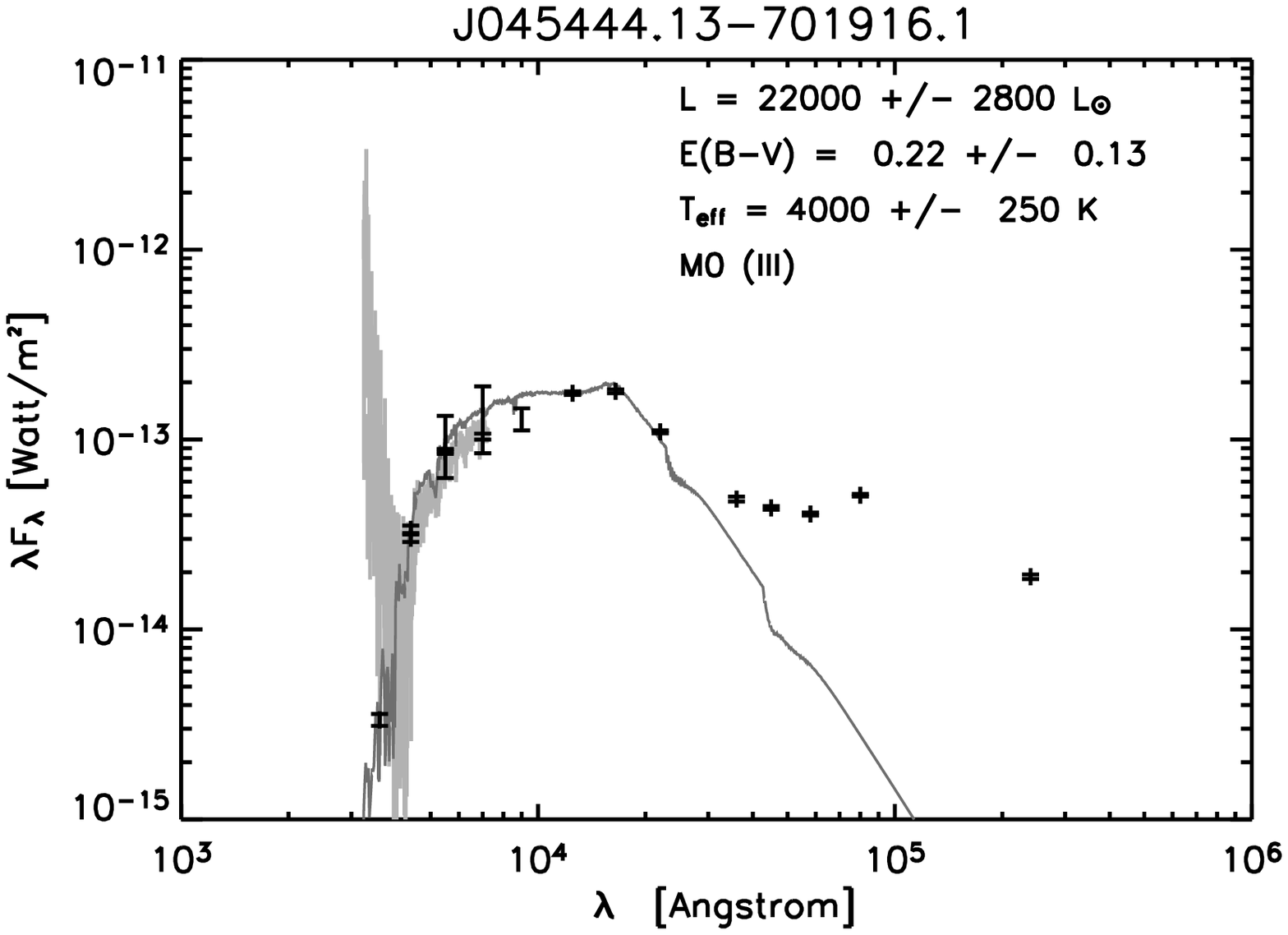}
\includegraphics{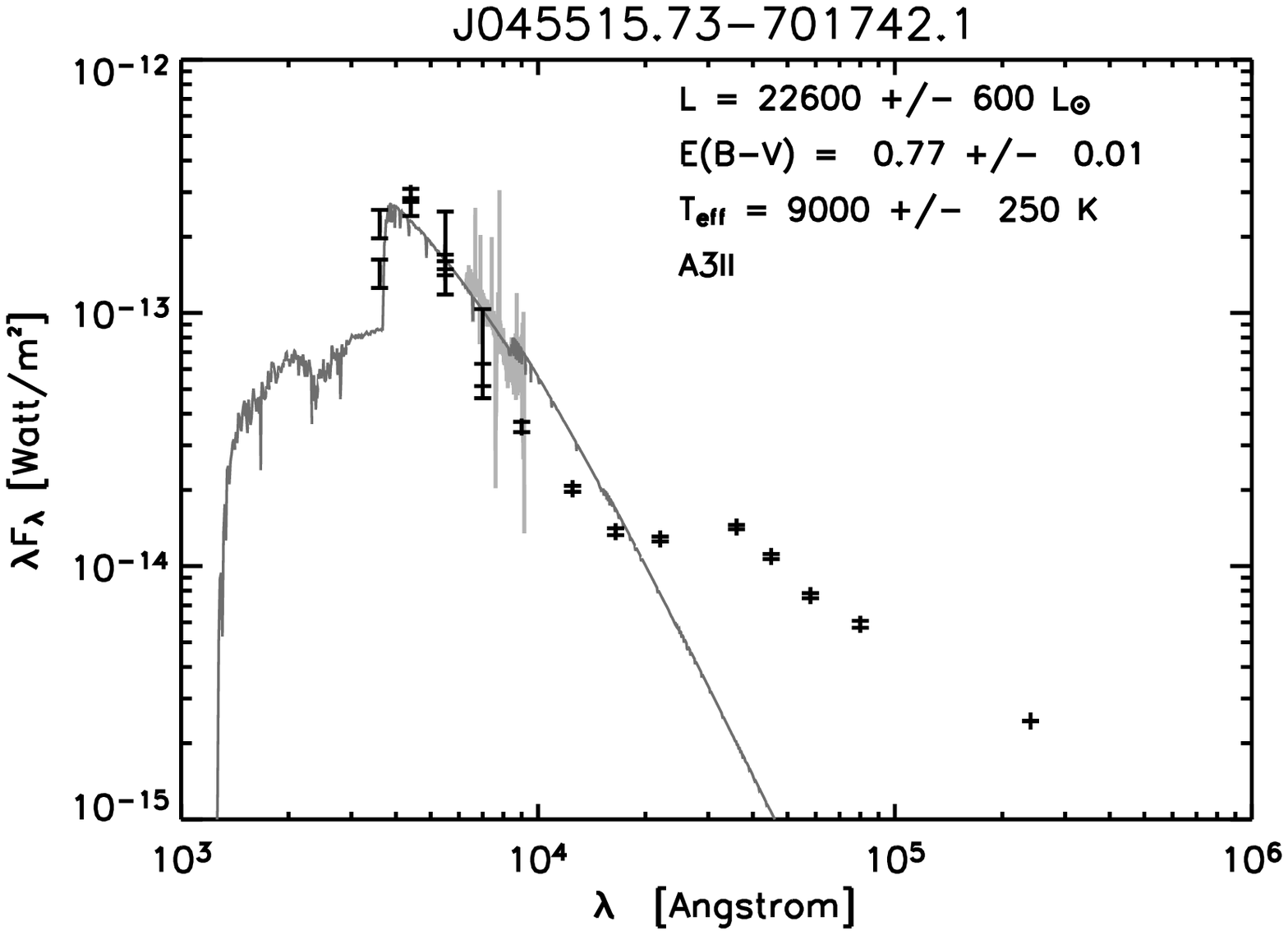}
\includegraphics{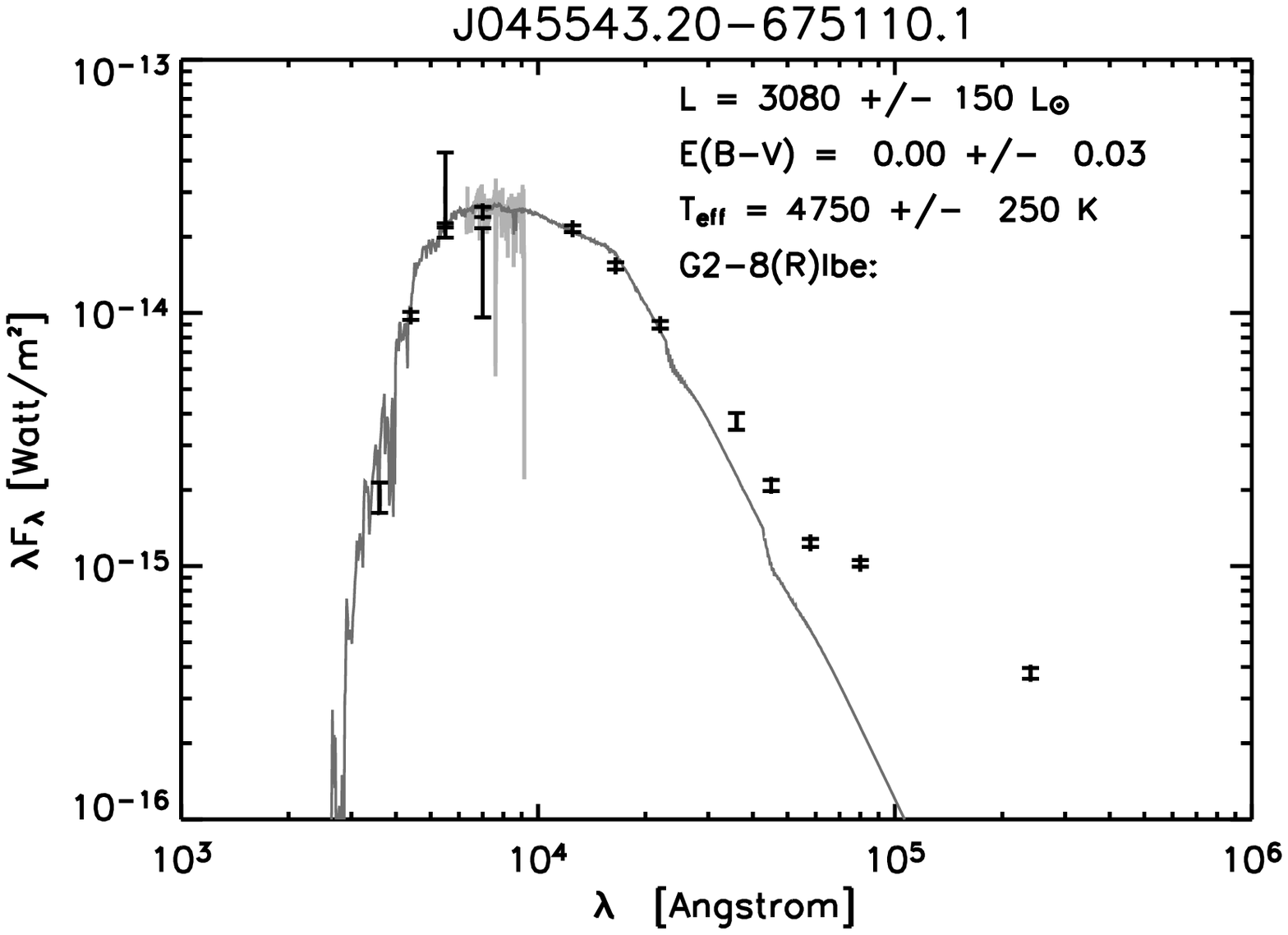}
}
\resizebox{\hsize}{!}{
\includegraphics{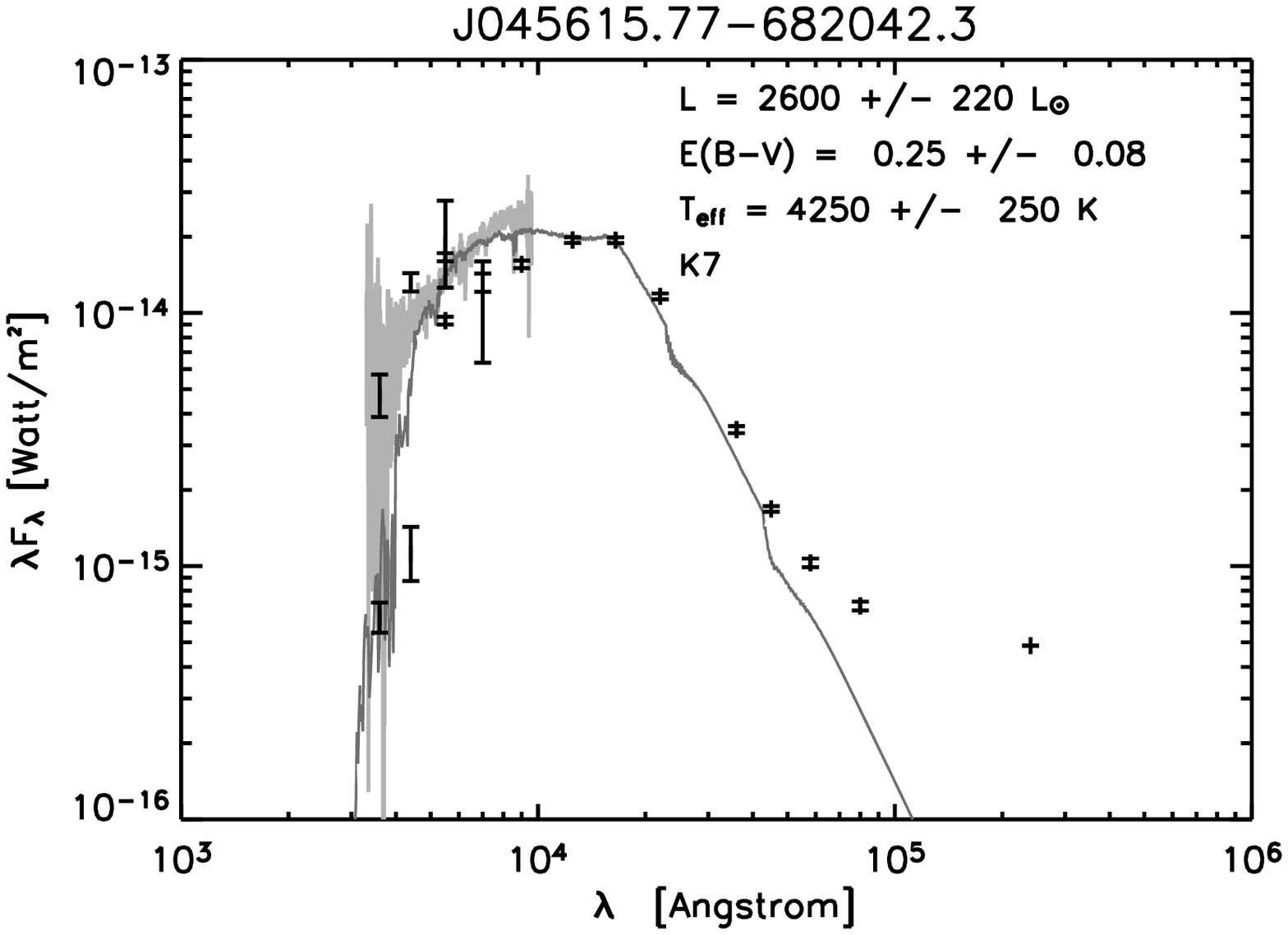}
\includegraphics{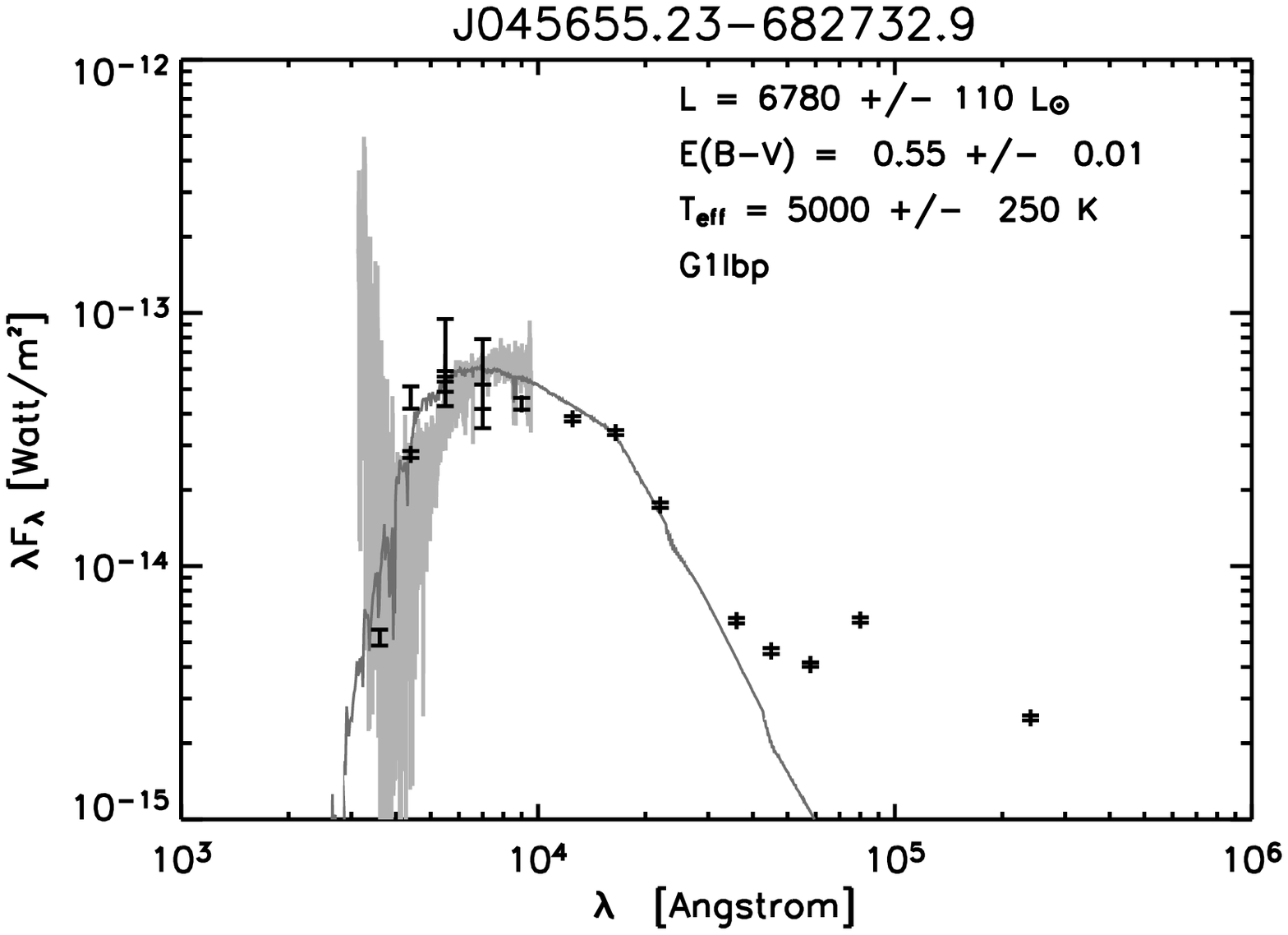}
\includegraphics{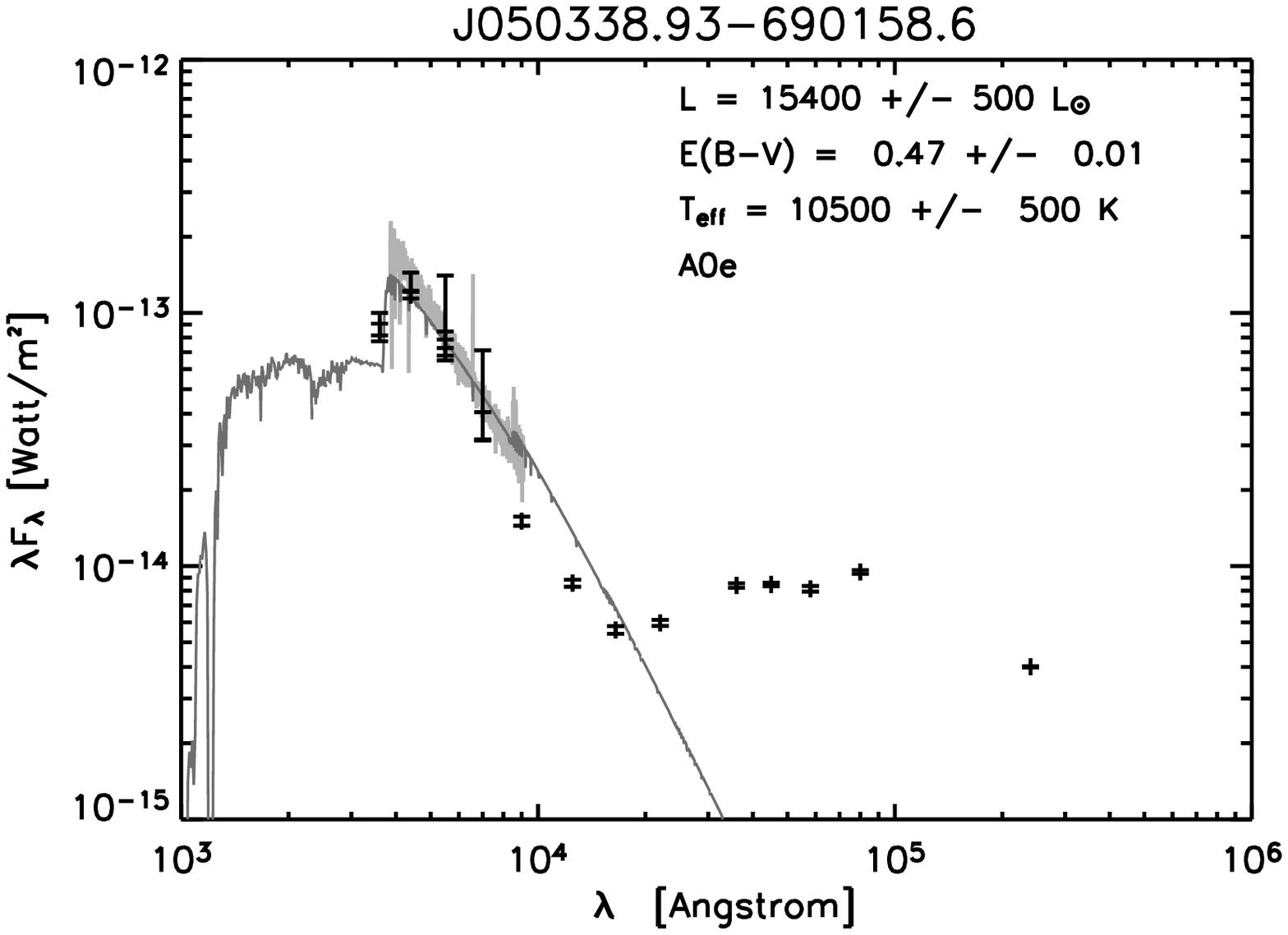}
}
\resizebox{\hsize}{!}{
\includegraphics{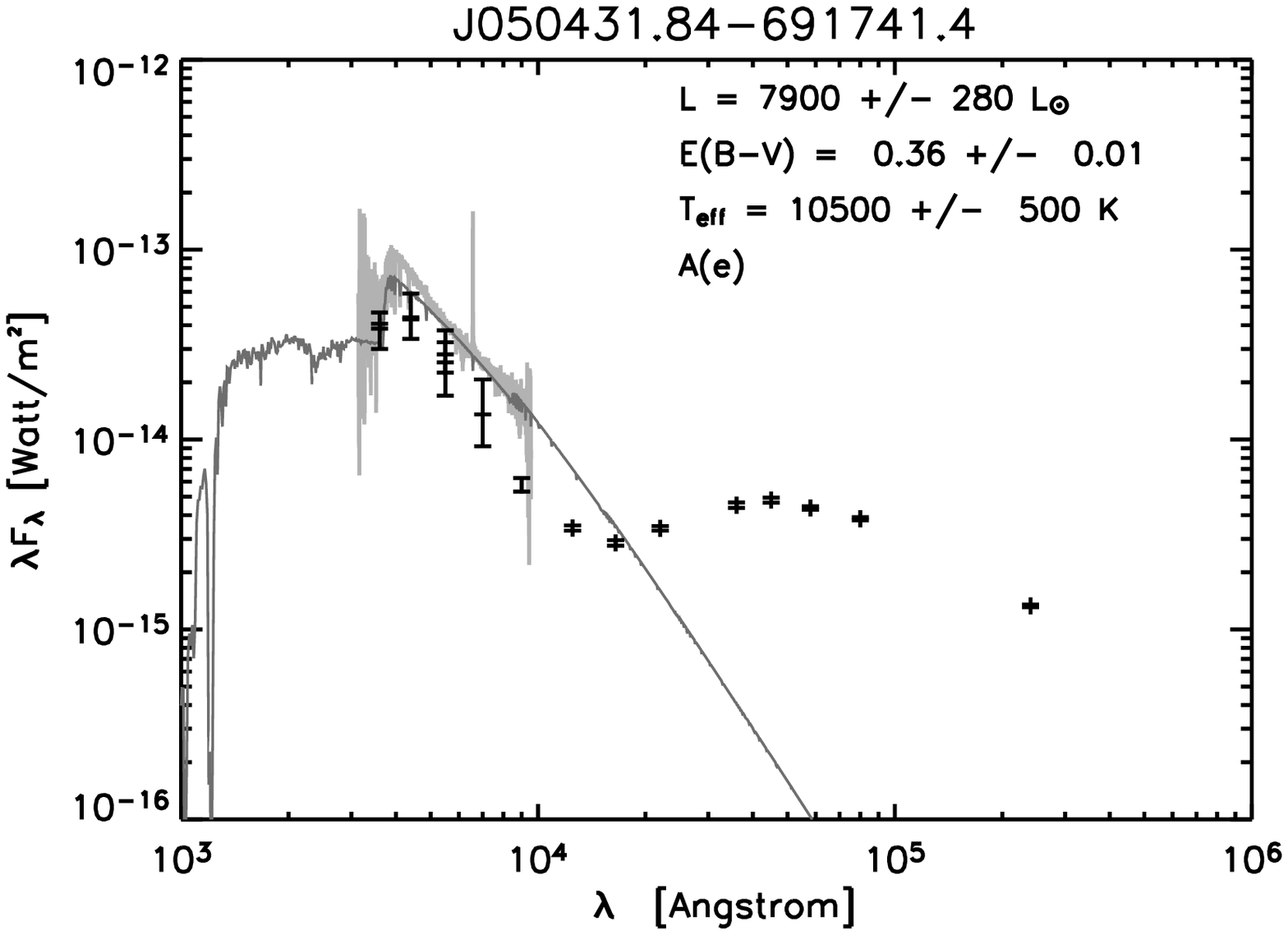}
\includegraphics{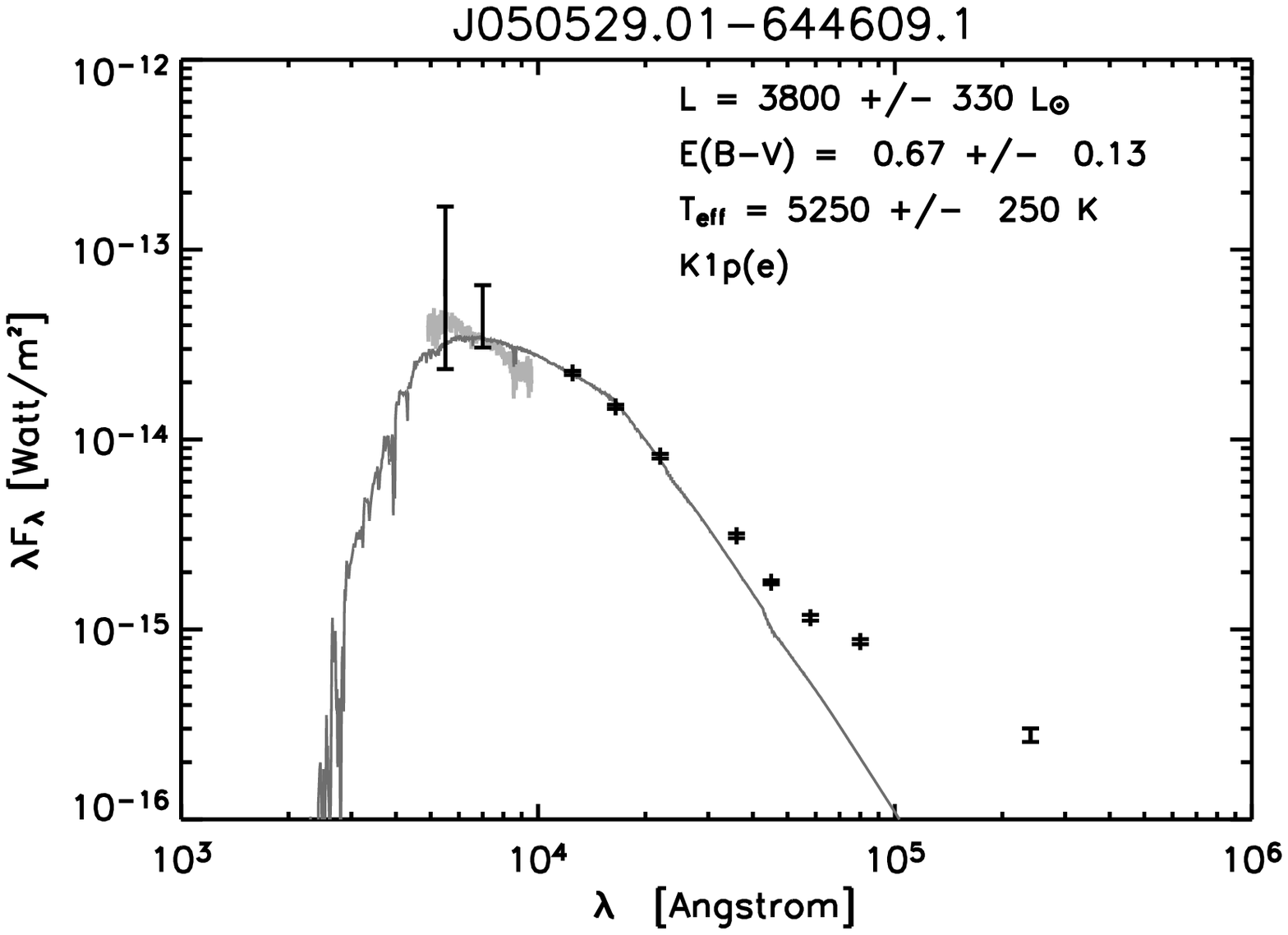}
\includegraphics{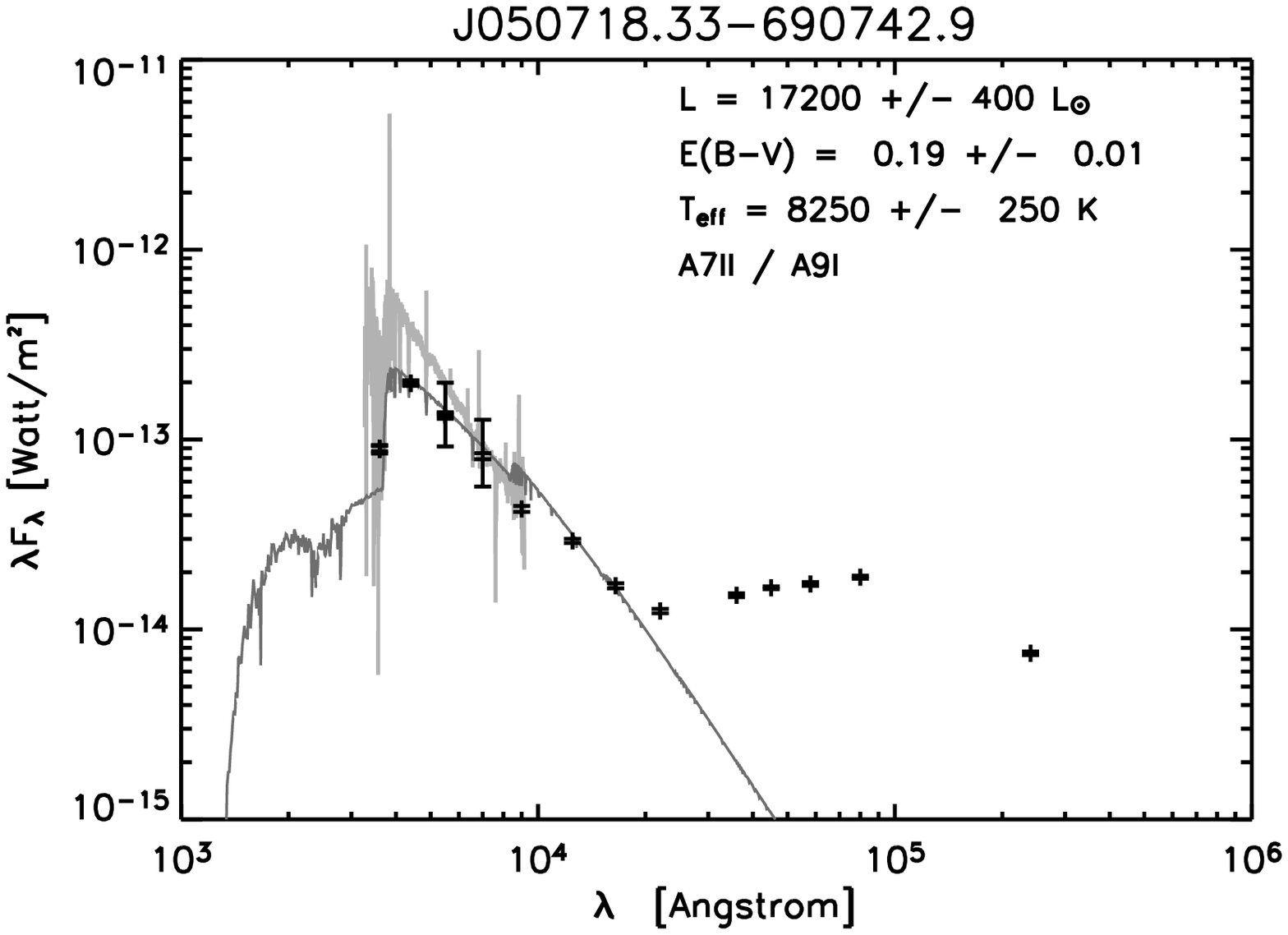}
}
\resizebox{\hsize}{!}{
\includegraphics{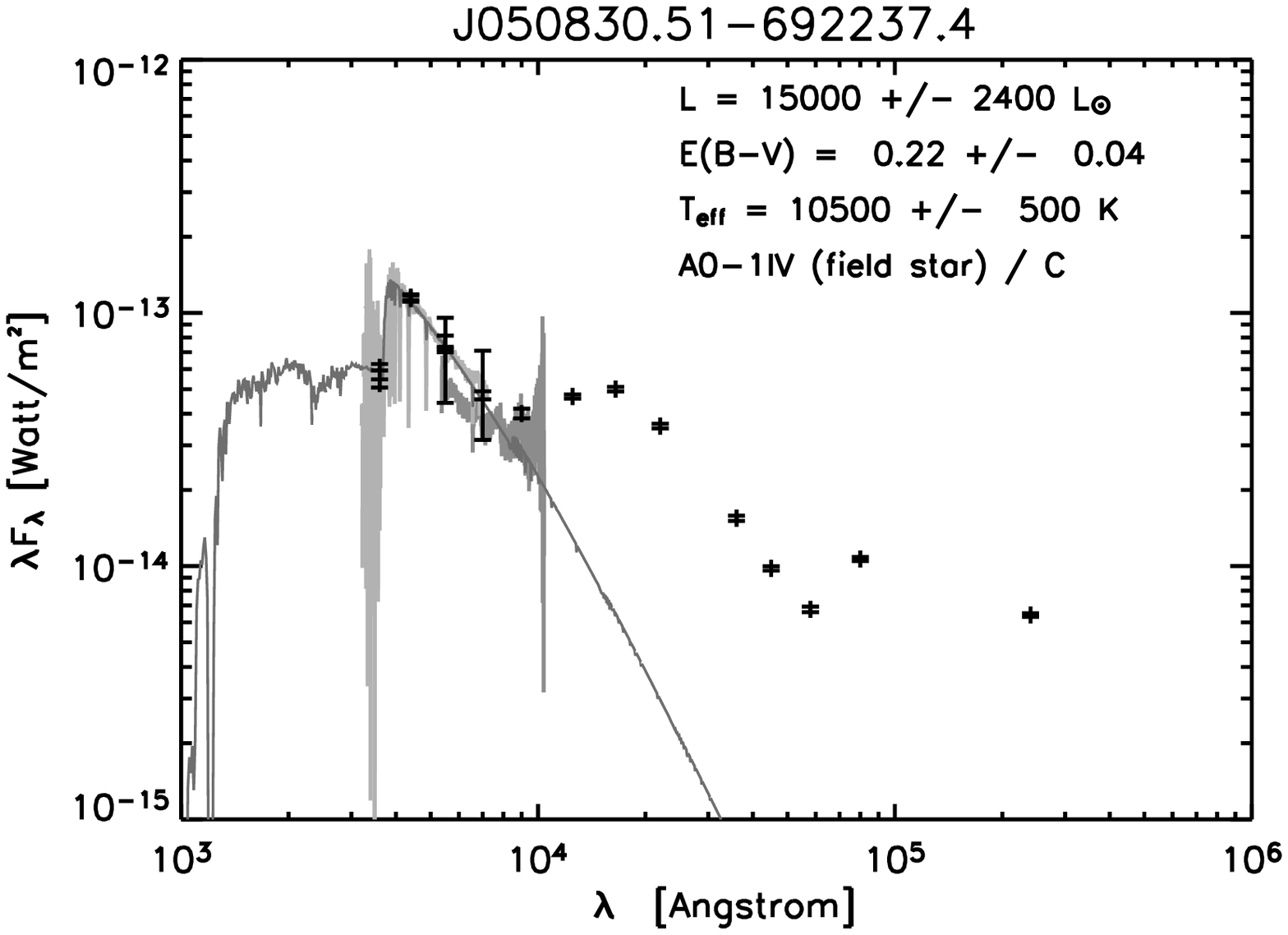}
\includegraphics{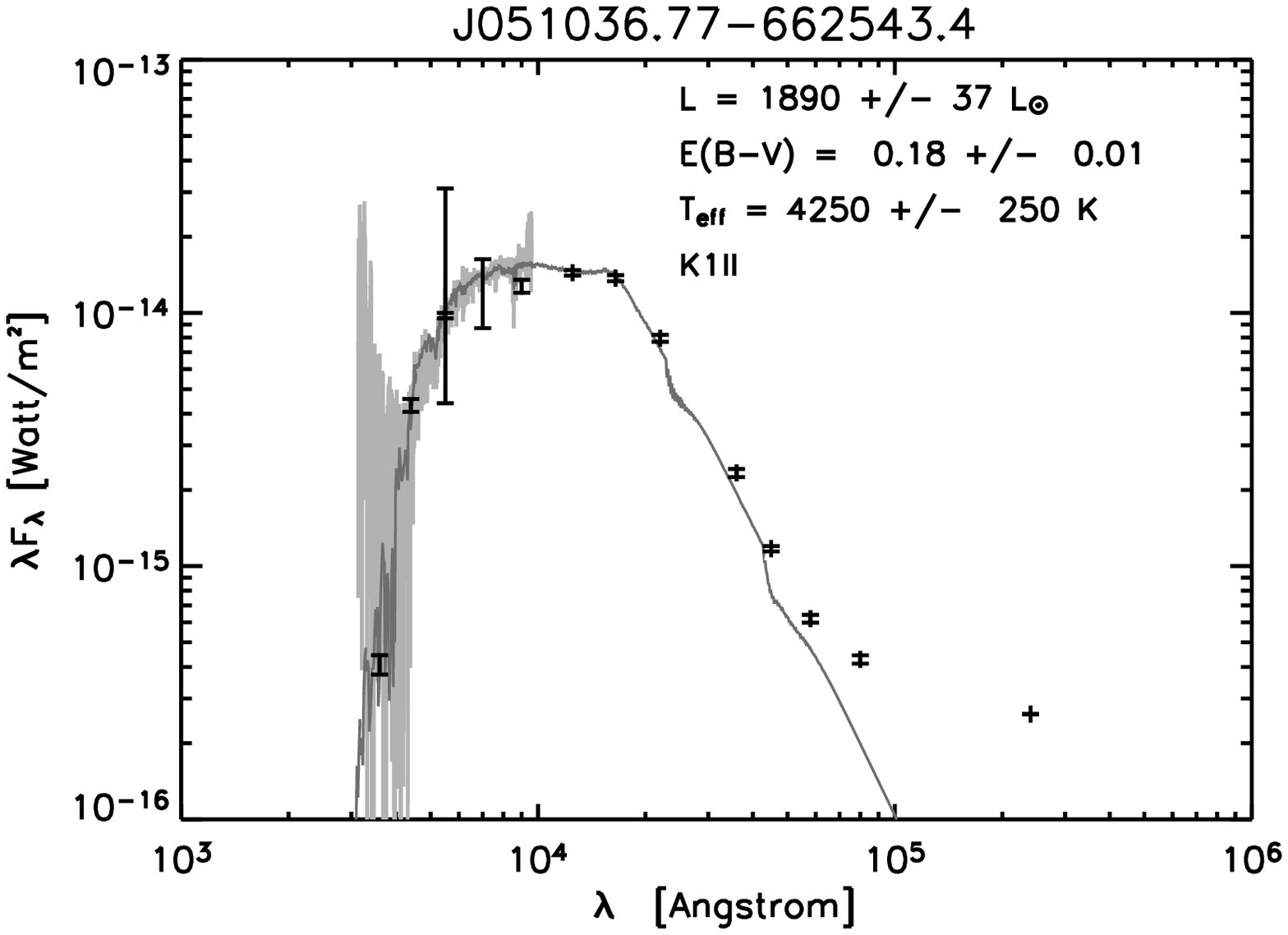}
\includegraphics{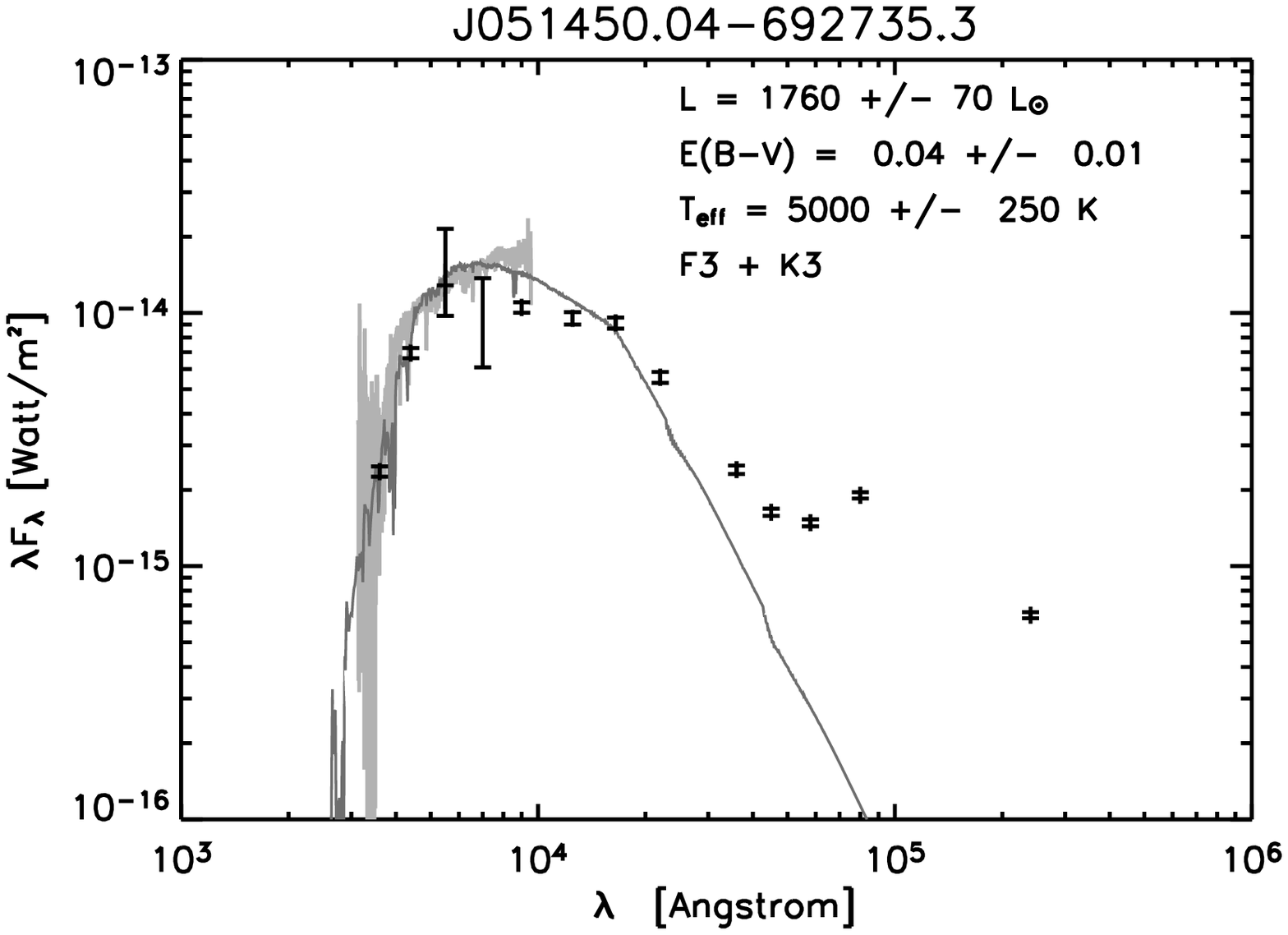}
}
\resizebox{\hsize}{!}{
\includegraphics{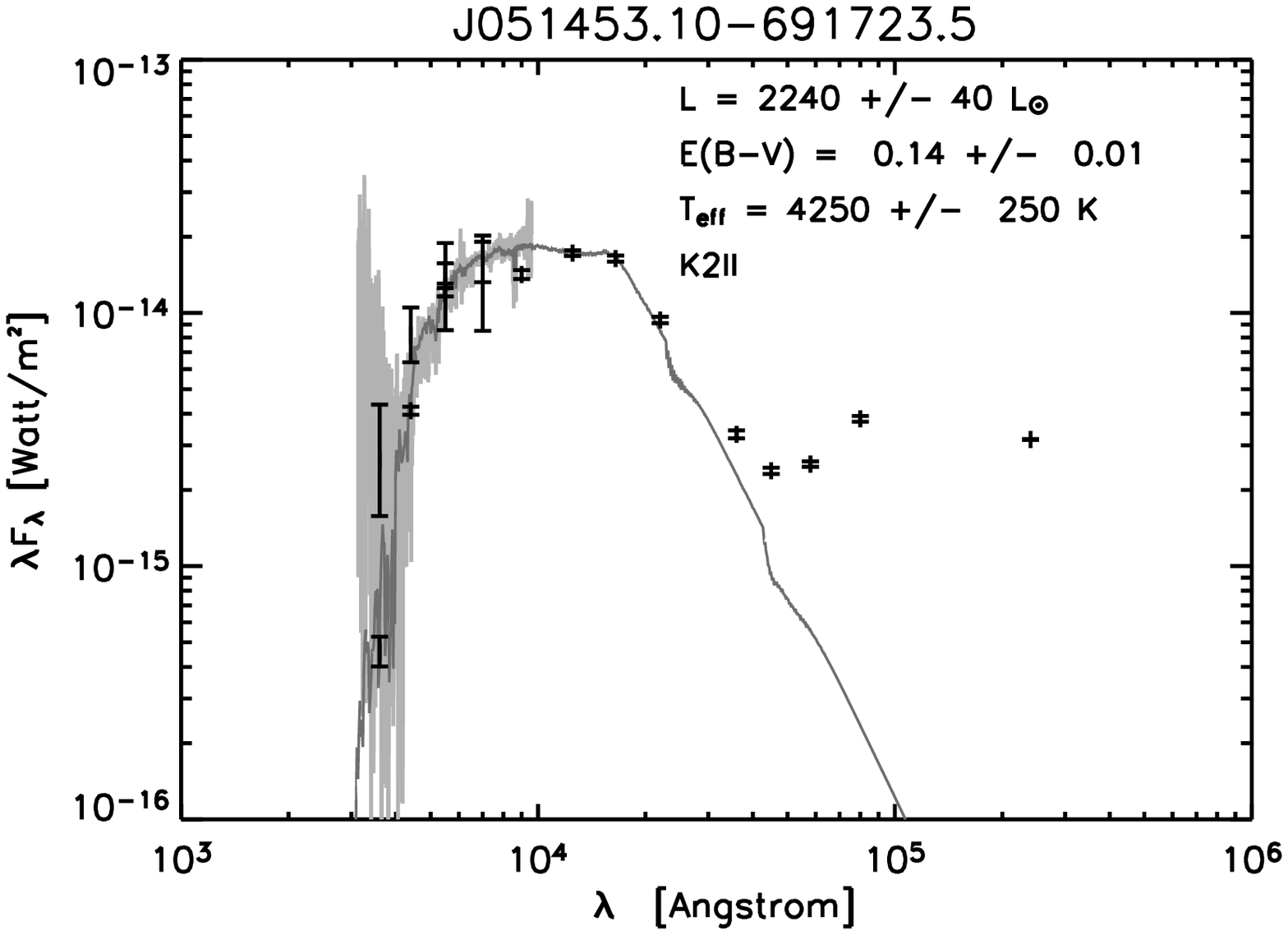}
\includegraphics{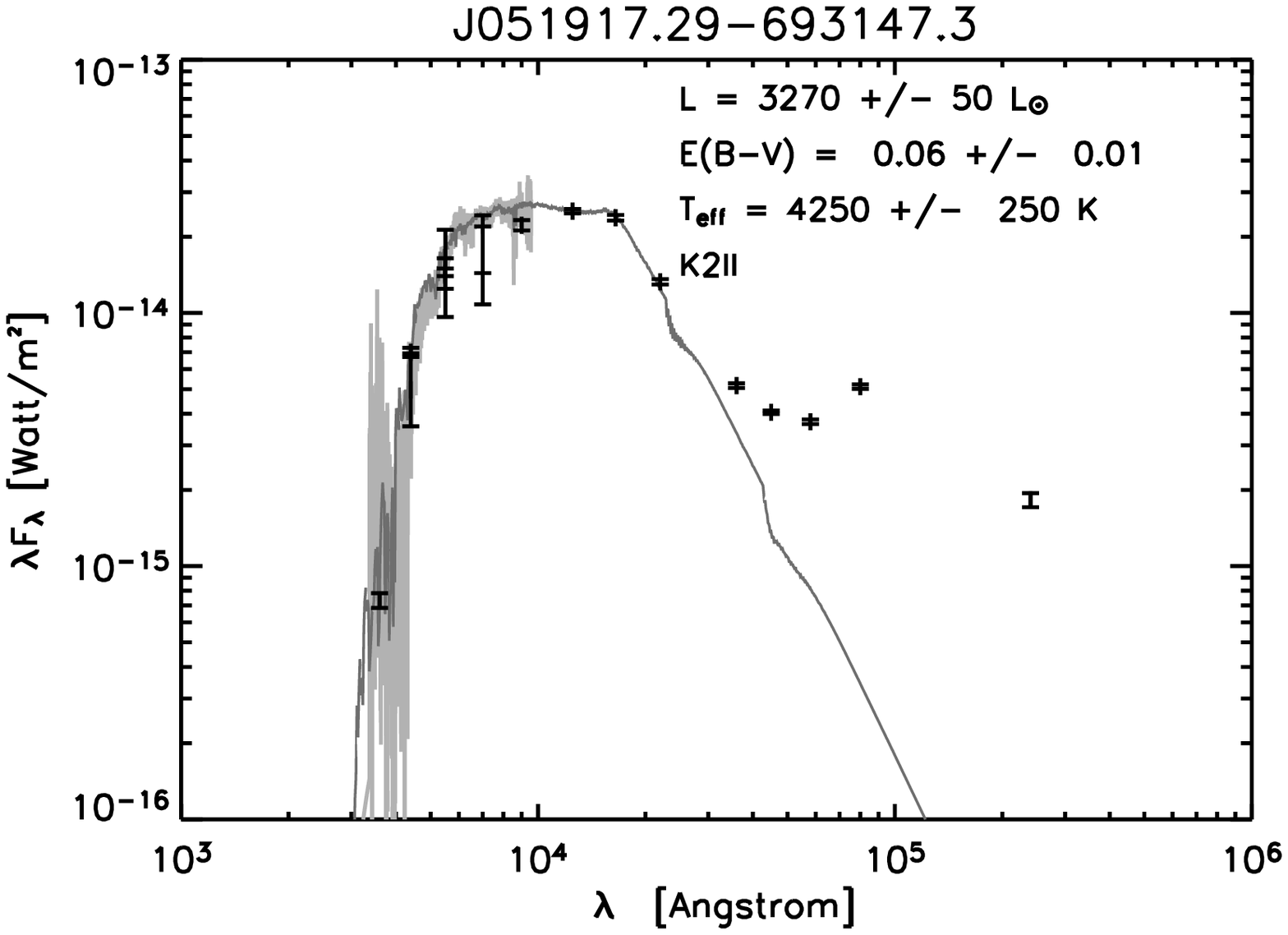}
\includegraphics{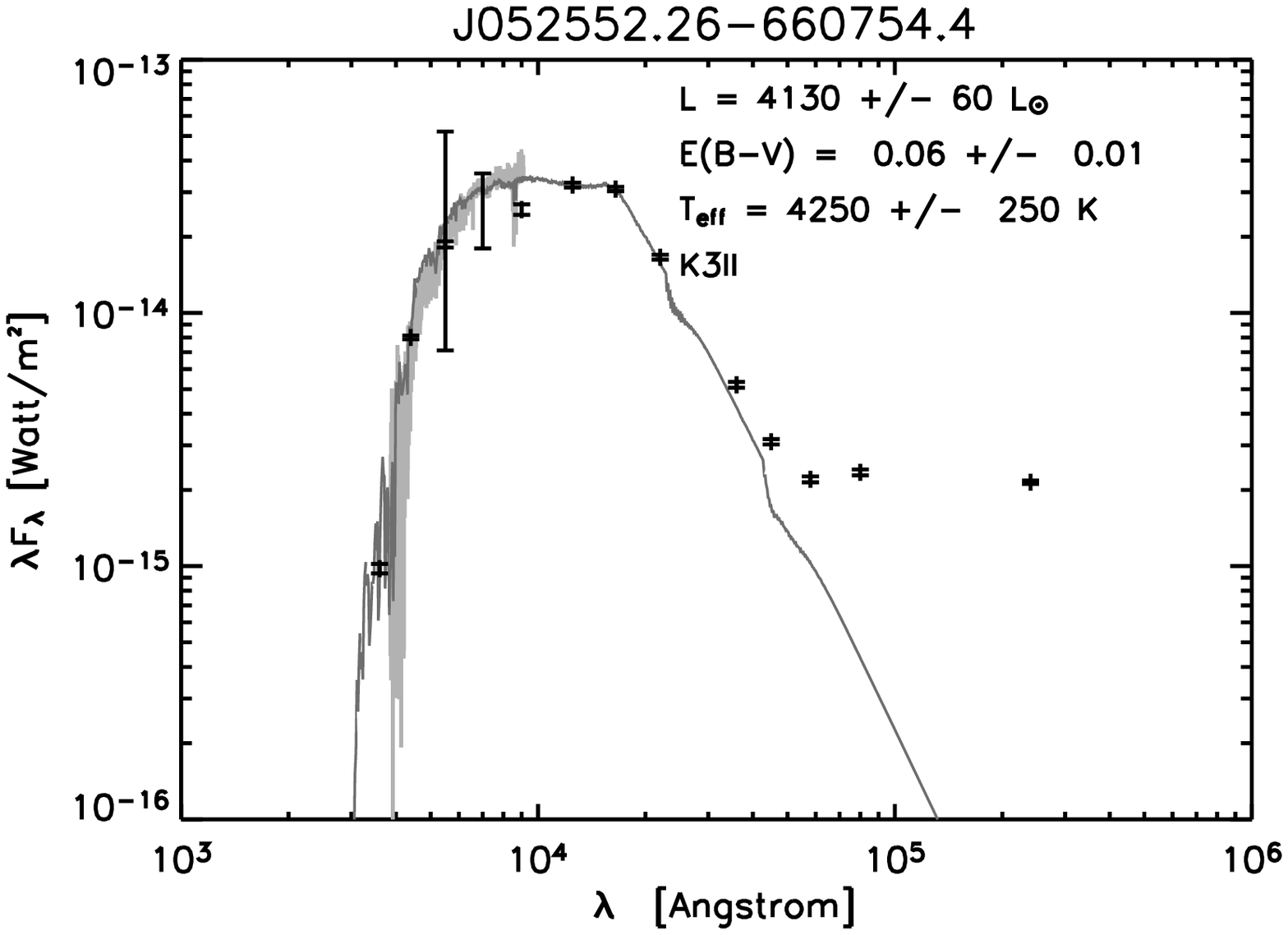}
}
\caption{Same as Fig.~D.1, but for objects of which the SED is inconclusive on whether they have a disc or a shell.}
\label{fig:unclearspec88}
\end{figure*}
}

\onlfig{6}{
\begin{figure*}
\resizebox{\hsize}{!}{
\includegraphics{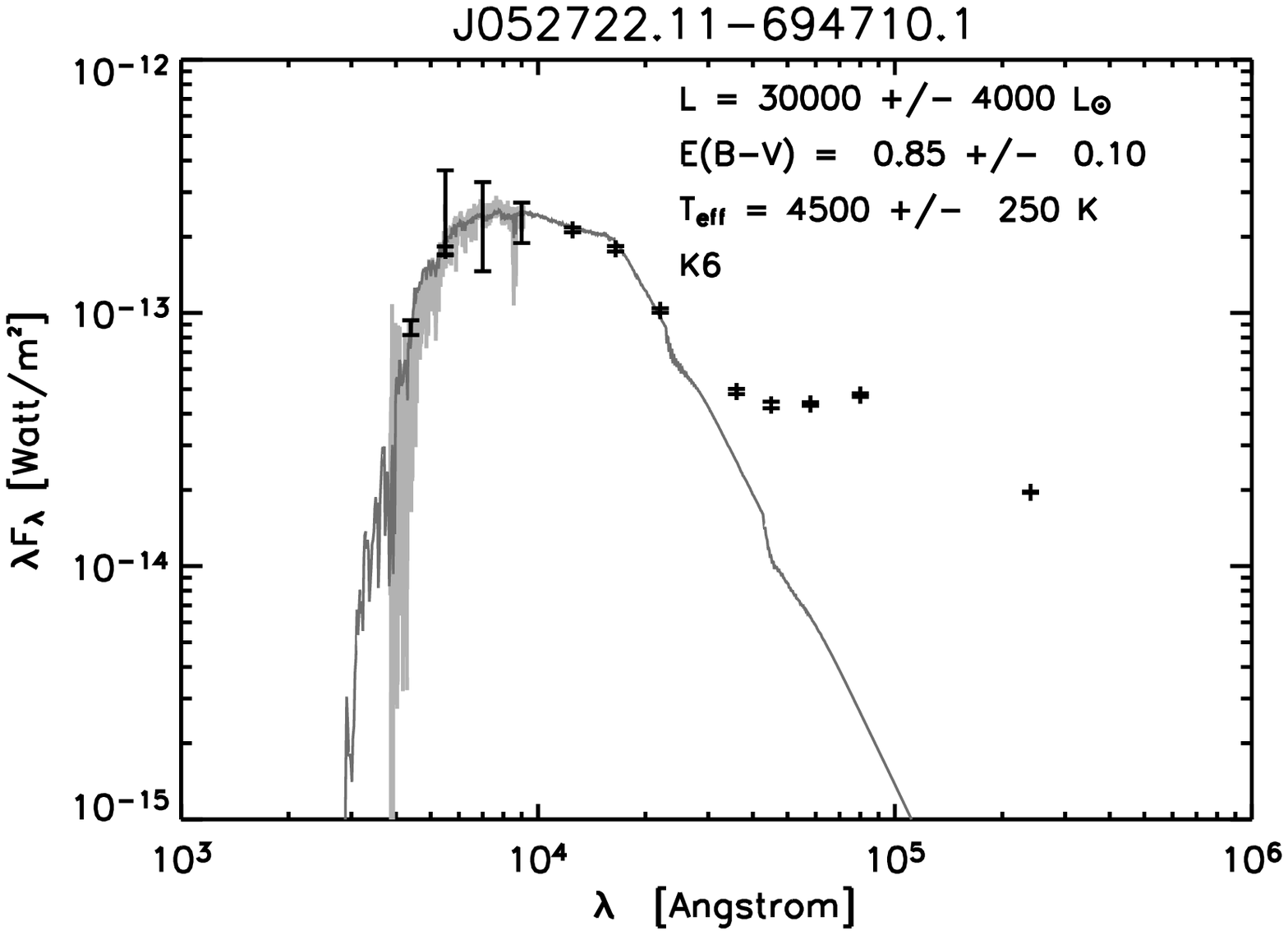}
\includegraphics{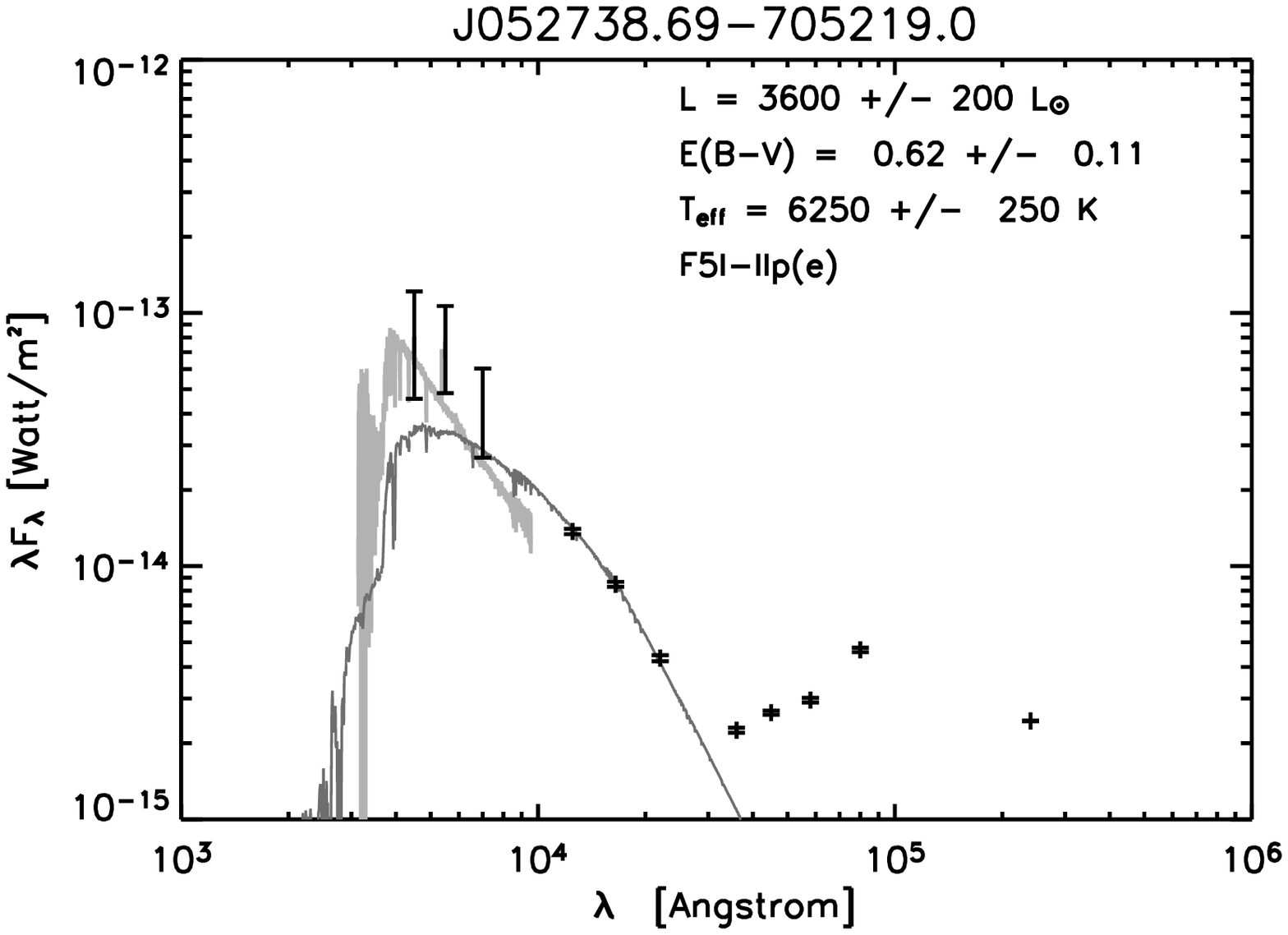}
\includegraphics{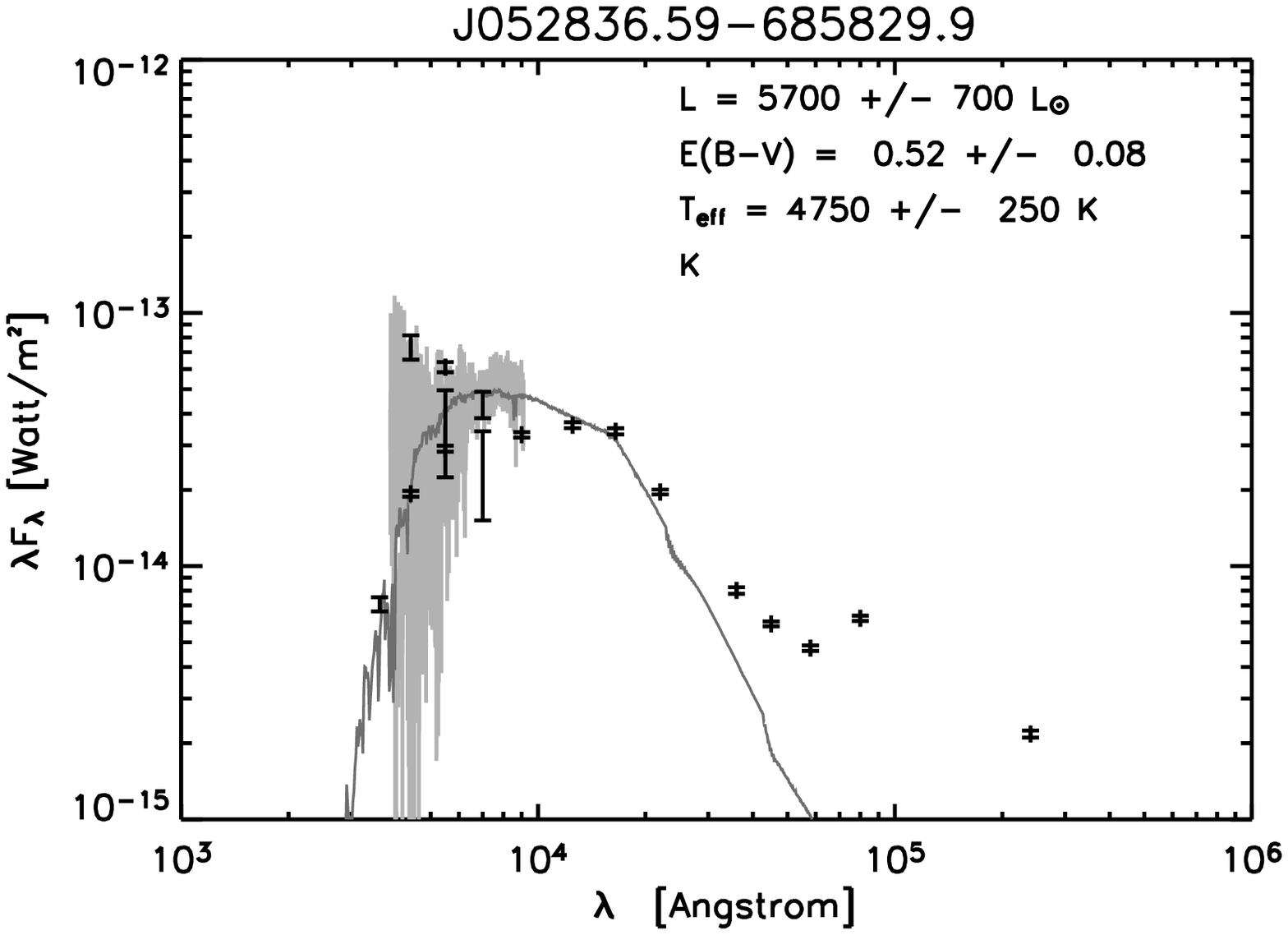}
}
\resizebox{\hsize}{!}{
\includegraphics{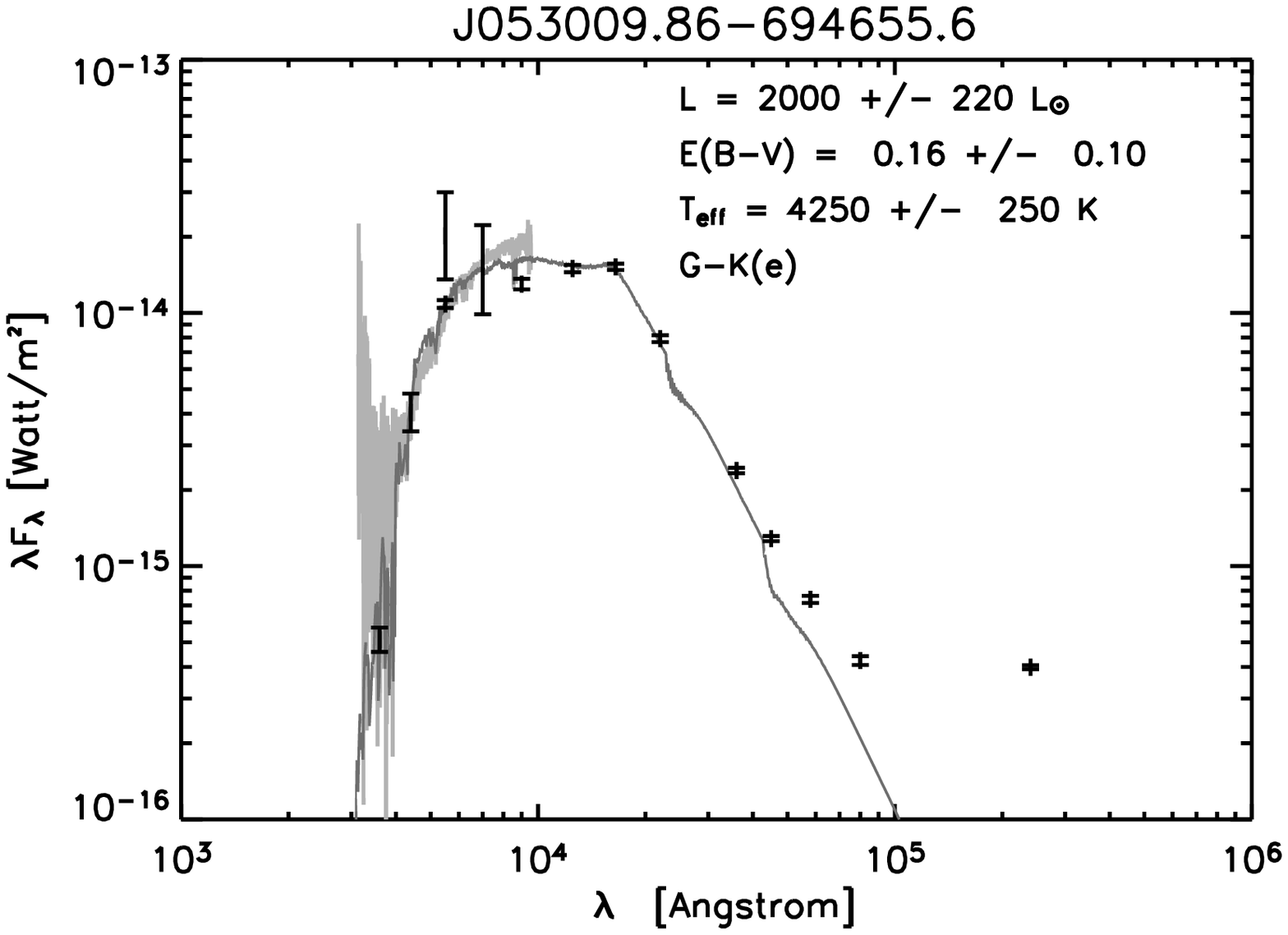}
\includegraphics{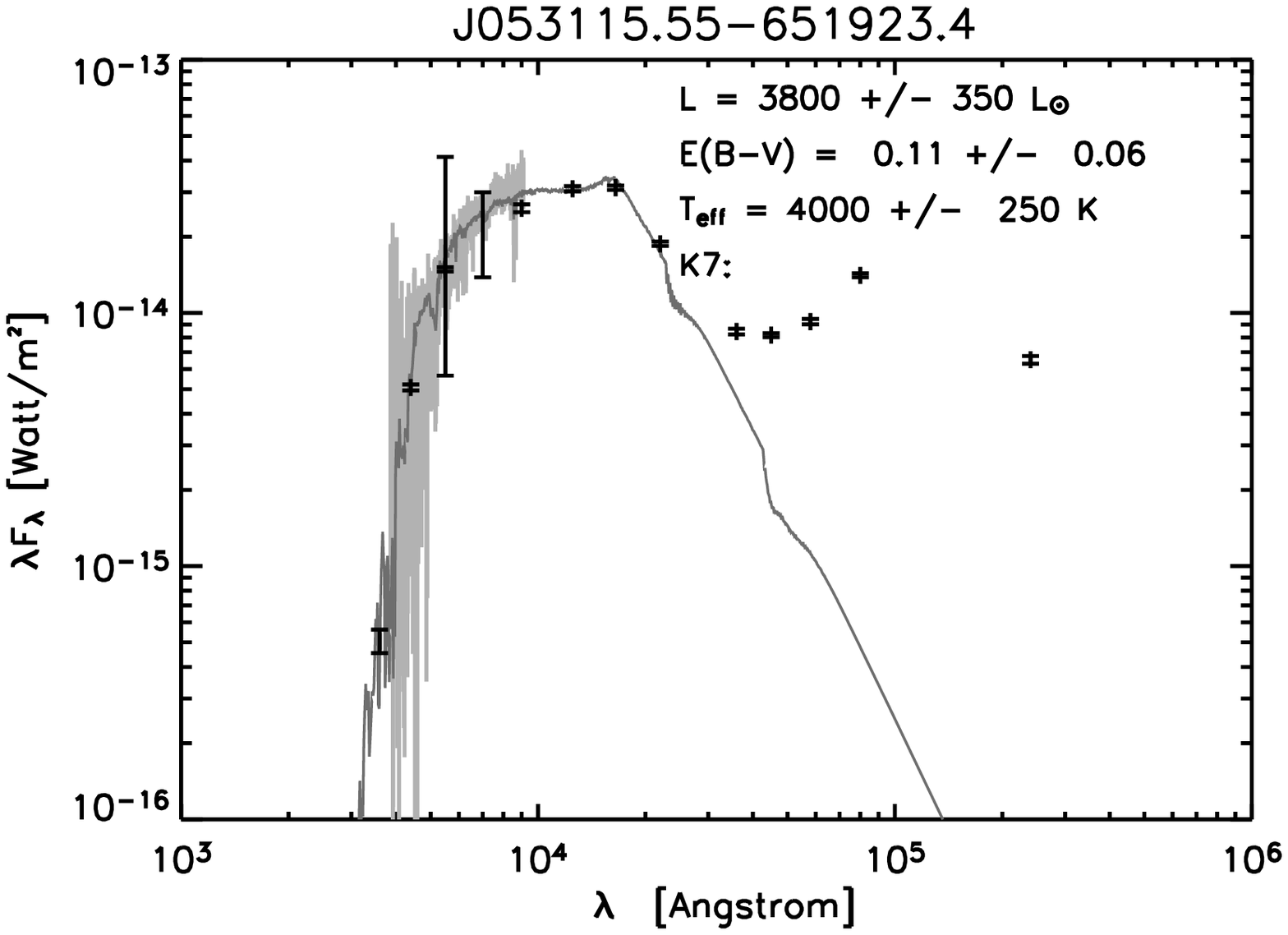}
\includegraphics{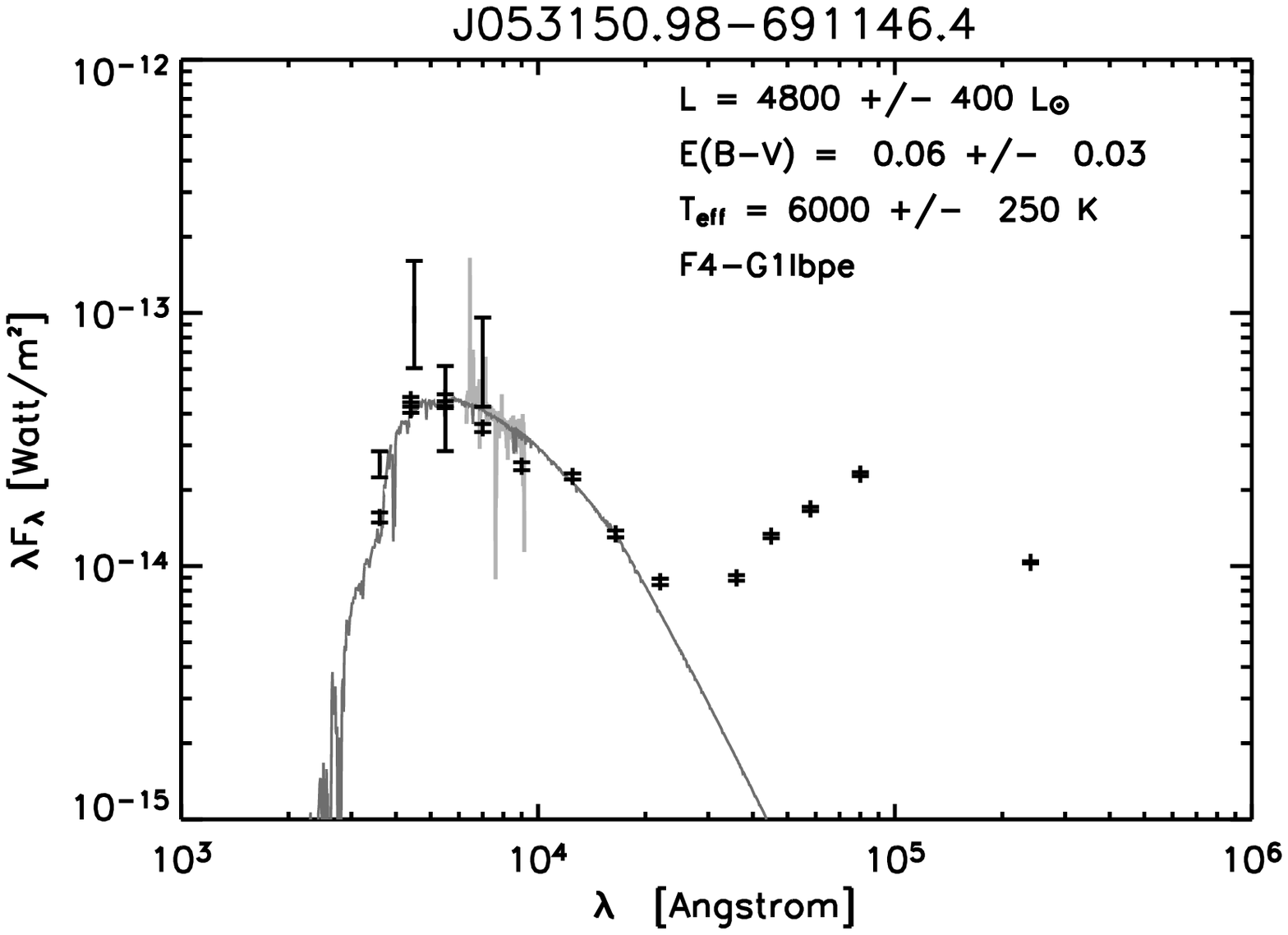}
}
\resizebox{\hsize}{!}{
\includegraphics{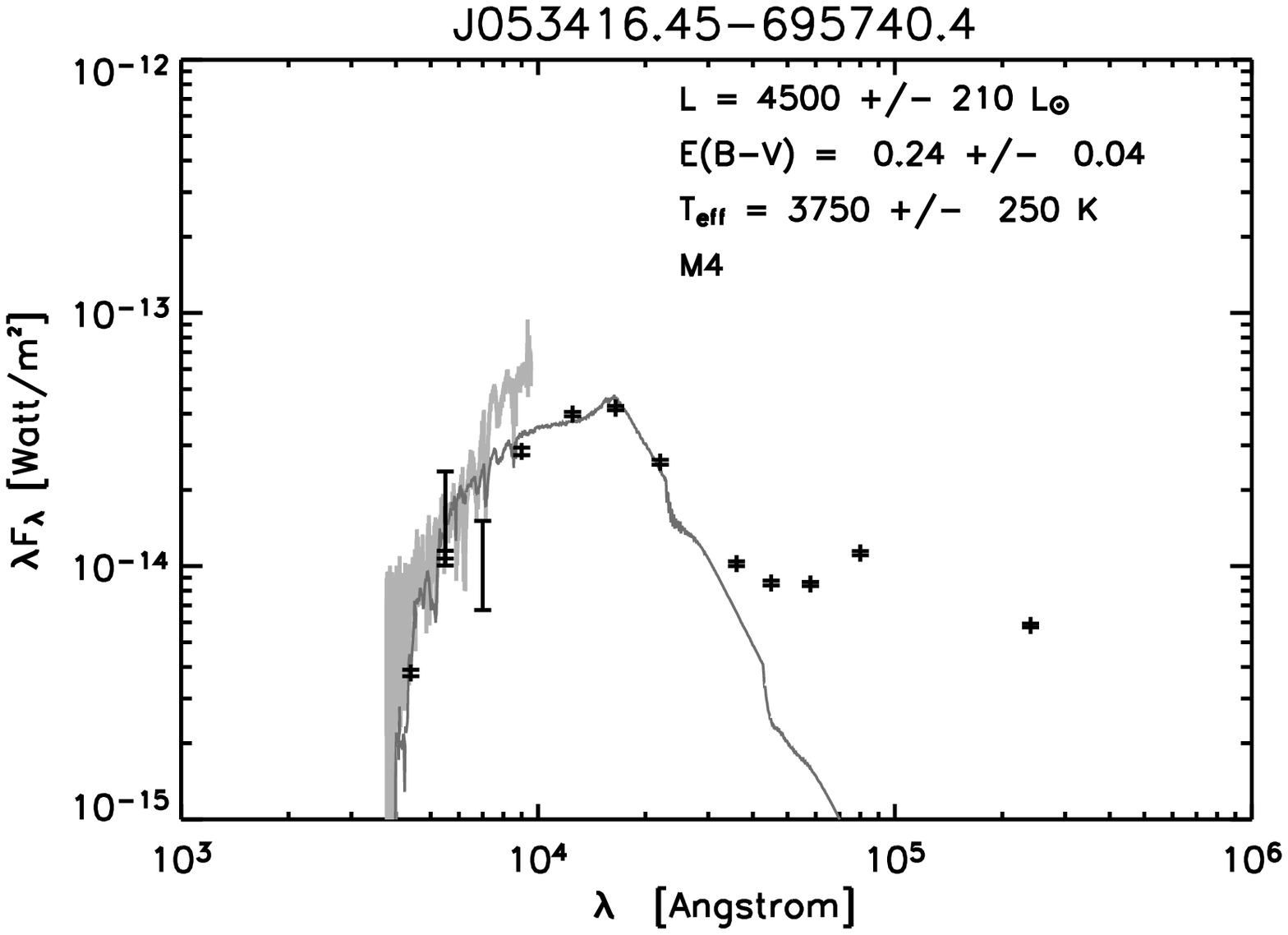}
\includegraphics{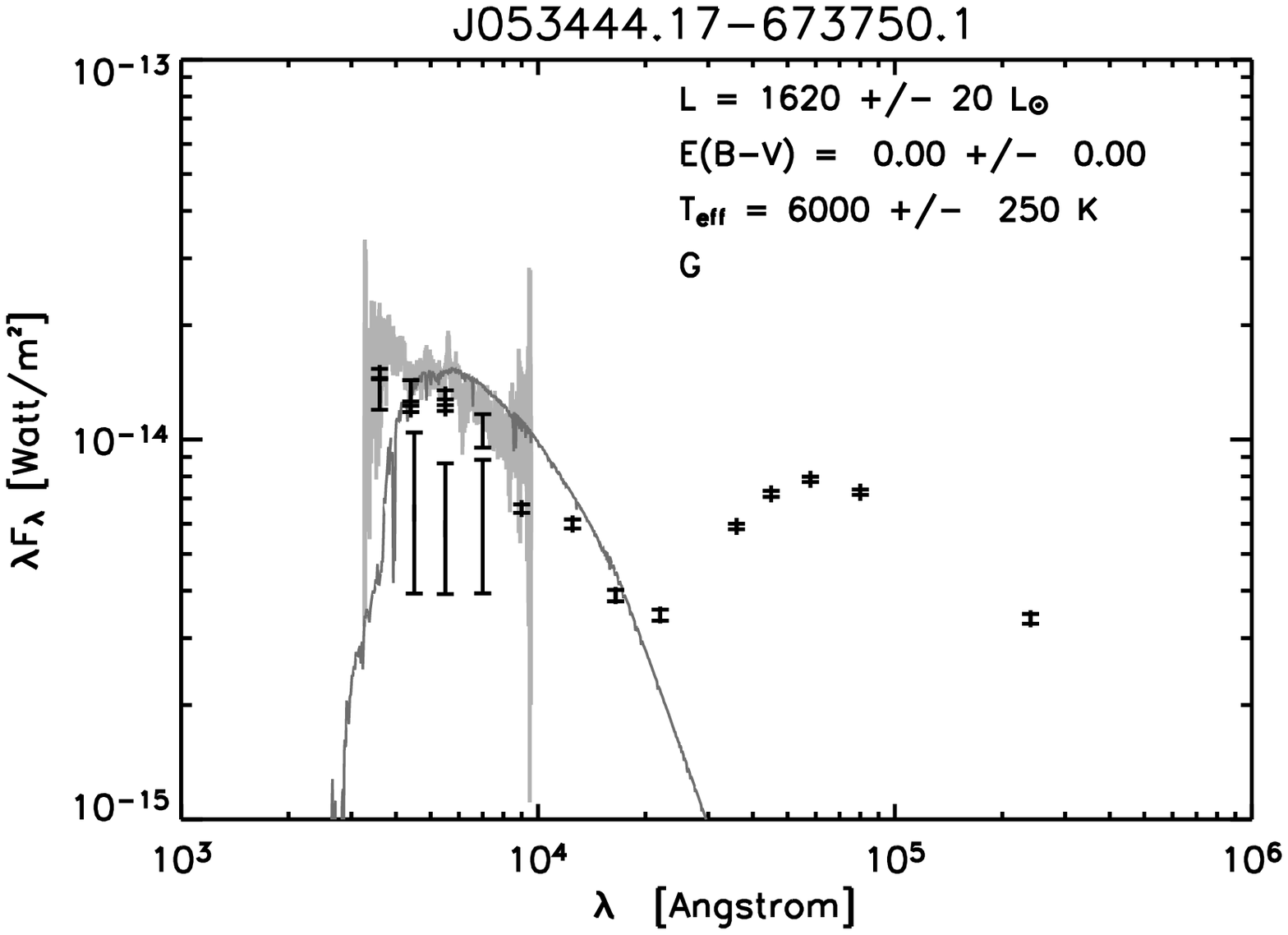}
\includegraphics{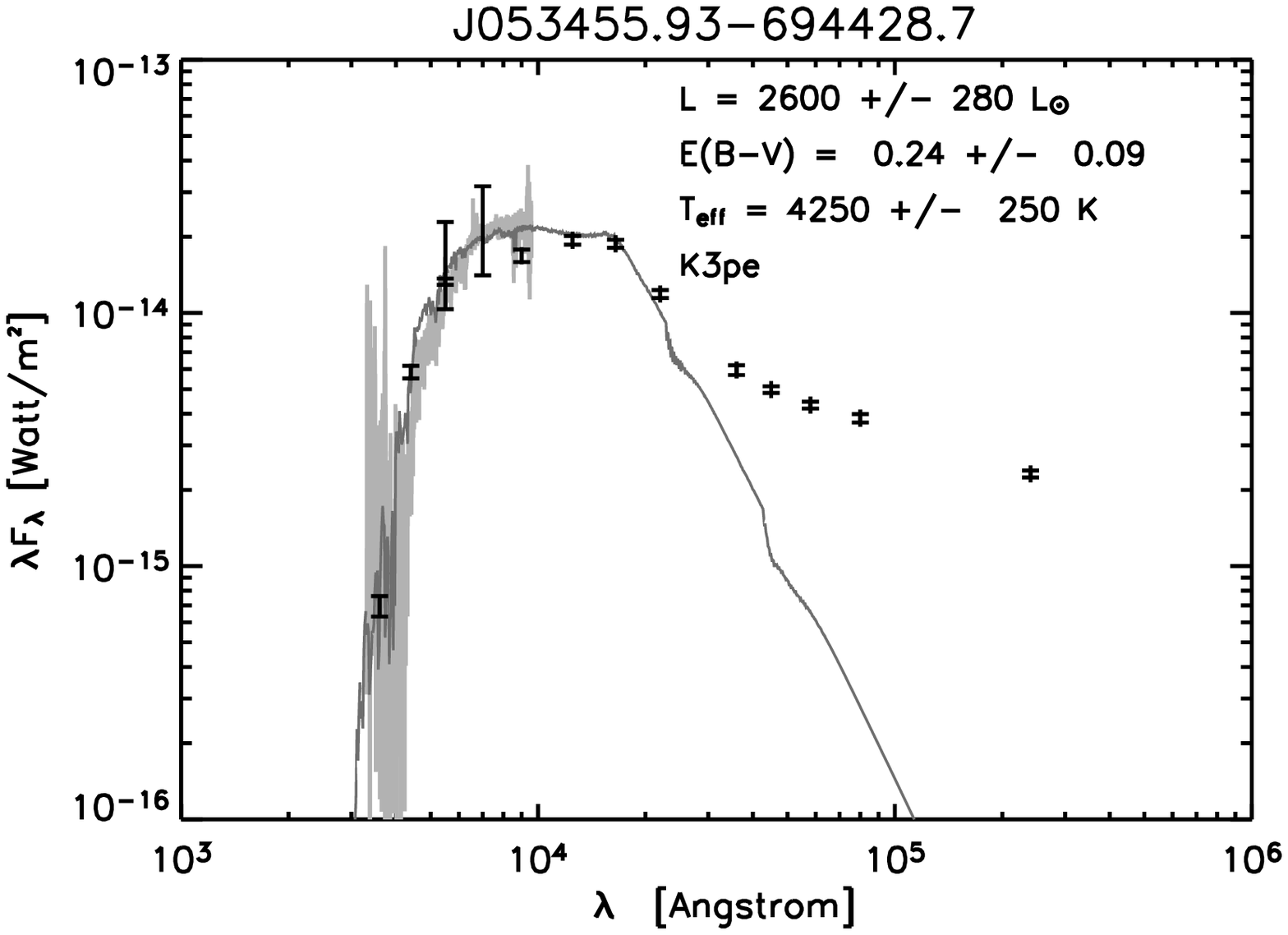}
}
\resizebox{\hsize}{!}{
\includegraphics{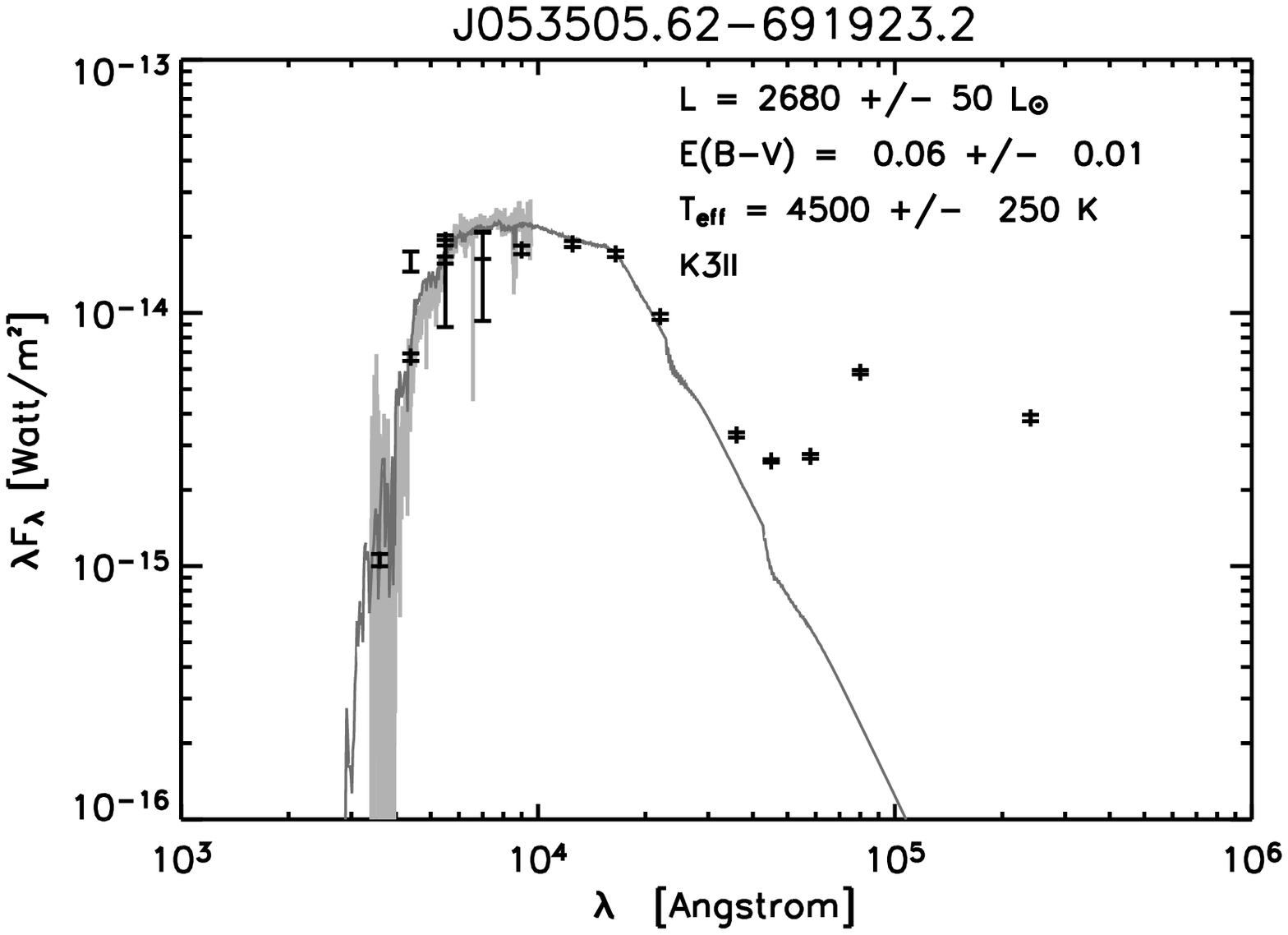}
\includegraphics{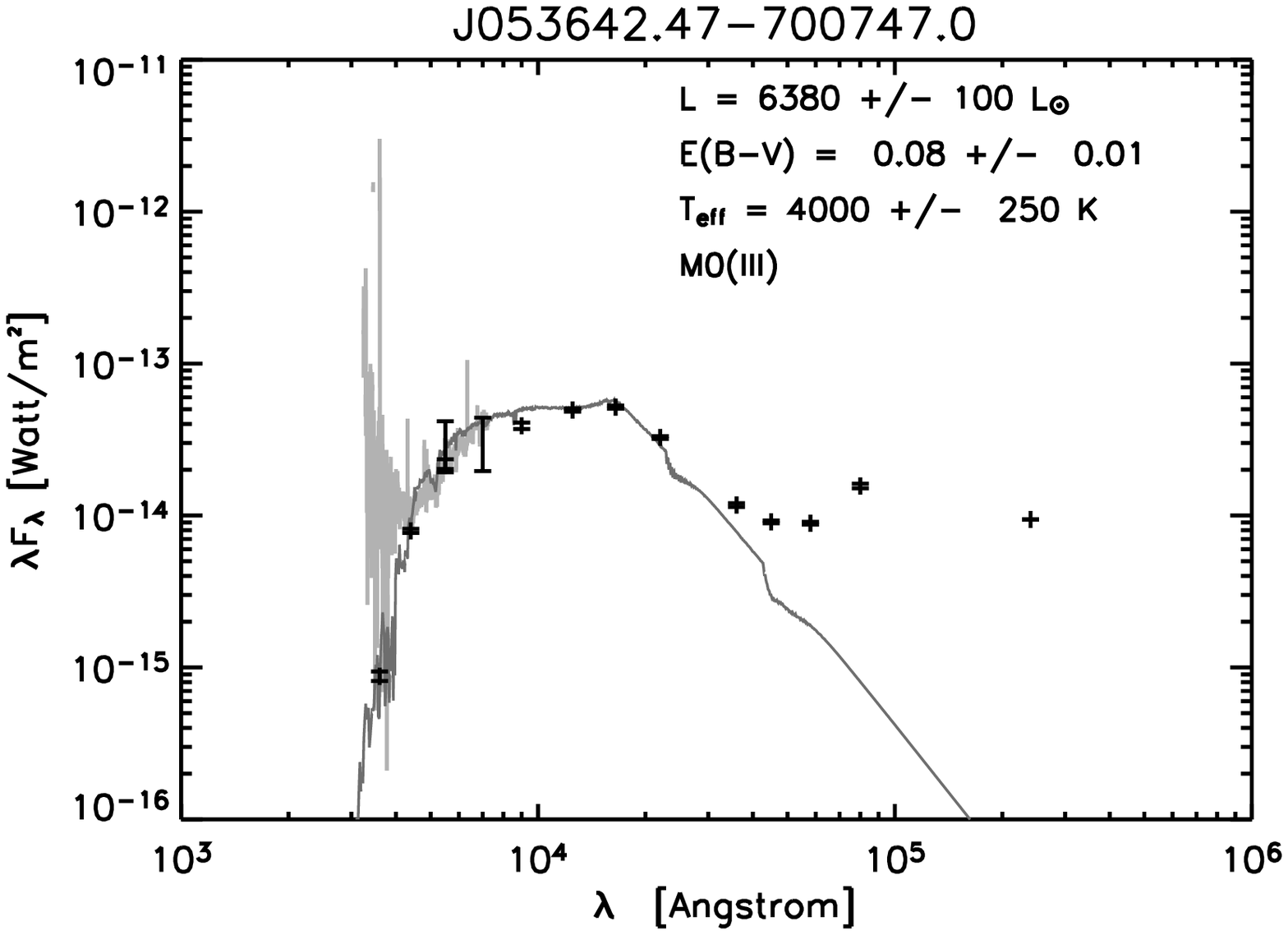}
\includegraphics{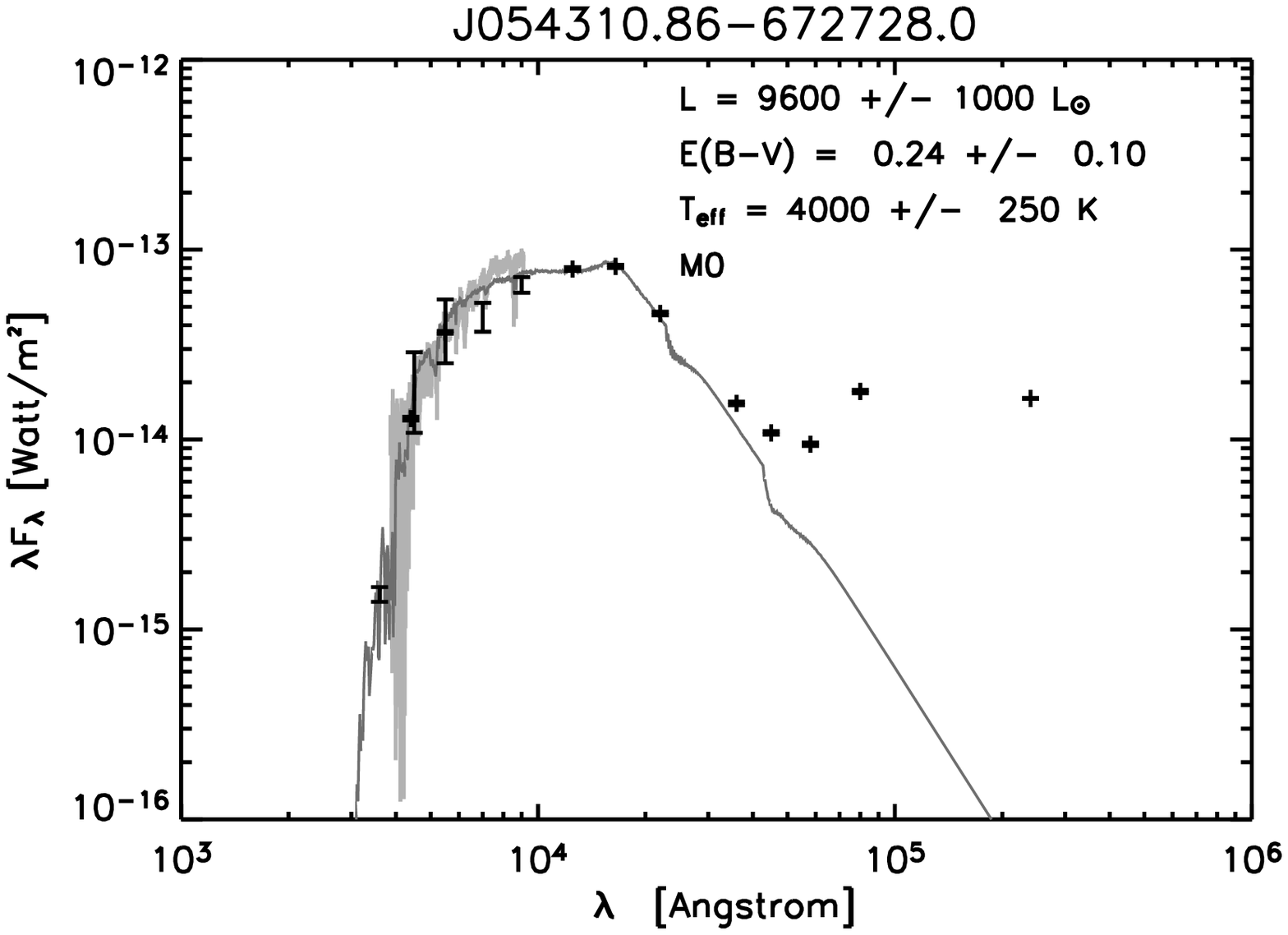}
}
\resizebox{\hsize}{!}{
\includegraphics{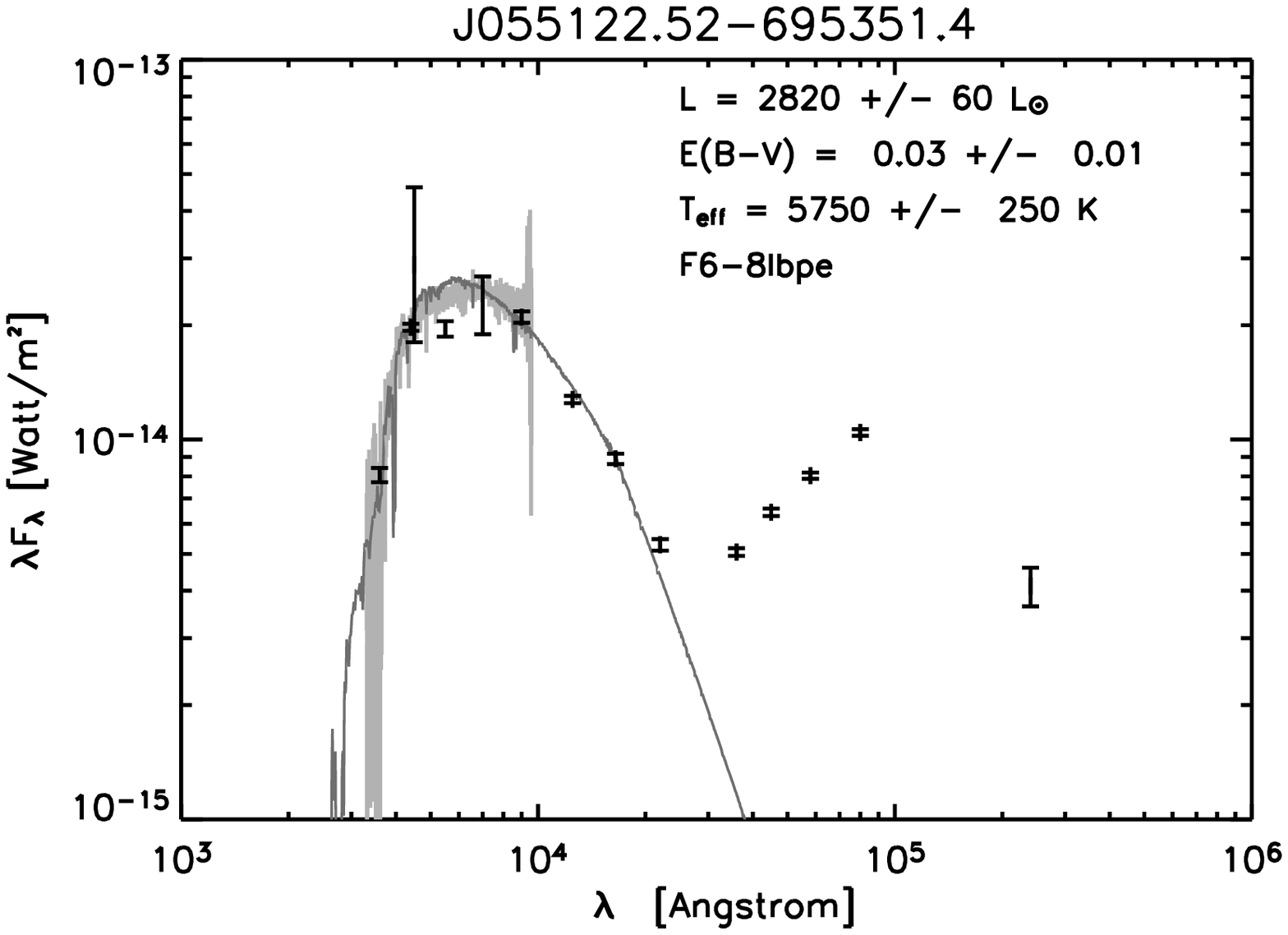}
\includegraphics{15834fgd2e.ps}
\includegraphics{15834fgd2e.ps}
}
\caption{Continuation of Fig.~\ref{fig:unclearspec88}.}
\label{fig:unclearspec89}
\end{figure*}
}

\end{document}